%% file: dps-mat-decomp-v3.tex
\documentclass[10pt,journal]{IEEEtran}

\input{include/preamble}
\input{include/acronyms}

\input{include/macros}

\usepackage{comment}

\title{Material Decomposition in Photon-Counting Computed Tomography with Diffusion Models: Comparative Study and Hybridization with Variational Regularizers }

\author{Corentin Vazia, Thore Dassow, Alexandre Bousse, Jacques Froment, Béatrice Vedel, Franck Vermet, Alessandro Perelli, Jean-Pierre Tasu and Dimitris Visvikis

    \thanks{
        C. Vazia and T. Dassow share first authorship.
    }
    \thanks{
       This work involved human subjects or animals in its research. The author(s) confirm(s) that all human/animal subject research procedures and protocols are exempt from review board approval.
    }
	\thanks{
       This work was supported by the French National Research Agency (ANR) under grant No ANR-20-CE45-0020, by France Life Imaging under grant No ANR-11-INBS-0006,  by CPER 2021--2027 IMAGIIS (INNOV-XS) and by the Royal Academy of Engineering under the RAEng / Leverhulme Trust Research Fellowship LTRF-2324-20-160.
    }
	\thanks{
        C. Vazia, J. Froment and B. Vedel are with Univ. Bretagne Sud, CNRS 6205, LMBA, F-56000 Vannes, France.
    }
    \thanks{
        T. Dassow, A. Bousse and D. Visvikis are with Univ. Brest, LaTIM, Inserm, U1101, 29238~Brest, France.
    }
    \thanks{
        Franck Vermet is with Univ. Brest, CNRS 6205, LMBA, 29000~Brest, France.
    }
    \thanks{
        T. Dassow is also with Siemens Healthcare, SHS EMEA FBL FRA S-SO\&ST SP, 92400 Courbevoie, France.
    }
    \thanks{
        A. Perelli was with the School of Science and Engineering, University of Dundee, DD1 4HN, Dundee, UK. He is now with the School of Cardiovascular and Metabolic Health, University of Glasgow, G12 8TA, Glasgow, UK.
    }
    \thanks{
        J.-P. Tasu is with Department of Radiology, University Hospital Poitiers, Poitiers, France.
    }
    \thanks{Corresponding authors: \texttt{bousse@univ-brest.fr}.}
}

\begin{document}
	
\maketitle

\input{content/abstract}

\input{content/intro}	
\input{./content/background}
\input{./content/diffusion}
\input{./content/results}	
\input{./content/discussion}

\input{./content/conclusion}

\section*{Acknowledgment}

All authors declare that they have no known conflicts of interest in terms of competing financial interests or personal relationships that could have an influence or are relevant to the work reported in this article. 

\AtNextBibliography{\footnotesize} 
\printbibliography

\end{document}

%% file: include/preamble.tex
\usepackage{amsmath,amsfonts,amssymb}
\usepackage[dvipsnames]{xcolor}
\usepackage{breqn}
\usepackage{algorithm}
\usepackage{array}
\usepackage{textcomp}
\usepackage{stfloats}
\usepackage{url}
\usepackage{verbatim}
\usepackage{graphicx}
\usepackage{stmaryrd}
\usepackage{microtype}
\usepackage{datatool}
\usepackage{siunitx}
\usepackage{pgf}
\usepackage{multirow}
\usepackage{breqn}
\breqnsetup{breakdepth={5000}}

\usepackage[mathscr]{euscript}  

\usepackage{bm} 
\usepackage{lipsum}

\usepackage{adjustbox}

\usepackage{acro}[=v2] 
\acsetup{single}
\usepackage{bm}
\usepackage{amsfonts}
\usepackage{soul}

\usepackage{pgfplots}
\usepgfplotslibrary{colorbrewer}
\pgfplotsset{compat = 1.12, cycle list/Set1-8} 
\usetikzlibrary{pgfplots.statistics, pgfplots.colorbrewer} 
\usepackage{pgfplotstable}

\usepackage[
	style=ieee,
	dashed=false,
	maxbibnames=20,
	minbibnames=2,
	maxcitenames=1,
	mincitenames=1,
	backend=biber,
	sorting=none,
	useprefix=false
	]{biblatex}

\usepackage{overpic}

\usetikzlibrary{arrows.meta,shapes,intersections,calc,angles,decorations.pathreplacing,intersections,quotes,angles,positioning,fit,petri,through,shadows,plotmarks,spy}

\usepackage{pgfplotstable}
\usetikzlibrary{pgfplots.groupplots}

\usepackage{algpseudocode}

\usepackage{caption}
\usepackage[labelformat=simple,font=footnotesize]{subcaption}

\usetikzlibrary{datavisualization}

\usetikzlibrary{shapes.geometric, arrows}
\tikzstyle{startstop} = [rectangle, rounded corners, minimum width=3cm, minimum height=1cm,text centered, draw=black, fill=red!30]
\tikzstyle{process} = [rectangle, minimum width=3cm, minimum height=1cm, text centered, draw=black, fill=orange!30]\tikzstyle{arrow} = [thick,->,>=stealth]
\tikzstyle{intesitybox} = [rectangle, rounded corners, minimum height=1cm,text centered]
\tikzstyle{mubox} = [rounded corners, minimum width=2cm, minimum height=1cm, text centered, draw=black, fill=gray!30]
\tikzstyle{plusbox} = [circle, minimum width=1cm, minimum height=1cm, text centered, draw=black, fill=white!30]

\usepackage{hyperref}

\pdfoutput=1

%% file: include/acronyms.tex
\DeclareAcronym{2D}{
	short=2-D,
	long=two-dimensional,
}

\DeclareAcronym{3D}{
	short=3-D,
	long=three-dimensional,
}

\DeclareAcronym{dD}{
	short=$d$-D,
	long=$d$-dimensional,
}

\DeclareAcronym{CT}{
	short=CT, 
	long=computed tomography,
}

\DeclareAcronym{sCT}{
	short=sCT, 
	long=spectral computed tomography,
}

\DeclareAcronym{MD}{
	short=MD, 
	long=material decomposition,
}

\DeclareAcronym{PET}{
	short=PET, 
	long=positron emission tomography,
}

\DeclareAcronym{SPECT}{
	short=SPECT, 
	long=single-photon emission CT,
}

\DeclareAcronym{MRI}{
	short=MRI, 
	long=magnetic resonance imaging,
}

\DeclareAcronym{MR}{
	short=MR, 
	long=magnetic resonance,
}

\DeclareAcronym{PCCT}{
	short=PCCT, 
	long=photon-counting computed tomography,
}

\DeclareAcronym{DECT}{
	short=DECT, 
	long=dual-energy computed tomography,
}

\DeclareAcronym{MBIR}{
	short=MBIR, 
	long=model-based iterative reconstruction,
}

\DeclareAcronym{WLS}{
	short=WLS, 
	long=weighted least squares,
}
\DeclareAcronym{PWLS}{
	short=PWLS, 
	long=penalized weighted least squares,
}

\DeclareAcronym{PLS}{
	short=PLS, 
	long=parallel level set,
}

\DeclareAcronym{PML}{
	short=PML, 
	long=penalized maximum likelihood,
}

\DeclareAcronym{SPS}{
	short=SPS, 
	long=separable paraboloidal surrogates,
}

\DeclareAcronym{CS}{
	short=CS,
	long=compressive sensing,
}

\DeclareAcronym{TV}{
	short=TV,
	long=total variation,
}

\DeclareAcronym{TNV}{
	short=TNV, 
	long=total nuclear variation,
}

\DeclareAcronym{JTV}{
	short=JTV, 
	long=joint total variation,
}

\DeclareAcronym{DTV}{
	short=DTV, 
	long=directional total variation,
}

\DeclareAcronym{SQS}{
	short=SQS, 
	long=separable quadratic surrogate,
}

\DeclareAcronym{ADMM}{
	short=ADMM, 
	long=alternating direction method of multipliers,
}

\DeclareAcronym{ML}{
	short=ML,
	long= machine learning,
}

\DeclareAcronym{DL}{
	short=DL,
	long= deep learning,
}

\DeclareAcronym{DiL}{
	short=DiL,
	long= dictionary learning,
}

\DeclareAcronym{CDL}{
	short=CDL,
	long=convolutional DiL,
}

\DeclareAcronym{MCDL}{
	short=CDL,
	long=multichannel convolutional dictionary learning,
}

\DeclareAcronym{CAOL}{
	short=CAOL, 
	long=convolutional analysis operator learning,
}

\DeclareAcronym{MCAOL}{
	short=MCAOL, 
	long=multichannel convolutional analysis operator learning,
}

\DeclareAcronym{PDF}{
	short=PDF,
	long=probability density function,
}

\DeclareAcronym{PSNR}{
	short=PSNR, 
	long=peak signal-to-noise ratio,
}

\DeclareAcronym{SSIM}{
	short=SSIM, 
	long=structural similarity index measure,
}

\DeclareAcronym{LPIPS}{
    short = LPIPS,
    long = learned perceptual image patch similarity,
    long-plural-form = learned perceptual image patch similarities,
}

\DeclareAcronym{SNR}{
	short=SNR, 
	long=signal-to-noise ratio,
}

\DeclareAcronym{CNN}{
	short=CNN, 
	long=convolutional neural network,
}

\DeclareAcronym{NN}{
	short=NN, 
	long=neural network,
}

\DeclareAcronym{GAN}{
	short=GAN, 
	long=generative adversarial network,
}

\DeclareAcronym{WGAN}{
	short=W-GAN, 
	long=Wasserstein GAN,
}

\DeclareAcronym{VAE}{
	short=VAE, 
	long=variational autoencoder,
}

\DeclareAcronym{MVAE}{
	short=MVAE, 
	long=multi-branch VAE,
}

\DeclareAcronym{beta-VAE}{
	short=$\beta$-VAE, 
	long=beta-variational autoencoder,
}

\DeclareAcronym{LoR}{
	short=LoR,
	long=line of response,
	long-plural-form = lines of response,
}

\DeclareAcronym{EM}{
	short=EM,
	long=expectation-maximization,
}

\DeclareAcronym{MLEM}{
	short=MLEM,
	long=maximum-likelihood expectation-maximization,
}

\DeclareAcronym{OSEM}{
	short=OSEM, 
	long=ordered subsets expectation maximization,
}

\DeclareAcronym{MAPEM}{
	short=MAPEM, 
	long=maximum \textit{a posteriori} expectation maximization,
}

\DeclareAcronym{FBP}{
	short=FBP, 
	long=filtered backprojection,
}

\DeclareAcronym{IFFT}{
	short=IFFT, 
	long=inverse fast Fourier transform,
}

\DeclareAcronym{GT}{
	short=GT,
	long=ground truth,
}

\DeclareAcronym{HU}{
	short=HU,
	long=Houndsfield Units,
}

\DeclareAcronym{LAC}{
	short=LAC,
	long=linear attenuation coefficients,
}

\DeclareAcronym{MNIST}{
	short=MNIST,
	long=Modified National Institute of Standards and Technology,
}

\DeclareAcronym{LBFGS}{
	short=L-BFGS,
	long=limited-memory  Broyden-Fletcher-Goldfarb-Shanno,
}

\DeclareAcronym{KL}{
	short=KL,
	long=Kullback-Leibler
}

\DeclareAcronym{ReLU}{
	short=RelU,
	long=rectified linear unit
}

\DeclareAcronym{PSO}{
	short=PSO,
	long=particle swarm optimization
}

\DeclareAcronym{AI}{
	short=AI,
	long=artificial intelligence
}

\DeclareAcronym{DM}{
	short=DM,
	long=diffusion model
}

\DeclareAcronym{DPS}{
	short=DPS,
	long=diffusion posterior sampling
}

\DeclareAcronym{DDPM}{
	short=DDPM,
	long=denoising diffusion probabilistic model
}

\DeclareAcronym{DDIM}{
	short=DDIM,
	long=denoising diffusion implicit model
}

\DeclareAcronym{PCA}{
	short=PCA,
	long=principal component analysis 
}
\DeclareAcronym{MSE}{
	short=MSE,
	long=mean squared error 
}

\DeclareAcronym{LH}{
	short=LC-PET/HC-CT,
	long=low-count PET/high-count CT, 
}

\DeclareAcronym{HL}{
	short=HC-PET/LC-CT,
	long=high-count PET/low-count CT, 
}

\DeclareAcronym{VPET}{
	short=MVAE-PET,
	long=MVAE-reconstructed PET, 
}

\DeclareAcronym{VCT}{
	short=MVAE-CT,
	long=MVAE-reconstructed CT, 
}

\DeclareAcronym{PPET}{
	short=PLS-PET,
	long=PLS-reconstructed PET, 
}

\DeclareAcronym{PCT}{
	short=PLS-CT,
	long=PLS-reconstructed CT, 
}

\DeclareAcronym{EMPET}{
	short=MLEM-PET,
	long=MLEM-reconstructed PET, 
}

\DeclareAcronym{WCT}{
	short=WLS-CT,
	long=WLS-reconstructed CT, 
}

\DeclareAcronym{NIST}{
	short = NIST,
	long = National Institute of Standards and Technology}

\DeclareAcronym{SDE}{
	short = SDE,
	long = stochastic differential equation}

\DeclareAcronym{ODPS}{
	short = ODPS,
	long = one-step DPS}
	
\DeclareAcronym{TDPS}{
	short = TDPS,
	long = two-step diffusion posterior sampling}

\DeclareAcronym{ImTDPS}{
	short = im-TDPS,
	long = image-domain two-step DPS}

\DeclareAcronym{ProjTDPS}{
	short = proj-TDPS,
	long = projection-domain two-step DPS}

\DeclareAcronym{MAP}{
	short = MAP,
	long = maximization \emph{a posteriori}}	
	
\DeclareAcronym{OT}{
		short = OT,
		long = optimization transfer}

\DeclareAcronym{TDiL}{
	short=TDiL, 
	long=tensor DiL,
}

\DeclareAcronym{CDiL}{
	short=CDiL, 
	long=convolutional DiL,
}

\DeclareAcronym{VMI}{
	short=VMI, 
	long=virtual monoenergetic image,
}

%% file: include/macros.tex
\newcommand{\boldb}{\mathbf{b}}
\newcommand{\boldc}{\mathbf{c}}

\newcommand{\boldu}{\mathbf{u}}

\newcommand{\boldw}{\mathbf{w}}
\newcommand{\boldx}{\mathbf{x}}
\newcommand{\boldy}{\mathbf{y}}
\newcommand{\boldz}{\mathbf{z}}

\newcommand{\boldI}{\mathbf{I}}

\newcommand{\boldL}{\mathbf{L}}

\newcommand{\boldU}{\mathbf{U}}

\newcommand{\boldW}{\mathbf{W}}
\newcommand{\boldX}{\mathbf{X}}
\newcommand{\boldY}{\mathbf{Y}}
\newcommand{\boldZ}{\mathbf{Z}}

\newcommand{\calF}{\mathcal{F}}

\newcommand{\calN}{\mathcal{N}}

\newcommand{\boldcalA}{\bm{\mathcal{A}}}

\newcommand{\boldcalF}{\bm{\mathcal{F}}}

\newcommand{\boldcalS}{\bm{\mathcal{S}}}

\newcommand{\rmd}{\mathrm{d}}
\newcommand{\rme}{\mathrm{e}}

\newcommand{\R}{\mathbb{R}}
\newcommand{\transp}{^\top}

\newcommand{\ybar}{\bar{y}}

\newcommand{\boldwhat}{\hat{\boldw}}

\newcommand{\ptilde}{\tilde{p}}

\newcommand{\ycheck}{\check{y}}

\newcommand{\boldell}{\bm{\ell}}
\newcommand{\boldellhat}{\hat{\boldell}}

\renewcommand{\hbar}{\bar{h}}

\newcommand{\alphabar}{\bar{\alpha}}

\newcommand{\boldxhat}{\hat{\boldx}}

\newcommand{\boldzhat}{\hat{\boldz}}
\newcommand{\bolduhat}{\hat{\boldu}}

\newcommand{\boldztilde}{\tilde{\boldz}}
\newcommand{\boldZtilde}{\tilde{\boldZ}}

\newcommand{\boldtheta}{\bm{\theta}}

\newcommand{\Np}{n_{\mathrm{p}}}
\newcommand{\Nb}{n_{\mathrm{b}}}
\newcommand{\Ne}{n_{\mathrm{e}}}
\newcommand{\Nm}{n_{\mathrm{m}}}
\newcommand{\Nmtilde}{\tilde{n}_{m}}

\newcommand{\boldeps}{\bm{\epsilon}}

\newcommand{\argmax}{\operatornamewithlimits{arg\,max}}

\newcommand{\boldzero}{\bm{0}}

\algnewcommand{\Inputs}[1]{%
	\State \textbf{Inputs:}
	\Statex \hspace*{\algorithmicindent}\parbox[t]{.8\linewidth}{\raggedright #1}
}
\algnewcommand{\Initialize}[1]{%
	\State \textbf{Initialize:}
	\Statex \hspace*{\algorithmicindent}\parbox[t]{.8\linewidth}{\raggedright #1}
}

\newlength{\tempdima}
\newcommand{\rowname}[1]
{\rotatebox{90}{\makebox[\tempdima][c]{\scriptsize#1}}}

%% file: content/abstract.tex
\begin{abstract}
	\Ac{PCCT} has emerged as a promising imaging technique, enabling spectral imaging and \ac{MD}. However,  images typically suffer from a low \ac{SNR} due to constraints such as low photon counts and sparse-view settings which provoke artifacts. To prevent this, variational methods minimize a data-fit function coupled with handcrafted regularizers that mimic a prior by enforcing image properties such as gradient sparsity. In the last few years, \acp{DM} have become predominant in the field of generative models and have been used as a learned prior for image reconstruction. This work investigates the use of \acp{DM} as regularizers for \ac{MD} tasks in \ac{PCCT}, specifically using \ac{DPS} guidance. Three \ac{DPS}-based approaches---\ac{ImTDPS}, \ac{ProjTDPS}, and \ac{ODPS}---are evaluated. The first two methods achieve \ac{MD} in two steps by performing reconstruction and \ac{MD} separately. The last method, \ac{ODPS},  samples the material images directly from the measurement data. The results indicate that \ac{ODPS} achieves superior performance compared to \ac{ImTDPS} and \ac{ProjTDPS}, providing sharper, noise-free and crosstalk-free images. Furthermore, we introduce a novel hybrid method for scenarios involving materials absent from the training dataset which combines \ac{DM} priors with standard variational handcrafted regularizers for the materials unknown to the \ac{DM}.  This hybrid method demonstrates improved \ac{MD} quality compared to a standard variational method and does not require additional training of the \ac{DM} \ac{NN}.
\end{abstract}

\begin{IEEEkeywords}
	Diffusion Posterior Sampling, Photon-Counting Computed Tomography, Material Decomposition
\end{IEEEkeywords}

%% file: content/intro.tex
\section{Introduction}

\IEEEPARstart{X}{-ray} \ac{CT}, primarily used for medical applications, produces monochromatic \ac{LAC} images without considering the full energy spectrum of the crossing beam. However, in the last decade, \ac{PCCT}~\cite{Taguchi_PCCT}, and more generally spectral \ac{CT}, have enabled the consideration of the energy of the X-ray spectrum~\cite{bousse2023systematic}. This advancement allows for polychromatic imaging with energy-dependent \ac{LAC}. However, in spectral \ac{CT}, measurements are grouped into energy bins that have a lower source intensity compared to classical \ac{CT}. Furthermore, due to public health concerns, efforts must be made to reduce radiation exposure ``as low as reasonably achievable''. This can be accomplished either by reducing the photon count per projection (low count) or by lowering the number of projection angles (sparse view). This leads to a low \ac{SNR} in each energy bin and potential streak artifacts in the reconstructed images, and therefore spectral \ac{CT} image reconstruction requires regularization. Synergistic regularization is a way to leverage the structural similarities between the spectral images. Such regularizations include handcrafted regularizers~\cite{Rigie_2015} as well as learned regularizers~\cite{Wu2018,perelli2022multi}.

Using the energy dependence of the \acp{LAC}, \ac{MD} can be performed~\cite{Alvarez1976}. This approach enables the generation of material-specific maps, such as bones, soft tissues, or iodine tracers. \Ac{MD} provides insight into atherosclerotic plaque composition~\cite{cormode2010atherosclerotic}, classifies renal stones~\cite{primak2007noninvasive, graser2008dual, boll2009renal}, improves lesion detection and characterization~\cite{silva2011dual}, and offers a wide range of clinical applications~\cite{mccollough2015dual}. However, \ac{MD} is contingent on image reconstruction and therefore requires strong regularization. This can be achieved through a \ac{MAP} formulation with a handcrafted or a learned prior. Methods addressing the \ac{MD} problem typically fall into one of two categories: (i) one-step methods, where reconstruction and decomposition are performed simultaneously~\cite{long2014multi, Mory2018}, making the problem non-convex and highly ill-posed; or (ii) two-step methods~\cite{Wu2021, abascal, NDucros, Xu2021}, where each subproblem is solved sequentially. Section~\ref{sec:MD formulation} provides an overview of the main approaches to \ac{MD} formulation.

In recent years, \ac{DL}-based approaches have become state-of-the-art for image and text generation~\cite{ho2020denoising, song2020generative, song2021scorebased} and for solving inverse problems such as image reconstruction~\cite{arridge2019solving}, and more particularly in spectral \ac{CT}~\cite{chen2023soul,bousse2023systematic,wang2024uconnect}. Among them, \acp{DM} have shown promising results~\cite{song2022solving, chung2022improving, MCGdiff, chung2023diffusion}, particularly in medical image reconstruction~\cite{webber2024diffusion}. In high dimensions, using \ac{DPS} is usually computationally more feasible than other \ac{DM}-based methods. In a recent work, we proposed to use \acp{DM} through \ac{DPS} as a synergistic regularizer for spectral \ac{CT} image reconstruction~\cite{vazia2024spectral}. More recently, \citeauthor{jiang2025multi}~\cite{jiang2025multi} used a \ac{DPS} framework for one-step \ac{MD} in dual-energy \ac{CT}. At the same time, we independently proposed a similar one-step \ac{MD} framework for \ac{PCCT}~\cite{vazia2024material}.

In this paper, we investigate multiple ways of incorporating diffusion priors into the \ac{MD} problem. We evaluate three \ac{DPS}-based \ac{MD} approaches in a \ac{PCCT} setting, including two two-step \ac{DPS} approaches, namely \ac{ImTDPS} and \ac{ProjTDPS}, and an \ac{ODPS} approach. Dual material/attenuation image \ac{DM}-based priors were not considered as they are too computationally expensive. In addition, we only considered image-domain priors (either on material or spectral images) so that the \acp{NN} are not impacted by potential changes in the projection geometry such as the number of projection angles. We compared these methods with other state-of-the-art techniques in two low-dose settings (full-view and sparse-view).

In addition, we propose a new method consisting of mixing the diffusion prior with standard variational regularizers, which can be utilized in situations where some of the material maps are not present in the training database. For such materials, a pseudo-prior, derived from the variational regularization of inverse problems, is combined with the \ac{DM}-learned prior. This variational regularizer is chosen independently of the \ac{DM} and incorporated in a single line of the algorithm. Therefore, this hybrid heuristic retains the powerful regularization of the \ac{DM} while incorporating new materials such as iodine, providing a potential solution for real-world clinical and research scenarios and avoiding retraining.

The remainder of the paper is structured as follows. We first present the \ac{MD} problem in \ac{PCCT} in Section~\ref{sec:MD}, including the two-step and one-step paradigms. Then, Section~\ref{sec:DPS for MD} reviews the \ac{DPS} method to regularize this inverse problem and introduces our hybridization method with variational regularizers. Section~\ref{sec:results} shows the data used for the experiments. After a description of the methods used for comparison, we present and discuss the obtained results. Section~\ref{sec:discussion} discusses the proposed methods and presents perspectives for future work. Finally, Section~\ref{sec:conclusion} concludes this work.

\subsection*{Notations}

The superscripts `$^\transp$' and `$^\dag$' respectively denote the matrix transposition and pseudoinverse inverse.

Given a real-valued matrix $\boldc = \{c_{n,m}\}_{n,m=1}^{N,M} \in \R^{N\times M}$, $\boldc_{n,:}$ and $\boldc_{:,m}$ respectively denote the $n$-th row and $m$-th column of $\boldc$, i.e.,
\begin{equation*}
	\boldc_{n,:} = [c_{n,1},\dots,c_{n,M}] \, , \quad \boldc_{:,m} = [c_{1,m},\dots,c_{N,m}]\transp \, .
\end{equation*}
For a given real-valued vector $\boldc = \{c_n\}_{n=1}^{N} \in \R^{N}$, $[\boldc]_n$ denotes the $n$-th entry of $\boldc$, i.e., $[\boldc]_n = c_n$. The zero element and the identity matrices (with dimensions given by the context) are denoted $\boldzero$ and $\boldI$, respectively. Capital letters represent random vectors, and their lowercase equivalents represent their realizations. For example, for a given pair $(\boldX,\boldY)$ of random vectors, $p_{\boldX | \boldY = \boldy}(\boldx)$ represents the conditional \ac{PDF} of $\boldX$ evaluated at $\boldx$ given $\boldY = \boldy$. Bold calligraphic capital letters represent vector-valued operators.

%% file: content/background.tex
\section{Background on Material Decomposition} \label{sec:MD}

This section presents the forward measurement model and the three main (non \ac{DL}-based) approaches for material decomposition.

\subsection{Measurement Model} \label{sec:measurementmodel}

We assume that the imaged object is a vector sampled on a grid composed of $\Np$ pixels. The X-ray attenuation image takes the form of an energy-dependent random column vector $\boldU(e)$,
\begin{equation*}
	\boldU(e) = [U_1(e),\dots,U_\Np(e)]\transp
\end{equation*}
where $U_j(e)$ is the \ac{LAC} at pixel location $j$ for an incoming beam of energy $e \in \mathbb{R}_+^*$. 

The attenuation $\boldU$ is determined by the $\Nm$ materials that compose the object. For all pixels $j=1,\dots,\Np$ and materials $m=1,\dots,\Nm$, let us denote by $Z_{j,m}$ the concentration of material $m$ in pixel $j$, and by $\boldZ = \{Z_{j,m}\}_{j,m=1}^{\Np,\Nm}$ the $\Nm$-channel material image, which takes the form of a $\Np\times \Nm$ random matrix; $\boldZ_{:,m}$ corresponds to the $m$-th material image, and $\boldZ_{j,:}$ is the vector of materials in pixel $j$.

The relationship between $U_j$ and $\boldZ_{j,:}$ is given by the energy-dependent mapping $\calF(\cdot,e) \colon \R^\Nm \to \R$ as
\begin{align}
	U_j(e) = {} & \sum_{m=1}^\Nm f_m(e) Z_{j,m} \nonumber \\
	:= {} & \calF( \boldZ_{j,:},e )   \label{eq:F} \nonumber
\end{align}
where $f_m(e)$ is the known X-ray mass attenuation coefficient of material $m$ at energy $e$. Furthermore, $\calF$ can be generalized to a vector-valued mapping $\boldcalF$ of the entire material image $\boldZ$ and the energy $e$ as 
\begin{align}
	\boldcalF( \boldZ,e ) := {} & [\calF( \boldZ_{1,:},e ),\dots,\calF( \boldZ_{\Np,:},e ) ]\transp  \nonumber   \\
	= {} & \boldU(e)   \, .   \nonumber 
\end{align}

In \ac{PCCT}, the measurement represents the number of X-ray photons detected along each beam, categorized into energy bins. Let $\Nb = n_{\theta} \cdot n_{\rmd}$ denote the total number of beams, where $n_{\theta}$ is the number of projection angles, and $n_{\rmd}$ is the number of detectors. Furthermore, let $\Ne$ represent the number of energy bins, with each bin corresponding to an energy interval $[e_{k-1}, e_k[$, $k=1,\dots,\Ne$. Given a realization $\boldz \in \R^{\Np\times \Nm}$ of $\boldZ$, the number of detected photons along the $i$-th ray, $i=1,\dots,\Nb$, and in energy bin $k$ is modeled by a random variable $Y_{i,k}$ with conditional distribution
\begin{equation}\label{eq:poisson}
	\left( Y_{i,k} \mid \boldZ  = \boldz \right) \sim \mathrm{Poisson} \left( \ybar_{i,k} (\boldz)  \right)\, .
\end{equation} 
The expected number of counts $\ybar_{i,k} (\boldz)$ is given by applying the Beer--Lambert law to a realization of $\boldU(e)$ with $\boldZ = \boldz$ and integrating over energy, i.e.,
\begin{align}
	\ybar_{i,k} (\boldz) :=  {} & \int_0^{+\infty} h_k(e) \cdot \rme^{ - [\boldcalA( \boldcalF( \boldz,e ) )]_i  } \, \rmd e  \label{eq:forward}  \\
	= {}  &  \mathbb{E}[Y_{i,k} \mid \boldZ = \boldz ] \nonumber
\end{align}
where $\boldcalA \colon \R^\Np \to \R^\Nb$ is the operator that computes the line integrals along each of the $\Nb$ X-ray beams, and $h_k(e)$ is the photon flux for energy bin $k$ (see Figure~\ref{fig:binned_spectrum}). Finally, we denote by $\boldY = \{Y_{i,k}\}_{i,k=1}^{\Nb,\Ne}$ the entire measurement vector. Note that since the relationship between $\boldZ$ and $\boldU$ is deterministic, conditioning $\boldY$ on $\boldZ$ or $\boldU$ is equivalent, i.e., they are equal in distribution,
\begin{equation*}
	\boldY \mid \boldZ \sim \boldY \mid \boldU \, .
\end{equation*}
Moreover, the random variables $Y_{i,k}$ are conditionally independent given $\boldU$.

\input{figures/spectre}

\subsection{Material Decomposition}\label{sec:MD formulation}

\Ac{MD} is the task of finding a material image estimate $\boldzhat \in \R^{\Np \times \Nm}$ given a realization $\boldy$ of the random variable $\boldY$. We summarize the three main approaches from the literature, starting from the two-step approaches and then the one-step approach.

\subsubsection{Image-domain Material Decomposition}

This \ac{MD} approach finds an estimate $\boldzhat$ in a two-step process, i.e.,
\begin{align}
	\bolduhat \in  \argmax_{\boldu} & \, p_{\boldY | \boldU = \boldu }(\boldy) \cdot p_{\boldU} (\boldu)   \label{eq:cont_recon} \\
	\text{then finding $\boldzhat$ s.t.}\quad  &  \boldcalF(\boldzhat,\cdot) = \bolduhat(\cdot) \, .  \nonumber
\end{align}
However, solving the \ac{MAP} problem \eqref{eq:cont_recon} with respect to a continuous function of the energy $\boldu(e)$ is challenging, and a common practice for this approach is to utilize an energy-discretized version of $\boldU$, i.e., a multi-energy image $\boldX$ defined as
\begin{equation*}
	\boldX = [\boldX_1,\dots,\boldX_\Ne]  \, , \quad \boldX_k = \boldU\left(\bar{e}_k\right) \, \forall k \in \{1, 2, \dots, \Ne \}
\end{equation*}  
where $\bar{e}_k$ is the weighted mean of the photon flux for the $k$-th bin. Thus, the following approximated forward model can be used:
\begin{align}
	& \boldx = \boldcalF_{\mathrm{discr}} (\boldz) \, , \quad \ybar_{i,k} (\boldx_k) \approx \hbar_{k}  \cdot \rme^{ - \left[\boldcalA(\boldx_k )\right]_i  }\,, \nonumber \\
	& (Y_{i,k} \mid \boldX_k = \boldx_k)  \sim \mathrm{Poisson} \left( \ybar_{i,k} (\boldx_k)  \right)\,  \label{eq:forward2} 
\end{align}
where $\hbar_k := \int h_k(e) \rmd e$ is the total photon flux at bin $k$, and 
\begin{align*}
	\boldcalF_{\mathrm{discr}} : \: &\mathbb{R}^{\Np \times \Nm} \to \mathbb{R}^{\Np \times \Ne} \\ 
	&\boldz \mapsto [ \boldcalF\left(\boldz,\bar{e}_1\right),\dots, \boldcalF\left(\boldz,\bar{e}_{\Ne}\right) ]
\end{align*}
is the discretized version of $\boldcalF$. The two-step \ac{MAP} estimation therefore simplifies to
\begin{align}
	\boldxhat \in \argmax_{\boldx} & \, p_{\boldY | \boldX = \boldx}(\boldy) \cdot p_{\boldX}(\boldx)  \label{eq:image_domain1} \\
	\text{then finding $\boldzhat$ s.t.}\quad  &  \boldcalF_{\mathrm{discr}}(\boldzhat) = \boldxhat \, . \label{eq:image_domain2}
\end{align}
The conditional \ac{PDF} $p_{\boldY | \boldX}$ is given by the approximate model \eqref{eq:forward2}. The joint \ac{PDF} $p_{\boldX} = p_{\boldX_1, \dots, \boldX_\Ne}$ is unknown but can be replaced with synergistic regularizers. As an example of such regularizers, \ac{TNV} \cite{Rigie_2015} promotes structural similarities across channels or with a reference image \cite{Synergistic_DTV}. Trained regularizers include multichannel \ac{DiL} approaches such as \ac{TDiL} \cite{zhang2016tensor}, \ac{CDL} \cite{perelli2022multi}, or U-Nets \cite{wang2024uconnect}. Similarly, the \ac{MD} sub-problem \eqref{eq:image_domain2} can be solved with \ac{DiL} \cite{Wu2021}.

\subsubsection{Projection-domain Material Decomposition}

The operators $\boldcalF$ and $\boldcalA$ can be interchanged so that the forward model \eqref{eq:forward} can be defined through the material sinograms. More precisely, given a material image $\boldZ$ and the corresponding material sinograms $\boldL = [\boldcalA (\boldZ_{:,1}) ,\dots ,  \boldcalA(\boldZ_{:,\Nm})]$, we have 
\begin{align*}
	(Y_{i,k} \mid \boldZ = \boldz) \sim {} & (Y_{i,k} \mid \boldL = \boldell) \\
	\sim {} & \mathrm{Poisson}\left( \ycheck_{i,k} ( \boldell_{i,:}  ) \right)
\end{align*}
with
\begin{equation*}
	\ycheck_{i,k} ( \boldell_{i,:}  ) = \int_0^{+\infty} h_k(e) \cdot \rme^{  - \calF( \boldell_{i,:} , e )   } \, \rmd e  \, .
\end{equation*}
Projection-domain \ac{MD} consists in first finding an estimate $\boldellhat$ of $\boldell$ (projection-based \ac{MD} step), followed by deriving $\boldzhat$ from $\boldellhat$ (material reconstruction step): 
\begin{align}
	\boldellhat \in  \argmax_{\boldell} & \, p_{\boldY |  \boldL = \boldell }(\boldy)   \label{eq:projmethod1} \\
	\text{then finding $\boldzhat$ s.t.}\quad  & \boldcalA_{\mathrm{mat}}(\boldzhat) \approx \boldellhat    \label{eq:projmethod2} 
\end{align}
where  $\boldcalA_{\mathrm{mat}} \colon \R^{\Np\times \Nm} \to \R^\Nb \times \R^\Nm$ is the material projector defined as
\begin{equation*}
	\boldcalA_{\mathrm{mat}}(\boldz) := [\boldcalA(\boldz_{:,1}),\dots,\boldcalA(\boldz_{:,\Nm})] \, .
\end{equation*}
Note that solving \eqref{eq:projmethod1} does not involve a prior \ac{PDF} for $\boldL$, although a regularization can be added  \cite{NDucros}. 

In order to use the \ac{DPS} method (Section~\ref{sec:dps}), we need to formulate \eqref{eq:projmethod2} in a \ac{MAP} fashion. This requires us to assume that the relationship between $\boldZ$ and $\boldL$ is not deterministic. We choose the following inference model:
\begin{align}
	& (\boldL \mid \boldZ = \boldz) \sim \mathscr{N} \left(   \boldcalA_{\mathrm{mat}}(\boldz)  ,  \sigma^2 \boldI  \right) \, .\label{eq:gaussian}
\end{align}
The resulting \ac{MAP} estimation for $\boldz$ is
\begin{equation}\label{eq:map_proj}
	\boldzhat = \argmax_{\boldz} \,  p_{\boldL |\boldZ = \boldz}\left(\boldellhat\right) \cdot p_{\boldZ}(\boldz)
\end{equation}
where  $\boldellhat$ is a solution of \eqref{eq:projmethod1}, and the parameter $\sigma$ in \eqref{eq:gaussian} gives the weight of $p_{\boldL |\boldZ = \boldz}$.

\subsubsection{One-step Material Decomposition}

This \ac{MD} approach consists in finding an estimate $\boldzhat$ by solving a single problem that combines decomposition and reconstruction, generally derived from a \ac{MAP} estimation, i.e.,
\begin{equation}\label{eq:onestep}
	\boldzhat \in \argmax_{\boldz} \, p_{\boldY | \boldZ = \boldz }(\boldy) \cdot p_{\boldZ} (\boldz) \, . 
\end{equation}
The conditional \ac{PDF} $p_{\boldY | \boldZ = \boldz }(\boldy)$ corresponds to the data-fidelity term and is derived from \eqref{eq:poisson} and \eqref{eq:forward}, while the prior \ac{PDF} is unknown and is generally replaced by a regularizer that promotes piecewise smooth images \cite{long2014multi}. Solving \eqref{eq:onestep} is achieved with the help of iterative algorithms such as \ac{OT} \cite{long2014multi} or \ac{LBFGS} \cite{zhu1997algorithm}, \cite{lbfgs}, which we used in this work. 

\subsection{Summary of the Methods}

Two-step \ac{MD} approaches solve sequentially the tomographic reconstruction problem and the \ac{MD} problem. Image-domain \ac{MD} relies on the approximation \eqref{eq:forward2} for the reconstruction step \eqref{eq:image_domain1}, which assumes that the energy spectrum is a weighted sum of Dirac functions in each energy bin $k$. This results in \ac{MD} errors in the decomposition step \eqref{eq:image_domain2}, such as crosstalk between materials. Projection-domain \ac{MD} does not use this approximation in the decomposition step \eqref{eq:projmethod1}, but the final material image obtained in the reconstruction step \eqref{eq:projmethod2} is not a \ac{MAP} estimate given the measurement $\boldY = \boldy$. On the other hand, one-step \ac{MD} \eqref{eq:onestep} is an actual \ac{MAP} estimate given the measurement and is known to outperform image-domain and projection-domain \ac{MD} \cite{mechlem2017joint, Mory2018}. 

%% file: figures/spectre.tex
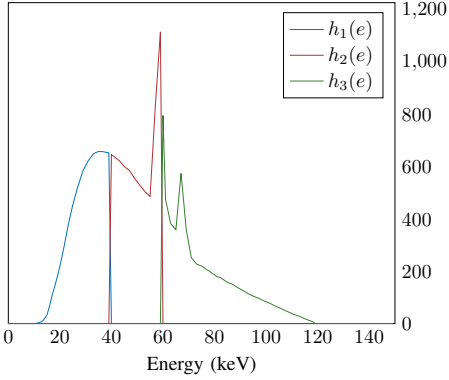
\begin{figure}
		\centering
		\begin{tikzpicture}[scale=0.75, trim axis left]
	\begin{axis}[
		xlabel={Energy (keV)},
		yticklabel pos=right,
		xmin=0, xmax=150,
		ymin=0,
		xtick={0,20,40,60,80,100,120,140},
		legend pos=north east,
		tick style={draw=none}
		]

		\addplot[
		color=RoyalBlue,
		]
		coordinates {
(0, 0.0)
(1, 0.0)
(2, 0.0)
(3, 0.0)
(4, 0.0)
(5, 0.0)
(6, 0.0)
(7, 0.0)
(8, 0.0)
(9, 0.0)
(10, 0.4250484108924866)
(11, 1.537368655204773)
(12, 3.9133431911468506)
(13, 7.877310276031494)
(14, 18.568937301635742)
(15, 33.20743179321289)
(16, 67.62055206298828)
(17, 109.39227294921875)
(18, 142.0759735107422)
(19, 179.96112060546875)
(20, 220.2608184814453)
(21, 264.88336181640625)
(22, 313.78973388671875)
(23, 366.1908264160156)
(24, 411.4814147949219)
(25, 455.17657470703125)
(26, 490.93927001953125)
(27, 526.7390747070312)
(28, 556.5011596679688)
(29, 587.0056762695312)
(30, 605.9227905273438)
(31, 624.3451538085938)
(32, 637.4859008789062)
(33, 650.6837768554688)
(34, 656.05517578125)
(35, 659.5780639648438)
(36, 659.5503540039062)
(37, 658.6022338867188)
(38, 656.5196533203125)
(39, 654.4552612304688)
(40, 0.0)
(41, 0.0)
(42, 0.0)
(43, 0.0)
(44, 0.0)
(45, 0.0)
(46, 0.0)
(47, 0.0)
(48, 0.0)
(49, 0.0)
(50, 0.0)
(51, 0.0)
(52, 0.0)
(53, 0.0)
(54, 0.0)
(55, 0.0)
(56, 0.0)
(57, 0.0)
(58, 0.0)
(59, 0.0)
(60, 0.0)
(61, 0.0)
(62, 0.0)
(63, 0.0)
(64, 0.0)
(65, 0.0)
(66, 0.0)
(67, 0.0)
(68, 0.0)
(69, 0.0)
(70, 0.0)
(71, 0.0)
(72, 0.0)
(73, 0.0)
(74, 0.0)
(75, 0.0)
(76, 0.0)
(77, 0.0)
(78, 0.0)
(79, 0.0)
(80, 0.0)
(81, 0.0)
(82, 0.0)
(83, 0.0)
(84, 0.0)
(85, 0.0)
(86, 0.0)
(87, 0.0)
(88, 0.0)
(89, 0.0)
(90, 0.0)
(91, 0.0)
(92, 0.0)
(93, 0.0)
(94, 0.0)
(95, 0.0)
(96, 0.0)
(97, 0.0)
(98, 0.0)
(99, 0.0)
(100, 0.0)
(101, 0.0)
(102, 0.0)
(103, 0.0)
(104, 0.0)
(105, 0.0)
(106, 0.0)
(107, 0.0)
(108, 0.0)
(109, 0.0)
(110, 0.0)
(111, 0.0)
(112, 0.0)
(113, 0.0)
(114, 0.0)
(115, 0.0)
(116, 0.0)
(117, 0.0)
(118, 0.0)
(119, 0.0)
(120, 0.0)
(121, 0.0)
(122, 0.0)
(123, 0.0)
(124, 0.0)
(125, 0.0)
(126, 0.0)
(127, 0.0)
(128, 0.0)
(129, 0.0)
(130, 0.0)
(131, 0.0)
(132, 0.0)
(133, 0.0)
(134, 0.0)
(135, 0.0)
(136, 0.0)
(137, 0.0)
(138, 0.0)
(139, 0.0)
(140, 0.0)
(141, 0.0)
(142, 0.0)
(143, 0.0)
(144, 0.0)
(145, 0.0)
(146, 0.0)
(147, 0.0)
(148, 0.0)
(149, 0.0)
		};
	
		\addplot[
color=Maroon,
]
coordinates {
(0, 0.0)
(1, 0.0)
(2, 0.0)
(3, 0.0)
(4, 0.0)
(5, 0.0)
(6, 0.0)
(7, 0.0)
(8, 0.0)
(9, 0.0)
(10, 0.0)
(11, 0.0)
(12, 0.0)
(13, 0.0)
(14, 0.0)
(15, 0.0)
(16, 0.0)
(17, 0.0)
(18, 0.0)
(19, 0.0)
(20, 0.0)
(21, 0.0)
(22, 0.0)
(23, 0.0)
(24, 0.0)
(25, 0.0)
(26, 0.0)
(27, 0.0)
(28, 0.0)
(29, 0.0)
(30, 0.0)
(31, 0.0)
(32, 0.0)
(33, 0.0)
(34, 0.0)
(35, 0.0)
(36, 0.0)
(37, 0.0)
(38, 0.0)
(39, 0.0)
(40, 647.7391357421875)
(41, 640.4197998046875)
(42, 632.4494018554688)
(43, 624.9588623046875)
(44, 614.193603515625)
(45, 602.1958618164062)
(46, 594.5499267578125)
(47, 586.7399291992188)
(48, 571.571044921875)
(49, 556.1884155273438)
(50, 543.7601318359375)
(51, 530.9984130859375)
(52, 518.4116821289062)
(53, 505.5843200683594)
(54, 496.6144104003906)
(55, 487.5590515136719)
(56, 658.0980224609375)
(57, 829.5460815429688)
(58, 974.23095703125)
(59, 1118.5255126953125)
(60, 0.0)
(61, 0.0)
(62, 0.0)
(63, 0.0)
(64, 0.0)
(65, 0.0)
(66, 0.0)
(67, 0.0)
(68, 0.0)
(69, 0.0)
(70, 0.0)
(71, 0.0)
(72, 0.0)
(73, 0.0)
(74, 0.0)
(75, 0.0)
(76, 0.0)
(77, 0.0)
(78, 0.0)
(79, 0.0)
(80, 0.0)
(81, 0.0)
(82, 0.0)
(83, 0.0)
(84, 0.0)
(85, 0.0)
(86, 0.0)
(87, 0.0)
(88, 0.0)
(89, 0.0)
(90, 0.0)
(91, 0.0)
(92, 0.0)
(93, 0.0)
(94, 0.0)
(95, 0.0)
(96, 0.0)
(97, 0.0)
(98, 0.0)
(99, 0.0)
(100, 0.0)
(101, 0.0)
(102, 0.0)
(103, 0.0)
(104, 0.0)
(105, 0.0)
(106, 0.0)
(107, 0.0)
(108, 0.0)
(109, 0.0)
(110, 0.0)
(111, 0.0)
(112, 0.0)
(113, 0.0)
(114, 0.0)
(115, 0.0)
(116, 0.0)
(117, 0.0)
(118, 0.0)
(119, 0.0)
(120, 0.0)
(121, 0.0)
(122, 0.0)
(123, 0.0)
(124, 0.0)
(125, 0.0)
(126, 0.0)
(127, 0.0)
(128, 0.0)
(129, 0.0)
(130, 0.0)
(131, 0.0)
(132, 0.0)
(133, 0.0)
(134, 0.0)
(135, 0.0)
(136, 0.0)
(137, 0.0)
(138, 0.0)
(139, 0.0)
(140, 0.0)
(141, 0.0)
(142, 0.0)
(143, 0.0)
(144, 0.0)
(145, 0.0)
(146, 0.0)
(147, 0.0)
(148, 0.0)
(149, 0.0)
};
		
		\addplot[
		color=OliveGreen,
		]
		coordinates {
			(0, 0.0)
			(1, 0.0)
			(2, 0.0)
			(3, 0.0)
			(4, 0.0)
			(5, 0.0)
			(6, 0.0)
			(7, 0.0)
			(8, 0.0)
			(9, 0.0)
			(10, 0.0)
			(11, 0.0)
			(12, 0.0)
			(13, 0.0)
			(14, 0.0)
			(15, 0.0)
			(16, 0.0)
			(17, 0.0)
			(18, 0.0)
			(19, 0.0)
			(20, 0.0)
			(21, 0.0)
			(22, 0.0)
			(23, 0.0)
			(24, 0.0)
			(25, 0.0)
			(26, 0.0)
			(27, 0.0)
			(28, 0.0)
			(29, 0.0)
			(30, 0.0)
			(31, 0.0)
			(32, 0.0)
			(33, 0.0)
			(34, 0.0)
			(35, 0.0)
			(36, 0.0)
			(37, 0.0)
			(38, 0.0)
			(39, 0.0)
			(40, 0.0)
			(41, 0.0)
			(42, 0.0)
			(43, 0.0)
			(44, 0.0)
			(45, 0.0)
			(46, 0.0)
			(47, 0.0)
			(48, 0.0)
			(49, 0.0)
			(50, 0.0)
			(51, 0.0)
			(52, 0.0)
			(53, 0.0)
			(54, 0.0)
			(55, 0.0)
			(56, 0.0)
			(57, 0.0)
			(58, 0.0)
			(59, 0.0)
			(60, 797.8550415039062)
			(61, 474.97125244140625)
			(62, 429.1647033691406)
			(63, 382.97772216796875)
			(64, 371.8099060058594)
			(65, 359.5016784667969)
			(66, 467.4542541503906)
			(67, 575.616943359375)
			(68, 470.2382507324219)
			(69, 362.0779724121094)
			(70, 307.42999267578125)
			(71, 252.84361267089844)
			(72, 240.9886474609375)
			(73, 228.2002410888672)
			(74, 224.1868896484375)
			(75, 220.7850799560547)
			(76, 213.9678192138672)
			(77, 206.1669158935547)
			(78, 201.08779907226562)
			(79, 193.1794891357422)
			(80, 187.78500366210938)
			(81, 179.1077117919922)
			(82, 177.85733032226562)
			(83, 171.54432678222656)
			(84, 164.77938842773438)
			(85, 157.9600067138672)
			(86, 154.26487731933594)
			(87, 150.6643524169922)
			(88, 144.13504028320312)
			(89, 137.5482635498047)
			(90, 132.31430053710938)
			(91, 126.94056701660156)
			(92, 121.2624282836914)
			(93, 115.10853576660156)
			(94, 110.2135238647461)
			(95, 105.34691619873047)
			(96, 101.4255599975586)
			(97, 97.54894256591797)
			(98, 92.78414916992188)
			(99, 88.01704406738281)
			(100, 84.01809692382812)
			(101, 79.8708267211914)
			(102, 75.1536865234375)
			(103, 69.82955169677734)
			(104, 65.5941390991211)
			(105, 61.43540573120117)
			(106, 56.54450225830078)
			(107, 51.75570297241211)
			(108, 47.59739685058594)
			(109, 43.192405700683594)
			(110, 38.63211441040039)
			(111, 34.20970916748047)
			(112, 30.898488998413086)
			(113, 27.86758804321289)
			(114, 23.396886825561523)
			(115, 18.565872192382812)
			(116, 14.973688125610352)
			(117, 11.24686050415039)
			(118, 7.5782318115234375)
			(119, 0.0)
			(120, 0.0)
			(121, 0.0)
			(122, 0.0)
			(123, 0.0)
			(124, 0.0)
			(125, 0.0)
			(126, 0.0)
			(127, 0.0)
			(128, 0.0)
			(129, 0.0)
			(130, 0.0)
			(131, 0.0)
			(132, 0.0)
			(133, 0.0)
			(134, 0.0)
			(135, 0.0)
			(136, 0.0)
			(137, 0.0)
			(138, 0.0)
			(139, 0.0)
			(140, 0.0)
			(141, 0.0)
			(142, 0.0)
			(143, 0.0)
			(144, 0.0)
			(145, 0.0)
			(146, 0.0)
			(147, 0.0)
			(148, 0.0)
			(149, 0.0)
		};
		
		\legend{$h_1(e)$, $h_2(e)$, $h_3(e)$}
		
	\end{axis}
\end{tikzpicture}
	\caption{Binned X-ray energy spectrum, with $\Ne = 3$ energy bins, that we used in our simulations.} 
	\label{fig:binned_spectrum}
\end{figure}

%% file: content/diffusion.tex
\section{Proposed Method: Material Decomposition by Diffusion Posterior Sampling} \label{sec:DPS for MD}

\Acp{DM} \cite{ho2020denoising, song2020generative} are the new state-of-the-art generative models for solving inverse problems \cite{song2022solving}, particularly in medical image reconstruction \cite{webber2024diffusion}. This section shows how to incorporate \acp{DM} into the three \ac{MD} approaches discussed in Section~\ref{sec:MD formulation}.

In previous work \cite{vazia2024spectral}, we proposed a two-step \ac{DPS} \ac{MD} framework consisting of (i) sampling the multi-energy image $\boldX = [\boldX_1,\dots,\boldX_\Ne]$ by \ac{DPS} using a \ac{DM}-trained prior $p_{\boldx}(\boldX)$ in the multi-energy image domain and guided by the approximated log-posterior gradient $\nabla_{\boldx}\log p_{\boldY |\boldX = \boldx}(\boldy)$ given by \eqref{eq:forward2}, followed by (ii) estimating the material image $\boldz$ by solving \eqref{eq:image_domain2}. This method is the \ac{DPS} analogue of \eqref{eq:image_domain1} and \eqref{eq:image_domain2} and is referred to as \ac{ImTDPS}. 

Another approach, evaluated in this paper, consists of sampling $\boldZ$ using a \ac{DM}-trained prior $p_{\boldZ}$ on the material image domain and guided by the pseudo log-posterior $\nabla_{\boldz} \log p_{\boldL |\boldZ = \boldz}(\boldellhat)$ given by \eqref{eq:gaussian}, where the estimated projected materials $\boldellhat$ are obtained by solving \eqref{eq:projmethod1}. This method is the analogue of \eqref{eq:map_proj} and is referred to as \ac{ProjTDPS}. 

Alternatively, $\boldZ$ can be directly sampled by \ac{DPS} from the data $\boldY=\boldy$, using the same \ac{DM}-trained prior $p_{\boldZ}$ as for \ac{ProjTDPS}, and the true Poisson log-posterior $\nabla_{\boldz} \log p_{\boldY |\boldZ = \boldz}(\boldy)$. This approach, which is the analogue of \eqref{eq:onestep}, is referred to as \ac{ODPS}.

Section~\ref{sec:dps_general} introduces the basics of \acp{DM} and \ac{DPS} in the context of \ac{ImTDPS} and \ac{ODPS}. We do not describe \ac{ProjTDPS} in detail, as it is implemented similarly to \ac{ODPS} by replacing the log-prior $\nabla_{\boldz} \log p_{\boldY |\boldZ = \boldz}(\boldy)$ with $\nabla_{\boldz} \log p_{\boldL |\boldZ = \boldz}(\boldellhat)$. Section~\ref{sec:dps_3mat} introduces a heuristic method derived from \ac{ODPS} to include additional materials not present in the training dataset. A variational regularizer is introduced for the added materials, while the \ac{DM} prior is kept for the other material images. We refer to this method as Hybrid-\ac{ODPS}. 

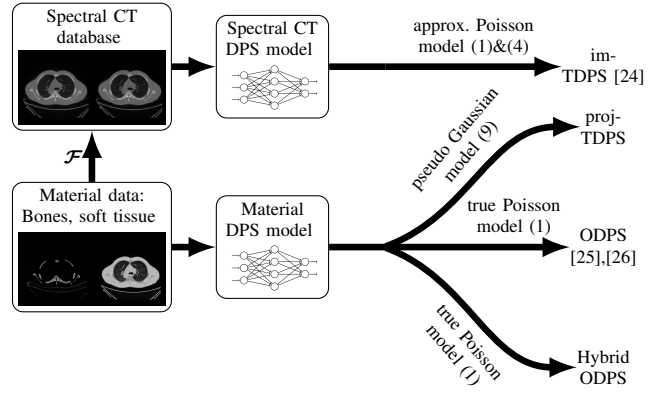
\begin{figure}
	\centering
	\input{figures/methods_flowchart}
	\caption{Overview of the described \ac{DPS}-based methods for \ac{MD}. \Ac{ODPS}, Hybrid-\ac{ODPS}, and \ac{ProjTDPS} share the same \ac{DM} trained on material images. They differ by the model used to solve \ac{MD}. The \ac{ImTDPS} method uses a \ac{DM} trained on spectral images for the two-step \ac{MD}.}
	\label{fig:flowchart}
\end{figure}

Figure~\ref{fig:flowchart} presents an overview of the proposed methods.

\input{figures/Bigfig2}

\subsection{Standard Approach with All Materials Present in the Training Dataset}\label{sec:dps_general}

We denote by $\boldW \in \{\boldX,\boldZ\}$ the random vector to be estimated by \ac{DPS}, i.e., the multi-energy image $\boldX$ for \ac{ImTDPS} (cf.~\eqref{eq:image_domain1}) or the material image $\boldZ$ for \ac{ODPS} (cf.~\eqref{eq:onestep}).

\subsubsection{Diffusion Models (DMs)}\label{sec:dm}

The training of a \ac{DM} (cf.~\citeauthor{ho2020denoising}~\cite{ho2020denoising}) involves a diffusion process that incrementally adds noise to an initial image $\boldW_0$, sampled from the training dataset with \ac{PDF} $p^{\mathrm{data}}$,
\begin{equation}\label{eq:wtfromw0}
	\boldW_t = \sqrt{\alphabar_t}\boldW_0 + \sqrt{1-\alphabar_t} \boldeps \, , \quad \boldeps \sim \mathscr{N}(\boldzero,\boldI)
\end{equation}
where $\alphabar_t = \prod_{s=1}^t \alpha_s$, $\alpha_t \in [0,1]$ is a monotonically decreasing sequence defined such that $\bar{\alpha}_T = 0$, and $T$ is the final step of the forward diffusion. An approximate reverse process, involving the score function $\nabla \log p_{\boldW_t}$, can be derived to sample $\boldW_{t-1}$ from $\boldW_t$ as
\begin{align}\label{eq:sampling}
	\boldW_{t-1}  = \frac{\sqrt{\alpha_t} (1-\alphabar_{t-1})}{1-\alphabar_t} & \boldW_t + \frac{\sqrt{\alphabar_{t-1}} (1-\alpha_t)}{1 - \alphabar_t} \boldwhat_0(\boldW_t) \nonumber \\
	& + \sigma_t \boldeps_t \, , \quad \boldeps_t \sim \calN(\boldzero,\boldI) \, ,
\end{align}
where $\sigma_t= (1-\alpha_t)(1-\alphabar_{t-1})/(1-\alphabar_{t})$ and $\boldwhat_0(\boldw_t)$ is given by Tweedie’s formula, i.e.,
\begin{align} 
	\boldwhat_0(\boldw_t,t) &:= \mathbb{E}[\boldW_0 | \boldW_t = \boldw_t] \label{eq:Tweedie} \\
	&= \frac{\boldw_t + (1 - \alphabar_t) \nabla \log p_{\boldW_t} (\boldw_t)}{\sqrt{\alphabar_t}} \nonumber \, .
\end{align}
The score function $\nabla \log p_{\boldW_t}$ is unknown and therefore approximated by a \ac{CNN} $\boldcalS_{\boldtheta} (\cdot,t)$ trained by score matching as
\begin{equation}\label{eq:score_matching}
	\min_{\boldtheta} \, \mathbb{E}_{t, \boldW_0,\boldW_t|\boldW_0} \left[ \left\| \boldcalS_{\boldtheta}(\boldW_t,t) - \nabla \log p_{\boldW_t|\boldW_0}(\boldW_t) \right\|_2^2 \right]
\end{equation}
where $\boldW_0 \sim p^{\mathrm{data}}$, $p_{\boldW_t|\boldW_0}$ is given by \eqref{eq:wtfromw0}, and $t$ is uniformly sampled on $[0,T]$.

\subsubsection{Diffusion Posterior Sampling (DPS)}\label{sec:dps}

It is possible to leverage the generative capability of a \ac{DM} to regularize an inverse problem; see, for instance, \cite{song2022solving, chung2022improving}. The idea is to condition the expectation \eqref{eq:Tweedie} on the measurements $\boldY=\boldy$. This leads to the conditional score $\nabla \log p_{\boldW_t|\boldY=\boldy}(\cdot)$, which, using Bayes’ rule, can be written as 
\begin{align*}
	\nabla \log p_{\boldW_t|\boldY=\boldy}(\boldw_t) = {} & \nabla \log p_{\boldW_t}(\boldw_t) \\ \nonumber 
	& + \nabla_{\boldw_t} \log p_{\boldY|\boldW_t=\boldw_t}(\boldy) 
\end{align*}
where the subscript $\boldw_t$ specifies the variable of differentiation. The first term $\nabla (\log p_{\boldW_t})(\boldw_t)$ is the unconditional score approximated by $\boldcalS_{\boldtheta} (\boldw_t,t)$. The second term $\nabla_{\boldw_t} (\log p_{\boldY|\boldW_t = \boldw_t})(\boldy)$ is intractable. 

In this work, we used the \ac{DPS} approximation \cite{chung2023diffusion}, i.e.,
\begin{equation} \label{eq:dps_approx}
	\nabla_{\boldw_t} \log p_{\boldY \mid \boldW_t = \boldw_t}(\boldy) \approx \nabla_{\boldw_t} \left(\log p_{\boldY \mid \boldW = \hat{\boldw}_0(\boldw_t)}\right)(\boldy)
\end{equation}
which is derived from the gradient of the log-likelihood $\log p_{\boldY|\boldW}$ given by the forward model \eqref{eq:poisson} (for $\boldW = \boldZ$) or \eqref{eq:forward2} (for $\boldW = \boldX$). The approximate conditional score is then added to the reverse process \eqref{eq:sampling} to generate a sample from $p_{\boldW|\boldY=\boldy}$. Therefore, \ac{DPS} alternates between sampling \eqref{eq:sampling} and performing a gradient ascent step using \eqref{eq:dps_approx}.

Using the \ac{DPS} method, we implemented \ac{ImTDPS} to sample $\boldW = \boldX$ from $p_{\boldX|\boldY=\boldy}$, followed by applying the pseudoinverse $\boldcalF_{\mathrm{discr}}^{\dag}$ to obtain material images $\boldz$ (cf.~Figure~\ref{fig:ImTDPS_Bigfig}), and \ac{ODPS} to directly sample $\boldW = \boldZ$ from $p_{\boldZ|\boldY=\boldy}$ (cf.~Figure~\ref{fig:ODPS_Bigfig}). The implementation of \ac{ProjTDPS} is similar to that of \ac{ODPS} and is shown in the same figure.

\input{figures/Bigfig}

\subsection{Hybridization with Variational Regularizer for Untrained Additional Materials}\label{sec:dps_3mat}

\Ac{DPS} methods are limited to \ac{MD} using models trained on databases containing the same materials. However, in some scenarios, \ac{MD} may need to include additional materials not present in the training dataset. For example, a model trained on bone and soft tissue images may need to be applied to a patient administered with an iodine tracer. In this section, we present a heuristic technique for \ac{ODPS}, namely Hybrid-\ac{ODPS}, to address this case, which resembles the \ac{DPS} approach for solving blind inverse problems \cite{dontas2024blind,depaepe2025solving}. 

We consider the random material image $[\boldZ, \boldZtilde]$, where $\boldZ$ represents the $\Nm$ materials present in the training database, and $\boldZtilde$ represents the $\Nmtilde$ materials unknown to the model. The trained model can be used to sample $\boldZ \sim p_{\boldZ}$. To sample $(\boldZ, \boldZtilde) \sim p_{\boldZ,\boldZtilde}$, the conditional \ac{PDF} $p_{\boldZtilde | \boldZ}$ is required but unknown. We therefore assume that $\boldZ$ and $\boldZtilde$ are independent, i.e.,
\begin{equation*}
	p_{\boldZ,\boldZtilde} \left(\boldz,\boldztilde\right) = p_{\boldZ}\left(\boldz\right) \cdot p_{\boldZtilde}\left(\boldztilde\right) \, .
\end{equation*}
This assumption is clearly incorrect (for example, iodine is only present in soft tissues), but we will see later that it leads to a reasonable algorithm. Another issue is that $p_{\boldZtilde}(\boldztilde)$ is also unknown. We therefore replace it with a pseudo \ac{PDF} $\ptilde$ of the form
\begin{equation*}\label{eq:pseudo_pdf}
	\tilde{p}\left(\boldztilde\right)  \propto  \rme^{- R\left(\boldztilde\right)}
\end{equation*}
where $R \colon \R^{\Np \times \Nmtilde} \to \R$ is a convex edge-preserving regularizer, such as the Huber penalty \cite{Elbakri2002,huber2011robust}. Note that $\ptilde$ is not a true \ac{PDF}, as it may integrate to infinity. As $\boldZ$ and $\boldZtilde$ are independent, the estimates $\boldZ_t$ and $\boldZtilde_t$ at time $t$ of the \ac{DM} are also independent. 
The conditional score at time $t$ can be written as
\begin{align*}
	\nabla_{\boldz_t,\boldztilde_t} {} & \log  p_{\boldZ_t,\boldZtilde_t|\boldY=\boldy}  (\boldz_t,\boldztilde_t) \nonumber \\ 
	= {} & \left[
	\begin{array}{c}
		\nabla \log p_{\boldZ_t} \left(\boldz_t\right)  +  \nabla_{\boldz_t} \log p_{\boldY \mid (\boldZ_t,\boldZtilde_t) = (\boldz_t,\boldztilde_t) }(\boldy)  \\ 
		\nabla \log p_{\boldZtilde_t}  \left(\boldztilde_t\right)  +  \nabla_{\boldztilde_t} \log p_{\boldY \mid (\boldZ_t,\boldZtilde_t) = (\boldz_t,\boldztilde_t) }(\boldy)   
	\end{array}	
	\right] \, .
\end{align*} 
The score function $\nabla \log p_{\boldZ_t} (\boldz_t)$, corresponding to the $\Nm$ materials present in the training dataset, is approximated by $\boldcalS_{\boldtheta}(\boldz_t,t)$ trained as in \eqref{eq:score_matching}, while the conditional score functions $\nabla_{\boldz_t} \log p_{\boldY \mid (\boldZ_t,\boldZtilde_t)}$ and $\nabla_{\boldztilde_t} \log p_{\boldY \mid (\boldZ_t,\boldZtilde_t)}$ can be approximated using the forward model (similarly to \eqref{eq:dps_approx}). The score function $\nabla \log p_{\boldZtilde_t}(\boldztilde_t)$ is unknown but can be approximated by $\nabla \log p_{\boldZtilde}(\boldztilde_t) =  -\nabla R(\boldztilde_t)$. This leads to the following update rule for $\boldztilde$ (with an added gradient step $\xi_t$ on the log-prior $R$):
\begin{equation}\label{eq:gradient_step}
	\boldztilde_{t-1} = \boldztilde_{t} + \nabla_{\boldztilde_t} \log p_{\boldY \mid (\boldZ,\boldZtilde) = (\boldz_t,\boldztilde_t)} - \xi_t \nabla R(\boldztilde_t) \, .
\end{equation} 

In summary, this approach is equivalent to performing an iterative algorithm to optimize an objective function consisting of a data fidelity term and a regularizer, in parallel with \ac{DPS}. Note that the gradient step \eqref{eq:gradient_step} can be replaced by any optimization algorithm on the penalized log-likelihood; in this work, we used an \ac{LBFGS} algorithm initialized from $\boldztilde_{t}$.

\subsection{Summary of the Algorithms}

The overall \ac{ODPS} method is summarized in Algorithm~\ref{algo:odps}. The hybridization case is controlled by the parameter $\tilde{n}_m$. The initial material image $\boldztilde^{\mathrm{init}}$ (corresponding to the material absent from the training dataset) can be obtained with any standard \ac{MD} technique. In this work, we use an image-based \ac{MD} on energy-discretized images $\boldxhat_k$ reconstructed from $\boldy_{:,k}$ by \ac{FBP}.

Algorithm~\ref{algo:imdps} presents the \Ac{ImTDPS} method. It is implemented in a similar fashion but with multi-energy images $\boldx_t$ instead of the material image $\boldz_t$, and with a multi-energy image score function $\tilde{\boldcalS}_{\tilde{\boldtheta}}$. After the reconstruction of the multi-energy image, an image-based \ac{MD} method is applied (here using the pseudoinverse of $\boldcalF_{\mathrm{discr}}$). 

Finally, the \ac{ProjTDPS} method (described in Algorithm~\ref{algo:projdps}) uses the same \ac{DPS} prior as \ac{ODPS} (same learned score $\boldcalS_{\boldtheta}$). 
Note that the hybridization method could be employed here, but we did not test it in this paper.

Furthermore, we use the ``jump-start strategy'' from \citeauthor{jiang2024strategies}~\cite{jiang2024strategies}, starting from $T' < T$ with a scout \ac{FBP}-reconstructed multi-energy image (followed by \ac{MD} using $\boldcalF_{\mathrm{discr}}^\dag$ for \ac{ODPS} and \ac{ProjTDPS}) diffused to time $T'$ using \eqref{eq:wtfromw0} (not shown in the presented pseudo-codes). Finally, the log-likelihood gradient \eqref{eq:dps_approx} is further approximated by differentiating with respect to $\boldzhat_0$ instead of $\boldz_t$, following the strategy of \citeauthor{jiang2024strategies}~\cite{jiang2024strategies}; this prevents differentiation through the \ac{NN}.

\let\OldComment\Comment
\renewcommand{\Comment}[1]{{\tiny\OldComment{#1}}}

\begin{algorithm}
	\caption{\Ac{ODPS} and Hybrid-\ac{ODPS}}\label{algo:odps}
	\scriptsize
	\begin{algorithmic}[1]
		\Require $T$, $\boldtheta$, $\boldy$, $\{ \zeta_t \}_{t=1}^T$, $\{ \xi_t \}_{t=1}^T$, $\{ \sigma_t \}_{t=1}^T$, $\{ \alpha_t \}_{t=1}^T$, $\boldztilde^{\mathrm{init}}$
		\State $\boldz_T  \gets \boldZ_t \sim \calN(\boldzero, \boldI)$ {\tiny\Comment{Initialization}}
		\If{$\Nmtilde >0$}
		\State $\boldztilde_T \gets  \boldztilde^{\mathrm{init}} $
		\EndIf
		\For{$t = T$ \textbf{to} $1$}
		\State $\boldzhat_0 \gets \frac{1}{\sqrt{\bar{\alpha}_t}} \left(\boldz_t + (1 - \bar{\alpha}_t) \boldcalS_{\boldtheta}(\boldz_t, t)\right)$ \Comment{Estimation of the clean material image}
		\State $\boldeps \sim \calN(\boldzero, \boldI)$
		\State $\boldz'_{t} \gets \frac{\sqrt{\alpha_t} (1-\bar{\alpha}_{t-1})}{1-\bar{\alpha}_t} \boldz_t + \frac{\sqrt{\bar{\alpha}_{t-1}} (1-\alpha_t)}{1 - \bar{\alpha}_t} \boldzhat_0 + \sigma_t \boldeps$ \Comment{Unconditional backward diffusion}
		\If{$\Nmtilde >0$} \Comment{Guidance with hybridization}
		\State $\boldz_{t-1} \gets \boldz'_{t} + \zeta_t \nabla_{\boldz_t} \log p_{\boldY \mid (\boldZ,\boldZtilde) = (\hat{\boldz}_0, \boldztilde_t) }(\boldy)$
		\State $\boldztilde_{t-1} \gets \boldztilde_{t} + \nabla_{\boldztilde_t} \log p_{\boldY \mid (\boldZ,\boldZtilde) = (\hat{\boldz}_0,\boldztilde_t)}(\boldy) - \xi_t \nabla R(\boldztilde_t)$
		\Else \Comment{Guidance without hybridization}
		\State $\boldz_{t-1} \gets \boldz'_{t} + \zeta_t \nabla_{\boldz_t} \log p_{\boldY \mid \boldZ = \hat{\boldz}_0 }(\boldy)$ 
		\EndIf
		\EndFor
	\end{algorithmic}
\end{algorithm}

\begin{algorithm}
	\caption{\Ac{ImTDPS}}\label{algo:imdps}
	\footnotesize
	\begin{algorithmic}[1]
		\Require $T$, $\tilde{\boldtheta}$, $\boldy$,$\{ \zeta_t \}_{t=1}^T$, $\{ \sigma_t \}_{t=1}^T$, $\{ \alpha_t \}_{t=1}^T$
		\State $\boldx_T  \sim \calN(\boldzero, \boldI)$ \Comment{Initialization}
		\For{$t = T$ \textbf{to} $1$}
		\State $\hat{\boldx}_0 \gets \frac{1}{\sqrt{\bar{\alpha}_t}} \left(\boldx_t + (1 - \bar{\alpha}_t) \tilde{\boldcalS}_{\tilde{\boldtheta}}(\boldx_t, t)\right)$ \Comment{Estimation of the clean multi-energy image}
		\State $\boldeps \sim \calN(\boldzero, \boldI)$
		\State $\boldx'_{t-1} \gets \frac{\sqrt{\alpha_t} (1-\bar{\alpha}_{t-1})}{1-\bar{\alpha}_t} \boldx_t + \frac{\sqrt{\bar{\alpha}_{t-1}} (1-\alpha_t)}{1 - \bar{\alpha}_t} \hat{\boldx}_0 + \sigma_t \boldz_t$ \Comment{Unconditional backward diffusion}
		\State $\boldx_{t-1} \gets \boldx'_{t-1} + \zeta_t \nabla_{\hat{\boldx}_0} \log p_{\boldY \mid \boldX = \hat{\boldx}_0}(\boldy)$ \Comment{Guidance}
		\EndFor 
		\State $\boldz \gets \boldcalF_{\textrm{discr}}^\dagger(\boldx_0)$ \Comment{\textrm{Material decomposition}}
	\end{algorithmic}
\end{algorithm}

\begin{algorithm}
	\caption{\Ac{ProjTDPS}}\label{algo:projdps}
	\footnotesize
	\begin{algorithmic}[1]
		\Require $T$, $\boldtheta$, $\boldy$, $\{ \zeta_t \}_{t=1}^T$, $\{ \xi_t \}_{t=1}^T$, $\{ \sigma_t \}_{t=1}^T$, $\{ \alpha_t \}_{t=1}^T$
		\State Compute $\hat{\boldell}$ from $\boldy$ using \eqref{eq:projmethod1} and \ac{LBFGS} method.
		\State $\boldz_T  \gets \boldZ_t \sim \calN(\boldzero, \boldI)$ \Comment{Initialization}
		\If{$\Nmtilde >0$}
		\State $\boldztilde_T \gets  \boldztilde^{\mathrm{init}} $
		\EndIf
		\For{$t = T$ \textbf{to} $1$}
		\State $\boldzhat_0 \gets \frac{1}{\sqrt{\bar{\alpha}_t}} \left(\boldz_t + (1 - \bar{\alpha}_t) \boldcalS_{\boldtheta}(\boldz_t, t)\right)$ \Comment{Estimation of the clean material image}
		\State $\boldeps \sim \calN(\boldzero, \boldI)$
		\State $\boldz'_{t} \gets \frac{\sqrt{\alpha_t} (1-\bar{\alpha}_{t-1})}{1-\bar{\alpha}_t} \boldz_t + \frac{\sqrt{\bar{\alpha}_{t-1}} (1-\alpha_t)}{1 - \bar{\alpha}_t} \boldzhat_0 + \sigma_t \boldeps$ \Comment{Unconditional backward diffusion}
		\State $\boldz_{t-1} \gets \boldz'_{t} + \zeta_t \nabla_{\boldz_t} \log p_{\boldL \mid \boldZ = \boldzhat_0 }(\hat{\boldell})$  \Comment{Guidance}
		
		\EndFor
	\end{algorithmic}
\end{algorithm}

%% file: figures/methods_flowchart.tex
\begin{tikzpicture}[scale=0.8]
	\scriptsize
	\node(trainingdata) at (0,0) [align=center, draw, rounded corners] {
		 Material data: \\ Bones, soft tissue  \vspace{.1cm}\\\includegraphics[width=0.11\linewidth]{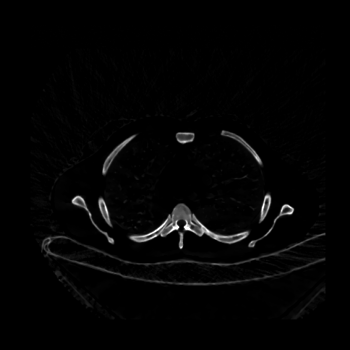}\includegraphics[width=0.11\linewidth]{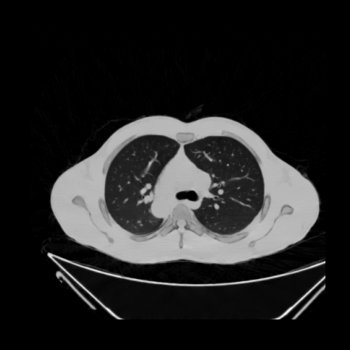}
	} ;
	
	\node(spectralct) at ([yshift=3cm]trainingdata)[align=center, draw, rounded corners] {
	Spectral CT \\ database \vspace{.1cm}\\
		\includegraphics[width=0.11\linewidth]{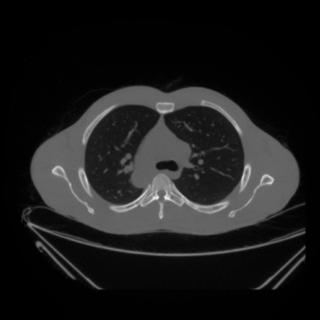}\includegraphics[width=0.11\linewidth]{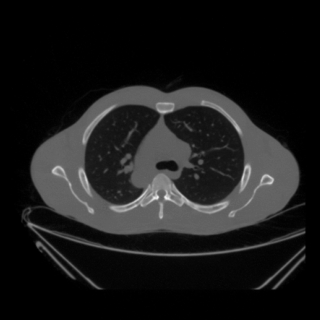}
	} ;
	\node(matmodel) at ([xshift=3cm]trainingdata) [align=center, draw, rounded corners] {
	Material \\ DPS model \vspace{.1cm} \\\includegraphics[width=0.15\linewidth]{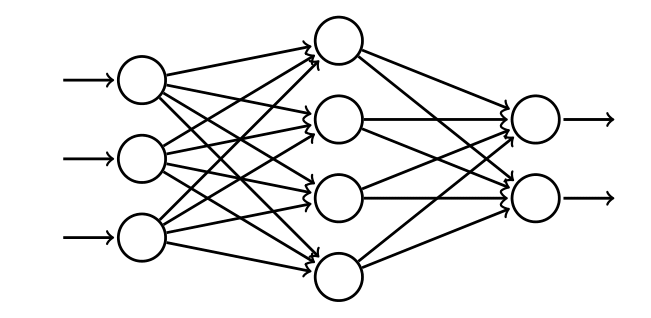}
	} ;
	
	\node(spectralmodel) at ([xshift=3cm]spectralct) [align=center, draw, rounded corners]{
		Spectral CT \\ DPS model \vspace{.1cm} \\\includegraphics[width=0.15\linewidth]{./figures/methodsflowchart/nnimage}
	} ;
	
	\node(im-TDPS) at ([xshift=5.5cm]spectralmodel)[align=center, inner sep=0pt]{
		im-\\TDPS \cite{vazia2024spectral}
	} ;
	\node(off1) at ([xshift=1.85cm]spectralmodel)[align=center, inner sep=0pt]{};
	\node(proj-TDPS) at ([xshift=5.5cm,yshift=2cm]matmodel)[align=center, inner sep=0pt]{
		proj-\\TDPS
	} ;
	\node(ODPS) at ([xshift=5.5cm]matmodel)[align=center, inner sep=0pt]{
		ODPS \\ \cite{jiang2025multi},\cite{vazia2024material}
	} ;
	\node(OODODPS) at ([xshift=5.5cm,yshift=-2cm]matmodel)[align=center, inner sep=0pt]{
		Hybrid \\ ODPS
	} ;
	\node(off2) at ([xshift=1.85cm]matmodel)[align=center, inner sep=0pt]{};

	\draw [->,>=latex,line width=2.5pt,draw,out=0,in=180] (off1) to node [midway](mid1) {} (im-TDPS) ;
	\node(im-TDPSlabel) at ([yshift=.5cm]mid1)[align=center, inner sep=0pt]{
	approx. Poisson \\ model \eqref{eq:poisson}\&\eqref{eq:forward2}
	} ;
	
	\draw [->,>=latex,line width=2.5pt,draw,out=20,in=180] (off2) to node [midway](mid1) {} (proj-TDPS) ;
	\node(proj-TDPSlabel) at ([xshift=-.3cm,yshift=.5cm]mid1)[align=center, inner sep=0pt,rotate=50]{
	pseudo Gaussian \\ model \eqref{eq:gaussian}
	} ;
	\draw [->,>=latex,line width=2.5pt,draw,out=0,in=180] (off2) to node [midway](mid1) {} (ODPS) ;
	\node(ODPSlabel) at ([yshift=.5cm,xshift=.6cm]mid1)[align=center, inner sep=0pt]{
	true Poisson \\ model  \eqref{eq:poisson}
	} ;
	
	\draw [->,>=latex,line width=2.5pt,draw,out=-20,in=180] (off2) to node [midway](mid1) {} (OODODPS) ;
	\node(OODODPSlabel) at ([xshift=-.3cm,yshift=-.5cm]mid1)[align=center, inner sep=0pt,rotate=-50]{
	true Poisson \\ model  \eqref{eq:poisson}
	} ;
	
	\draw [->,>=latex,line width=2.5pt,draw,out=0,in=180] (trainingdata.0) to node [midway](mid1) {} (matmodel) ;
	\draw [->,>=latex,line width=2.5pt,draw,out=0,in=180] (spectralct.0) to node [midway](mid1) {} (spectralmodel) ;
	\draw [->,>=latex,line width=2.5pt,draw] (trainingdata.90)  to node [midway](mid1) {} node[midway, left] {$\boldcalF$} (spectralct.270) ;
	
	\draw [-,>=latex,line width=2.5pt,draw,out=0,in=180] (spectralmodel.0) to node [midway](mid1) {} (off1.0) ;
	\draw [-,>=latex,line width=2.5pt,draw,out=0,in=180] (matmodel.0) to node [midway](mid1) {} (off2.0) ;
	

\end{tikzpicture}

%% file: figures/Bigfig2.tex
\begin{figure*}
	\begin{center}
		\begin{tikzpicture}[label distance = -0.2cm,
	     gateway/.style={circle, fill=blue!20!white, draw=blue, thick, inner sep=0pt,
	                     minimum size=6mm, label=below:Gateway},
	     root/.style={circle, fill=red!20!white, draw=red, thick, inner sep=0pt,
	                  minimum size=6mm, label=below:Root},
	     node/.style={circle, fill=gray!20!white, draw=gray, thick, inner sep=0pt,
	                  minimum size=6mm, label=below:Node},
	     every edge/.style={<->, semithick, draw},
	     background fill/.store in=\bgfill,
	     background fill=green!20!white,   
	     fill background/.style={
	       local bounding box=bbox,
	       execute at end scope={
	         \begin{pgfonlayer}{background}
	           \coordinate[rectangle, rounded corners, fill=\bgfill,fit=(bbox)];
	         \end{pgfonlayer}}},       
	     fill /.style={
	       local bounding box=bbox,
	       execute at end scope={
	         \begin{pgfonlayer}{background}
	           \coordinate[rectangle, rounded corners, fill=\bgfill, fit=(bbox)];
	         \end{pgfonlayer}}
	    }]    
	
			\node [opacity=0.9] (x_tprime) {\includegraphics[scale=0.08]{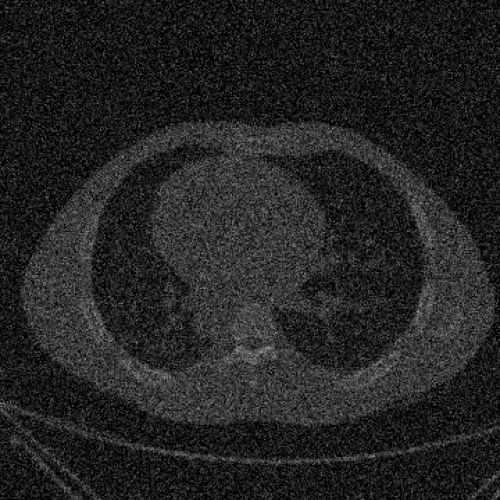}};
			\node[opacity=0.9] (bis1) [above of= x_tprime] {\includegraphics[scale=0.08]{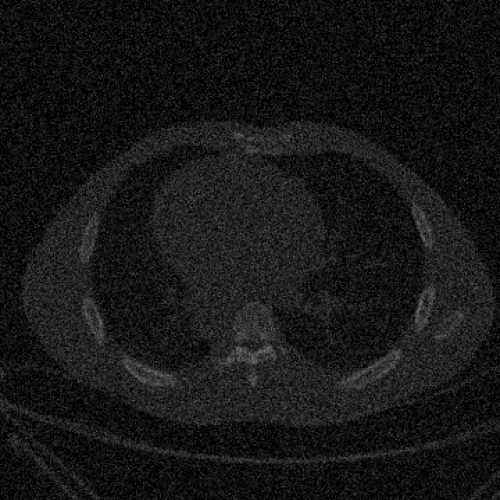}};
	        \node[opacity=0.9, label={-90: {$\boldx_{T'}$}}] (ter1) [below of= x_tprime] {\includegraphics[scale=0.08]{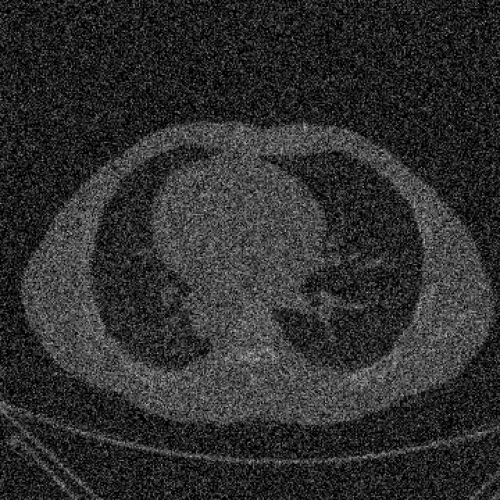}};
	        
	        \node [opacity=0.9, label={-90: {$ \cdots $}}] (dots1) [right of=x_tprime, xshift=0.5cm] {} ; 
	        
	        \node [opacity=0.9] (x_t) [right of=dots1, xshift=1cm]  {\includegraphics[scale=0.08]{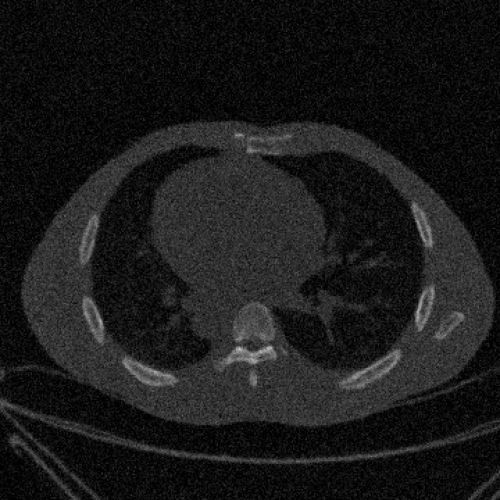}}; 
			\node[opacity=0.9] (bis) [below left of= x_t, yshift=0.5cm, xshift=0.5cm] {\includegraphics[scale=0.08]{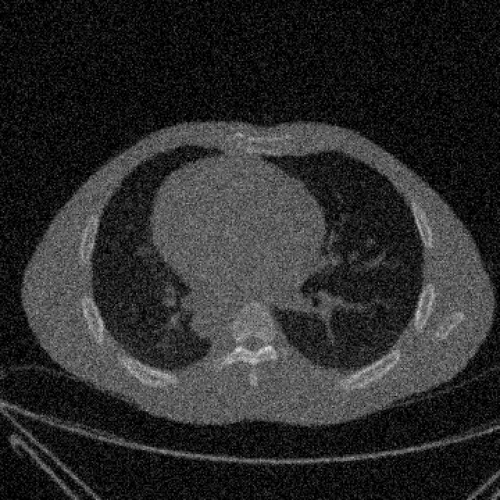}};
	        \node[opacity=0.9, label={-90: {$\boldx_{t}$}}] (ter) [below left of= bis, yshift=0.5cm, xshift=0.5cm] {\includegraphics[scale=0.08]{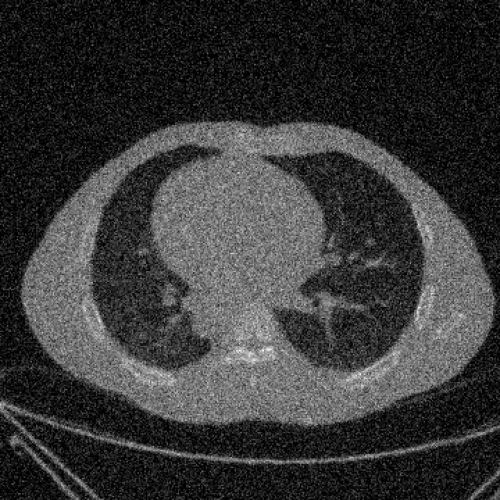}};

			\draw [->] (  x_tprime.east) -- ([xshift=-0.2cm]  dots1.west);
	        \draw [->] ([xshift=0.2cm] dots1.east) -- ([xshift=-0.4cm] x_t.west);

			\begin{scope}[fill background, background fill = green!20!white]
				\node [opacity = 0.7] (x'_t-1) [right of=x_t, xshift=1cm, yshift=1cm] {\includegraphics[scale=0.08]{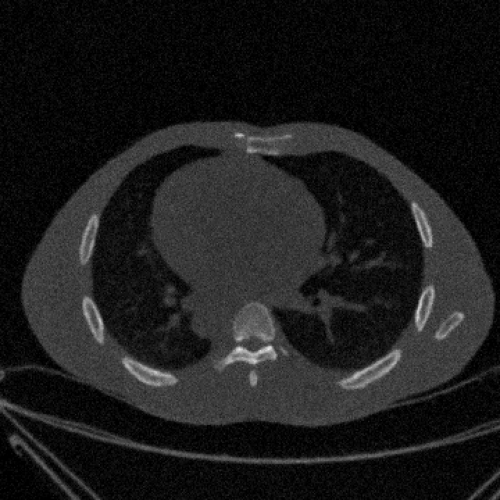}};
				\node[opacity=0.9] (bis) [below left of= x'_t-1, yshift=0.5cm, xshift=0.5cm] {\includegraphics[scale=0.08]{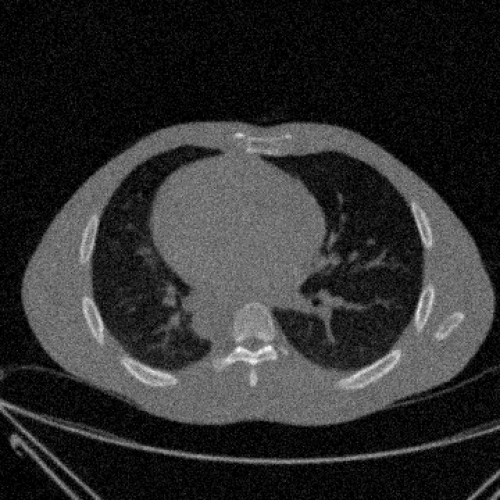}};
	            \node[opacity=0.9, label={-90: {$\boldx'_{t-1}$}}] (ter) [below left of= bis, yshift=0.5cm, xshift=0.5cm] {\includegraphics[scale=0.08]{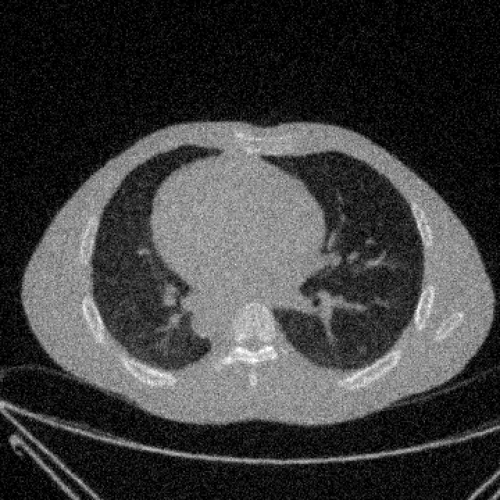}};
	            \node[right of=ter, xshift=0.3cm] {};
			\end{scope}	
	        \node [below of = ter, xshift=1cm, yshift=0.1cm] {\scriptsize Diffusion};
			\begin{scope}[fill background, background fill = blue!20!white]	
	
				\node [opacity = 0.7] (xhat_0) [below of=x'_t-1, yshift = -1.2cm] {\includegraphics[scale=0.08]{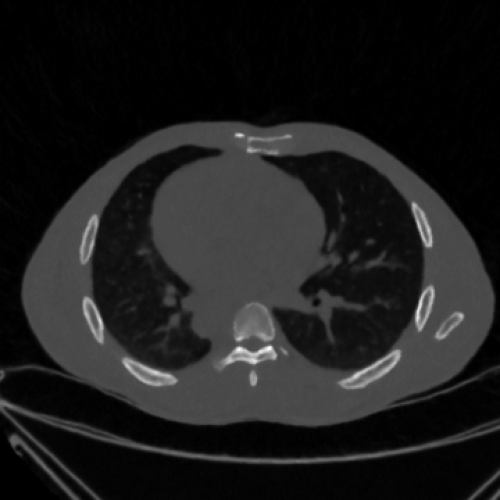}};
				\node[opacity=0.9] (bis) [below left of= xhat_0, yshift=0.5cm, xshift=0.5cm] {\includegraphics[scale=0.08]{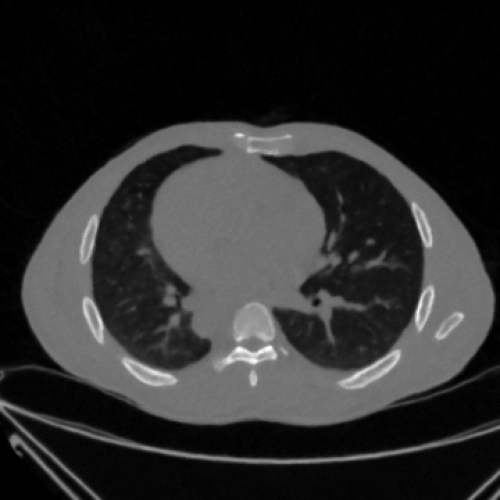}};
				\node[opacity=0.9, label={-90:  {$\hat{\boldx}_{0}$}}] (ter) [below left of= bis, yshift=0.5cm, xshift=0.5cm] {\includegraphics[scale=0.08]{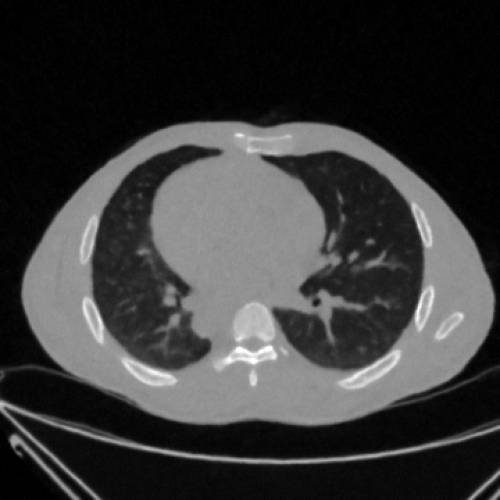}};
				
				\node [opacity=0.9] (y) [below of=xhat_0, yshift = -1cm]{\includegraphics[scale=0.08]{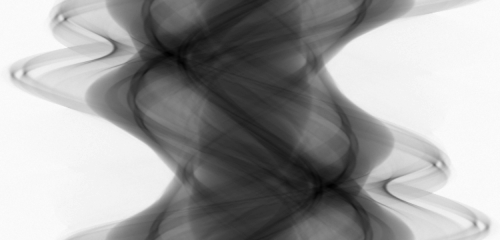}};
	            
			      \node[opacity=0.9] (bis) [below left of= y, yshift=0.5cm, xshift=0.5cm] {\includegraphics[scale=0.08]{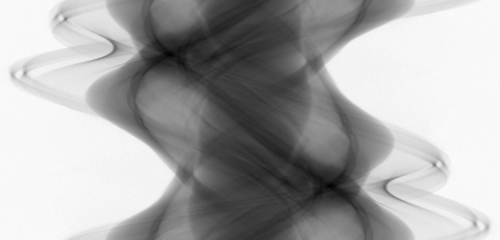}};
			      \node[opacity=0.9, label={-90: {$\boldy$}}] (bis) [below left of= bis, yshift=0.5cm, xshift=0.5cm] {\includegraphics[scale=0.08]{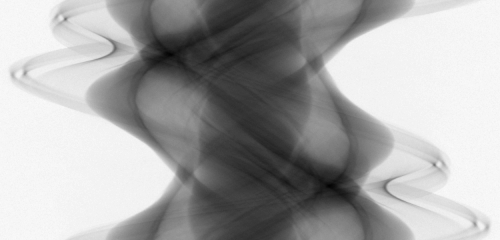}};
				\node (nablalog) [mubox, right of=xhat_0, xshift = 2cm, yshift = -1cm] {$\nabla_{\boldx_t} \log p_{\boldY \mid \boldX = \hat{\boldx}_0(\boldx_t)}(\boldy)$};
			\end{scope}
		
			\draw[->, rounded corners] (x_t.east) |- ([xshift=-0.4cm] x'_t-1.west);
			\draw[->, rounded corners] (x_t.east) |- ([xshift=-0.4cm] xhat_0.west);
			west
			\draw[->, rounded corners]  (xhat_0.east) -- ++(0.5,0) -- ([xshift=-1cm]nablalog.north);
			\draw[->, rounded corners]  (y.east) -- ++(0.5,0) -- ([xshift=-1cm]nablalog.south);
			\node (plus) [plusbox, right of =x'_t-1, xshift= 2 cm, yshift=-1cm] {$+$};
			\draw[->, rounded corners] (x'_t-1.east)  -|  (plus.north);
			\draw[->]  (nablalog.north) |- (plus.south);
	
	        \node [opacity=0.9] (x_t-1) [right of=plus, xshift=1cm]  {\includegraphics[scale=0.08]{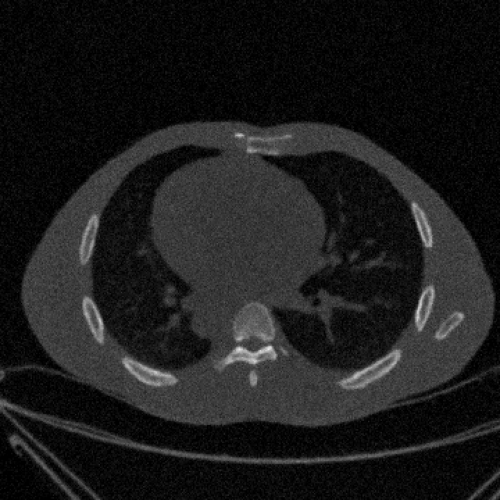}}; 
			\node[opacity=0.9] (bis) [below left of= x_t-1, yshift=0.5cm, xshift=0.5cm] {\includegraphics[scale=0.08]{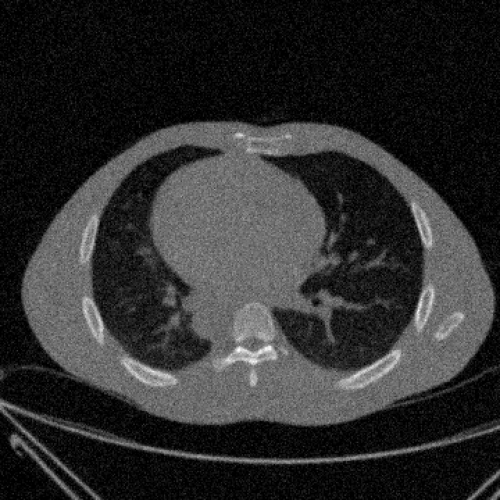}};
			\node[opacity=0.9, label={-90: {$\boldx_{t-1}$}}] (ter) [below left of= bis, yshift=0.5cm, xshift=0.5cm] {\includegraphics[scale=0.08]{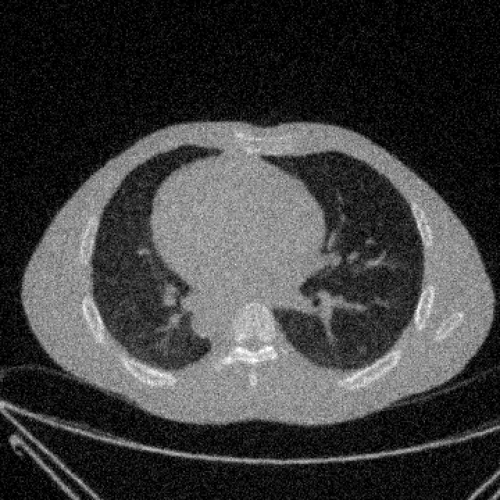}};
	        
	        \node [opacity=0.9, label={-90: {$ \cdots $}}] (dots2) [right of=x_t-1, xshift=0.5cm] {} ; 
	        
	        \node [label={-90: {$\boldx_{0}$}}] (x_t0)[right of= dots2, xshift=0.5cm]  {\includegraphics[scale=0.08]{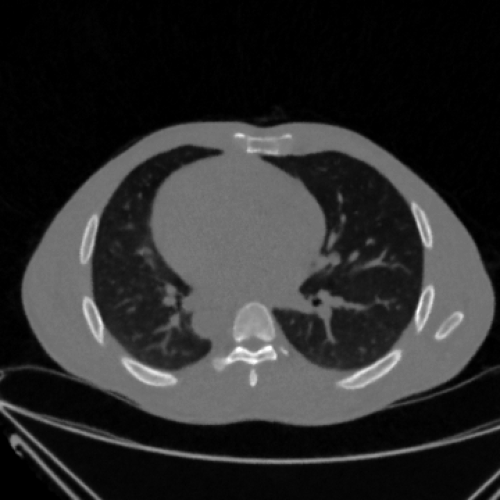}}; 
			\node [label={-90: {$\boldx_{0}$}}] (bis) [below of= x_t0] {\includegraphics[scale=0.08]{ImagesBigfig2/xtd_1.png}};
			\node (ter2) [above of= x_t0] {\includegraphics[scale=0.08]{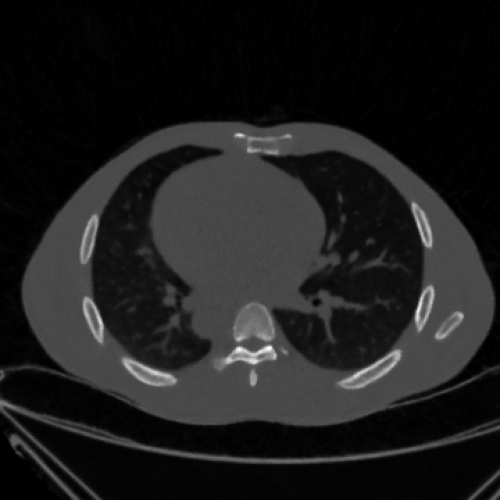}};
	
			\draw [->] ( x_t-1.east) --  ([xshift=-0.2cm] dots2.west);
	        \draw [->] ([xshift=0.2cm] dots2.east) -- (x_t0.west);
	
			\draw[->] (plus.east) -- ([xshift=-0.4cm] x_t-1.west); 
			\node[below of = nablalog, yshift = -1cm, xshift=1.3cm] {\scriptsize Data Fidelity};
		    \node[below of =y, yshift = 0.1cm] (ysouth) {};
			\node[draw = black, rounded corners, fit=(x_tprime) (xhat_0) (ysouth) (nablalog) (x'_t-1) (plus) (x_t) (x_t-1) (x_t0) (ter2) (ter1) , label={[yshift=-0.2cm]-90: {Reconstruction}}]  (fit1) {};
	
	        \node (z_0) [right of= x_t0, xshift=1cm, yshift=0.5cm]  {\includegraphics[scale=0.08]{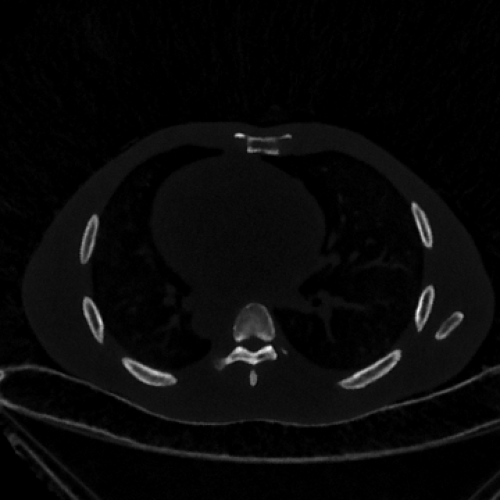}}; 
	        \node (z_1) [right of= x_t0, xshift=1cm, yshift=-0.5cm, label={-90: {$\boldz_0$}}]  {\includegraphics[scale=0.08]{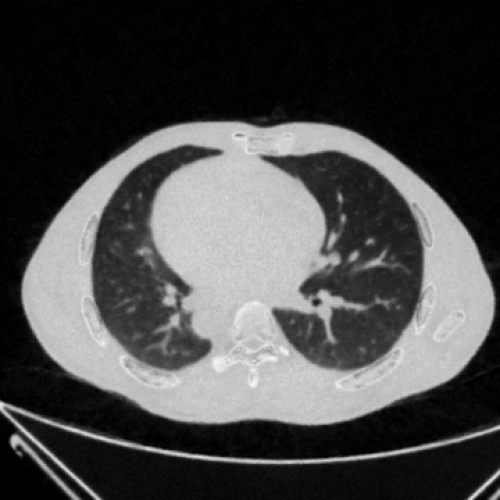}}; 
	
	        \draw[->, rounded corners] (x_t0.east) -- ([yshift=-0.5cm] z_0.west) node[midway, below]{$\boldcalF_{\mathrm{discr}}^{\dagger}$};
	
	        \node[right of = ysouth, xshift=7cm] (ysouth2) {};
	        \node[draw=black, rounded corners, fit=(z_0) (z_1) (x_t0) (bis) (ter2) (ysouth2), label={[yshift=-0.2cm]-90: {Decomposition}}] (fit2) {};

		\end{tikzpicture}
	    \caption{
	    	Summary of the two-step \ac{ImTDPS} method. The first step is the reconstruction of spectral images using the image domain formulation \eqref{eq:image_domain1} with a prior learned by a diffusion model. Afterward, the material decomposition \eqref{eq:image_domain2} is obtained with the pseudoinverse $\boldcalF_{\mathrm{discr}}^\dagger$. In order to show the overall effect of the algorithm, images presented in this figure are 200 steps apart.
	    	}
	    \label{fig:ImTDPS_Bigfig}   
	
	\end{center}

\end{figure*}

%% file: figures/Bigfig.tex
\begin{figure*}
	\begin{center}
		\begin{tikzpicture}[label distance = -0.2cm,
	     gateway/.style={circle, fill=blue!20!white, draw=blue, thick, inner sep=0pt,
	                     minimum size=6mm, label=below:Gateway},
	     root/.style={circle, fill=red!20!white, draw=red, thick, inner sep=0pt,
	                  minimum size=6mm, label=below:Root},
	     node/.style={circle, fill=gray!20!white, draw=gray, thick, inner sep=0pt,
	                  minimum size=6mm, label=below:Node},
	     every edge/.style={<->, semithick, draw},
	     background fill/.store in=\bgfill,
	     background fill=green!20!white,   
	     fill background/.style={
	       local bounding box=bbox,
	       execute at end scope={
	         \begin{pgfonlayer}{background}
	           \coordinate[rectangle, rounded corners, fill=\bgfill,fit=(bbox)];
	         \end{pgfonlayer}}},       
	     fill /.style={
	       local bounding box=bbox,
	       execute at end scope={
	         \begin{pgfonlayer}{background}
	           \coordinate[rectangle, rounded corners, fill=\bgfill, fit=(bbox)];
	         \end{pgfonlayer}}
	    }]    
			\node (z_t+1) {\includegraphics[scale=0.1]{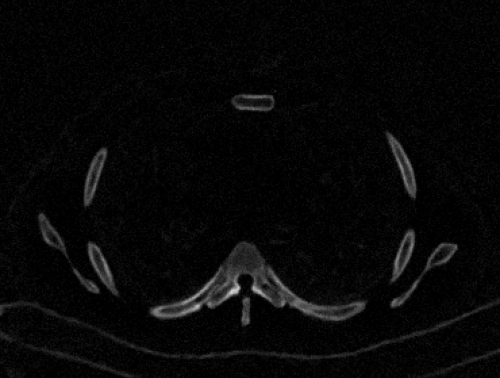}};
			\node (bis) [label={-90: {$\boldz_{t+1}$}}, below of= z_t+1] {\includegraphics[scale=0.1]{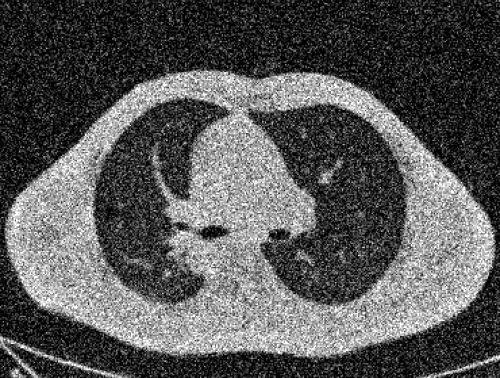}};
	
	        \node [label={-90: {$\cdots$}}] (dots1) [left of =z_t+1, xshift=-0.7cm, yshift = -0.5cm] {};
	        
			\node (z_t) [right of=z_t+1, xshift=1cm]  {\includegraphics[scale=0.1]{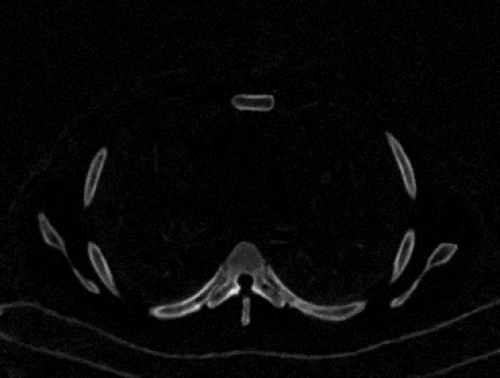}}; 
			\node[label={-90: {$\boldz_{t}$}}] (zt_bis) [below of= z_t] {\includegraphics[scale=0.1]{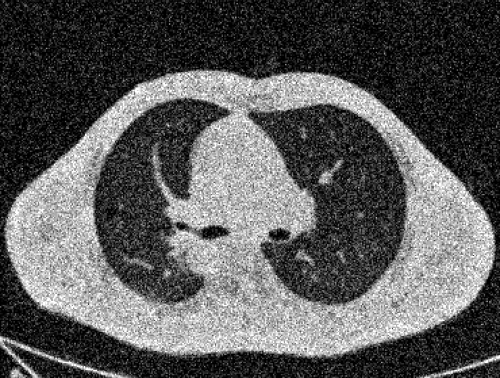}};
	        
			\node (z_t-1) [right of=z_t, xshift=5cm]  {\includegraphics[scale=0.1]{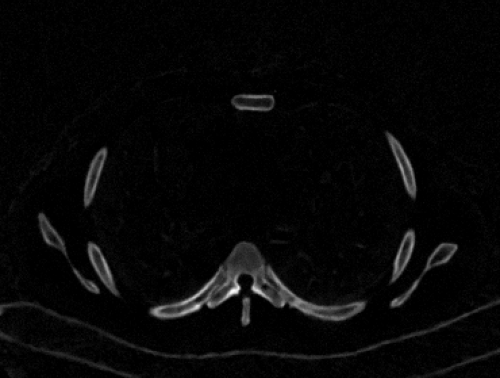}}; 
			\node[label={-90: {$\boldz_{t-1}$}}] (zt-1_bis) [below of= z_t-1] {\includegraphics[scale=0.1]{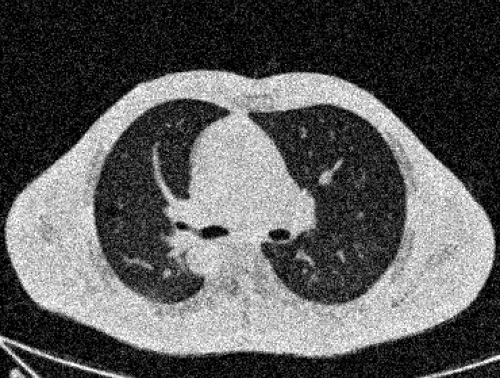}};
	        
			\node (z_t-2) [right of=z_t-1, xshift=1cm]  {\includegraphics[scale=0.1]{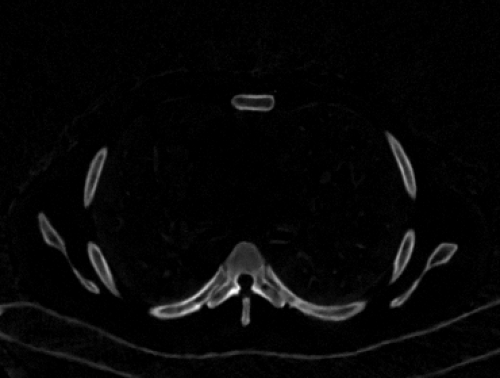}}; 
			\node[label={-90: {$\boldz_{t-2}$}}] (bis) [below of= z_t-2] {\includegraphics[scale=0.1]{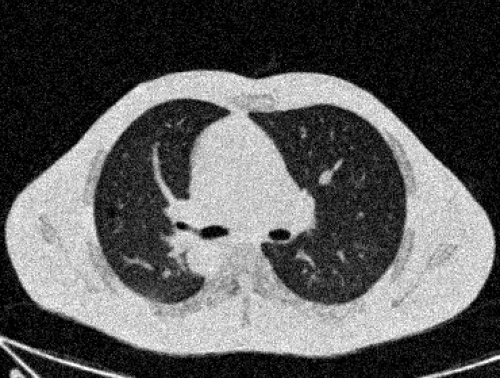}};
	        
	        \node [label={-90: {$\cdots$}}] (dots2) [right of =z_t-2, xshift=0.7cm, yshift = -0.5cm]  {};
	        
			\draw [->] ( [yshift =-0.5cm] z_t+1.east) -- ( [yshift =-0.5cm] z_t.west);
			\draw [->] ([yshift =-0.5cm] z_t.east) --  ([yshift =-0.5cm] z_t-1.west);
			\draw [->] ([yshift =-0.5cm] z_t-1.east) --  ([yshift =-0.5cm] z_t-2.west);
	        \draw [->] ([xshift=0.2cm] dots1.east) -- ([yshift =-0.5cm] z_t+1.west);
	        \draw [->] ([yshift=-0.5cm] z_t-2.east) -- ([xshift=-0.2cm] dots2.west);
			\node[below of =zt_bis, yshift=0.3cm] (ztsouth) {};
			\node[below of =zt-1_bis, yshift=0.3cm] (zt-1south) {};
			\node[draw = black, rounded corners, fit=(z_t) (z_t-1) (ztsouth) (zt-1south)] (boxtop) {};

			\begin{scope}[fill background, background fill = green!20!white]
				\node  (z'_t-1) [below of=z_t+1, yshift = -2.5cm, xshift=-1cm] {\includegraphics[scale=0.1]{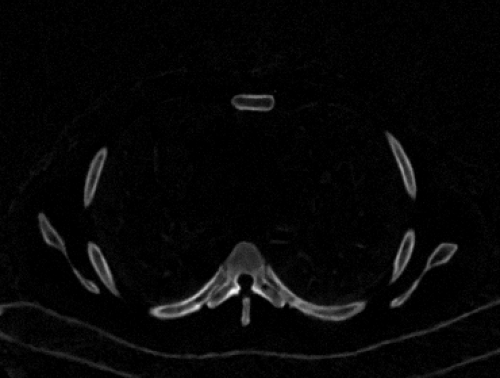}};
				\node[label={-90: {$\boldz'_{t-1}$}}] (fitnorth) [below of= z'_t-1] {\includegraphics[scale=0.1]{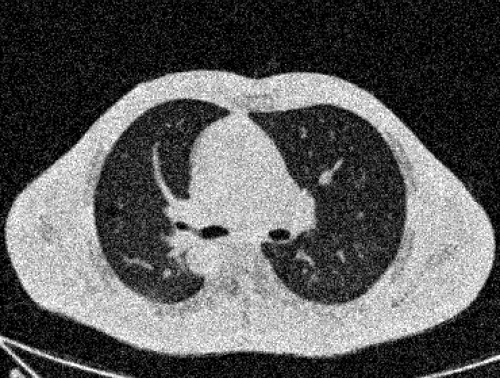}};
	            \node[right of = z'_t-1, xshift=0.5cm]{};
			\end{scope}	
	        \node [below of = fitnorth, xshift=1.2cm, yshift = 0.3cm] {\scriptsize Diffusion};
			\begin{scope}[fill background, background fill = blue!20!white]	
				\node  (zhat_0) [below of=z'_t-1, yshift = -1.6cm] {\includegraphics[scale=0.1]{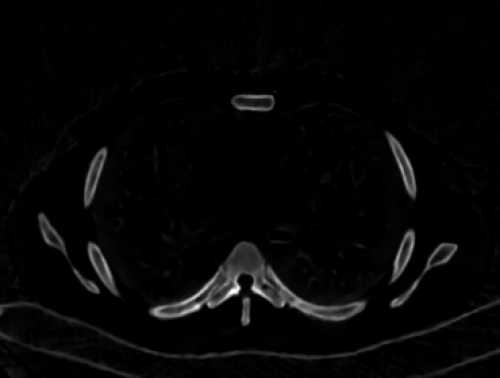}};
				\node[label={-90:  {$\hat{\boldz}_{0}(\boldz_t)$}}] (bis) [below of= zhat_0] {\includegraphics[scale=0.1]{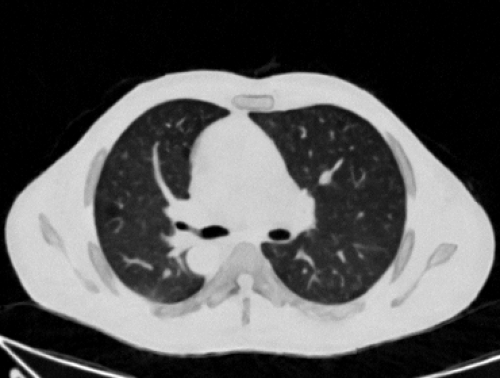}};
				
				\node [opacity=0.6, label={-90: {$\boldy$}}] (y) [below of=zhat_0, yshift = -1.8cm]{\includegraphics[scale=0.1]{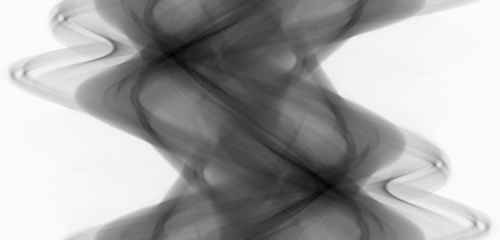}};
	
			\node[opacity=0.6] (bis) [above right of= y, yshift=-0.5cm, xshift=-0.5cm] {\includegraphics[scale=0.1]{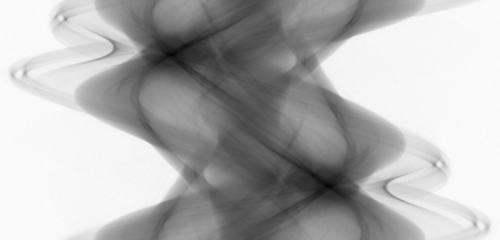}};
			\node[opacity=0.6] (bis) [above right of= bis, yshift=-0.5cm, xshift=-0.5cm] {\includegraphics[scale=0.1]{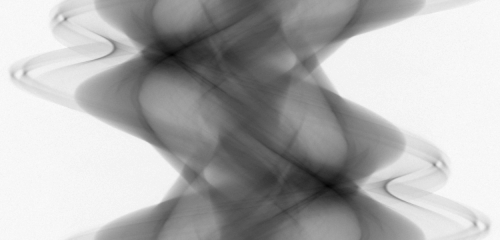}};
				\node (nablalog) [mubox, right of=zhat_0, xshift = 2cm, yshift = -1.5cm] {$\nabla_{\boldz_t} \log p_{\boldY \mid \boldZ = \hat{\boldz}_0(\boldz_t)}(\boldy)$};
	        \node [below of = y, yshift=0.2cm] (ODPSfill) {};
			\end{scope}
			
			\node  (z_t_odps) [left of=z'_t-1, xshift = -1cm] {\includegraphics[scale=0.1]{ImagesBigfig/ztb_0.png}};
			\node [ label={-90: {$\boldz_{t}$}}] (bis) [below of= z_t_odps] {\includegraphics[scale=0.1]{ImagesBigfig/ztb_1.png}};
					
			\draw[-latex] ([yshift =-0.5cm] z_t_odps.east) -- ([yshift =-0.5cm] z'_t-1.west);
			\draw[-latex, rounded corners] ([yshift =-0.5cm] z_t_odps.east) |- ([yshift =-0.5cm] zhat_0.west);
			west
			\draw[-latex, rounded corners]  ([yshift =-0.5cm] zhat_0.east) -- ++(0.5,0) -- ([xshift=-1cm]nablalog.north);
			\draw[-latex, rounded corners]  ([xshift=0.3cm] y.east) -- ++(0.5,0) -- ([xshift=-1cm]nablalog.south);
			\node (plus) [plusbox, right of =z'_t-1, xshift= 2 cm, yshift =-0.5cm] {$+$};
			\draw[-latex] ([yshift =-0.5cm] z'_t-1.east)  --  (plus.west);
			\draw[-latex]  (nablalog.north) -- (plus.south);
	
			\node  (z_t-1odps) [right of=plus, xshift = 1cm, yshift=0.5cm] {\includegraphics[scale=0.1]{ImagesBigfig/ztc_0.png}};
			\node[ label={-90: {$\boldz_{t-1}$}}] (bis) [below of= z_t-1odps] {\includegraphics[scale=0.1]{ImagesBigfig/ztc_1.png}};
			
			\draw[-latex] (plus.east) -- ([yshift =-0.5cm] z_t-1odps.west); 
			\node[below of = nablalog, yshift = -1.2cm, xshift=1cm] {\scriptsize Data Fidelity};
		    \node[below of =y, yshift = 0.2cm] (ysouth) {};
			\node[draw = black, rounded corners,  fit=(zhat_0) (ysouth) (nablalog) (z'_t-1) (plus) (z_t_odps) (z_t-1odps) (fitnorth), label={[yshift=-0.2cm]-90: {(a) ODPS}}]  (fit1) {};

			\begin{scope}[fill background, background fill = green!20!white]
				\node  (z'_t-1) [right of=z_t-1odps, xshift=3cm] {\includegraphics[scale=0.1]{ImagesBigfig/ztprime_0.png}};
				\node[ label={-90: {$\boldz'_{t-1}$}}] (fitnorth) [below of= z'_t-1] {\includegraphics[scale=0.1]{ImagesBigfig/ztprime_1.png}};
	            \node[right of = z'_t-1, xshift=0.5cm]{};
			\end{scope}	
	        \node [below of = fitnorth, xshift=1.2cm, yshift = 0.3cm] {\scriptsize Diffusion};
			\begin{scope}[fill background, background fill = blue!20!white]	
				\node  (zhat_0) [below of=z'_t-1, yshift = -1.6cm] {\includegraphics[scale=0.1]{ImagesBigfig/zthat_0.png}};
				\node[ label={-90: {$\hat{\boldz}_{0}(\boldz_t)$}}] (bis) [below of= zhat_0] {\includegraphics[scale=0.1]{ImagesBigfig/zthat_1.png}};
				\node [label={-90: {$\boldellhat(\boldy)$}}] (l_y) [below of=zhat_0, yshift = -1.8cm] {\includegraphics[scale=0.1]{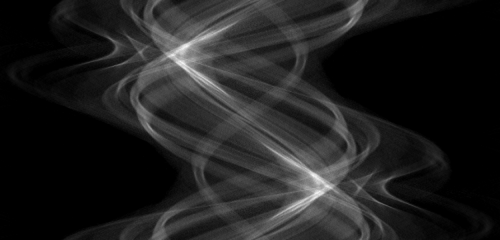}};
				\node[opacity=0.8] (bis) [above right of= l_y, yshift=-0.5cm, xshift=-0.5cm] {\includegraphics[scale=0.1]{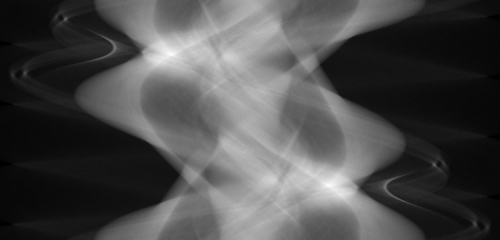}};
				\node (nablalog) [mubox, right of=zhat_0, xshift = 2cm, yshift = -1.5cm] {$\nabla_{\boldz_t} \log p_{\boldL \mid \boldZ = \hat{\boldz}_0(\boldz_t)}\left(\boldellhat\right)$};
			
				\node [opacity=0.8, label={-90: {$\boldy$}}] (y) [left of=l_y, xshift = -1cm]{\includegraphics[scale=0.1]{ImagesBigfig/y_0.png}};
				\node[opacity=0.8] (bis) [above right of= y, yshift=-0.5cm, xshift=-0.5cm] {\includegraphics[scale=0.1]{ImagesBigfig/y_1.png}};
				\node[opacity=0.8] (bis) [above right of= bis, yshift=-0.5cm, xshift=-0.5cm] {\includegraphics[scale=0.1]{ImagesBigfig/y_2.png}};
			\end{scope}

			\node  (z_t_odps) [left of=z'_t-1, xshift = -1cm] {\includegraphics[scale=0.1]{ImagesBigfig/ztb_0.png}};
			\node[ label={-90: {$\boldz_{t}$}}](bis) [below of= z_t_odps] {\includegraphics[scale=0.1]{ImagesBigfig/ztb_1.png}};
			\draw[-latex] ([yshift =-0.5cm] z_t_odps.east) -- ([yshift =-0.5cm] z'_t-1.west);
			\draw[-latex, rounded corners] ([yshift =-0.5cm] z_t_odps.east) |- ([yshift =-0.5cm] zhat_0.west);
			\draw[-latex, rounded corners]  ([yshift =-0.5cm] zhat_0.east) -- ++(0.5,0) -- ([xshift=-1cm]nablalog.north);
			\draw[-latex, rounded corners]  ([xshift =0.1cm] y.east) -- (l_y.west);
			\draw[-latex, rounded corners]  ([xshift=0.1cm] l_y.east) -- ++(0.5,0) -- ([xshift=-1cm]nablalog.south);
			\node (plus) [plusbox, right of =z'_t-1, xshift= 2 cm, yshift =-0.5cm] {$+$};
			\draw[-latex] ([yshift =-0.5cm] z'_t-1.east)  --  (plus.west);
			\draw[-latex]  (nablalog.north) -- (plus.south);
	
			\node  (z_t-1odps) [right of=plus, xshift = 1cm, yshift =0.5cm] {\includegraphics[scale=0.1]{ImagesBigfig/ztc_0.png}};
			\node[label={-90: {$\boldz_{t-1}$}}] (bis) [below of= z_t-1odps] {\includegraphics[scale=0.1]{ImagesBigfig/ztc_1.png}};
			\draw[-latex] (plus.east) -- ([yshift =-0.5cm] z_t-1odps.west); 
			\node[below of = nablalog, yshift = -1.2cm, xshift=0.9cm] {\scriptsize Data Fidelity};
		    \node[below of =y, yshift=0.2cm] (ysouth) {};
			\node[draw = black, rounded corners,
	        fit=(zhat_0) (ysouth) (nablalog) (z'_t-1) (plus) (z_t_odps) (z_t-1odps) (fitnorth) , label={[yshift=-0.2cm] -90: {(b) proj-DPS}}]  (fit2) {};
	
	        \draw[-, thick] (boxtop.south) -- (fit1.north);
	        \draw[-, thick] (boxtop.south) -- (fit2.north);

		\end{tikzpicture}
	    \caption{Summary of the \ac{ODPS} and \ac{ProjTDPS} methods. First row is the global reverse diffusion process. Second row presents one iteration of (a) \ac{ODPS}  and (b) \ac{ProjTDPS}. For \ac{ProjTDPS}, the data fidelity term is applied on an approximation $\hat{\boldsymbol{\ell}}(\boldy)$ of the material sinograms obtained by solving \eqref{eq:projmethod1}. In order to show the overall effect of both algorithms, images presented in this Figure are $200$ steps apart.}
	    \label{fig:ODPS_Bigfig}   
	
	\end{center}
\end{figure*}

%% file: content/results.tex
\section{Results}\label{sec:results}

All reconstruction methods and simulations were implemented in Python. The models were implemented and trained using PyTorch, and we used TorchRadon \cite{torch_radon} for the \ac{2D} \ac{CT} fan-beam projector $\mathcal{A}$.

Our projector incorporates 750 detectors, each with a width of 1.2~mm, with a source-to-origin distance of 624~mm and an origin-to-detector distance of 325~mm \cite{li2023coronaryatheroscleroticplaquecharacterization}. 

We used the \ac{SSIM} and the \ac{PSNR} from the \texttt{skimage.metrics} library as metrics for evaluation. In addition, we used \ac{LPIPS} from \citeauthor{zhang2018perceptual}~\cite{zhang2018perceptual} (version~0.1, AlexNet \ac{NN}), where each image was linearly scaled to $[-1,1]$. The metrics were computed with respect to a reference material image obtained during data preparation.

\subsection{Data Preparation}

We considered $\Nm = 2$ materials: soft tissues and bones. The mass attenuation coefficients of the materials used to define $\boldcalF$, with a 1-keV energy discretization, were derived using the Spektr toolkit \cite{spektr}. 

The dataset used for this experiment consists of 11 \ac{3D} chest \ac{CT} patient images for five full-dose energy bins (40, 60, 80, 100, and 120~keV) acquired on a Philips~IQon~Spectral~CT from Poitiers University Hospital, France. We performed three partitions of the dataset, where, for each partition, one patient is kept for testing while the others are used for training all \ac{NN} models. Therefore, for each partition, models were trained on nine patients and tested on one patient. Each \ac{3D} volume comprises approximately 350 slices of dimension 512\texttimes{}512 with 1-mm pixel size, leading to a testing dataset of roughly 1,000 images. The results presented in Tables~\ref{tab:TableLowCount}, \ref{tab:TableSparseView}, and \ref{tab:TableHyb}, as well as in Figures~\ref{fig:boxplot_lc}, \ref{fig:boxplot_sv}, and \ref{fig:boxplot_hyb}, were averaged over all slices of all testing patients.

Material images for training and reference were obtained by applying $\boldcalF_{\mathrm{discr}}^\dag$ to the clean attenuation images. The \ac{PCCT}-simulated $\boldy_k$ measurements were generated from the reference material images $\boldz$ using the forward model described in \eqref{eq:poisson} and \eqref{eq:forward}. 

For Experiment~1 (Section~\ref{sec:no_iodine}), the data were generated from bones and soft tissues using the energy spectrum shown in Figure~\ref{fig:binned_spectrum} with $\Ne=3$ energy bins: 10--40~keV, 40--60~keV, and 60--120~keV. We considered two cases: (i) full-view (360 angles) and (ii) sparse-view (60 angles). The total expected number of photons is $\sum_k\int h_{k}(e)\,\rmd e \approx 55,000$ for both (i) and (ii), to simulate a low-dose setting.

For Experiment~2 (Section~\ref{sec:iodine}), we considered additional artificial iodine distributions, with the same energy spectrum as in Experiment~1 but with $\Ne=6$ energy bins obtained by splitting the previous bins into two intervals of equal length. The artificial iodine distributions were created as a random number (between 2 and 5) of randomly generated shapes within the soft tissue, with a peak value at the center of each shape and decreasing towards the edge. The hybrid version of \ac{ODPS}, Hybrid-\ac{ODPS}, was implemented with the Huber regularizer for iodine (cf.~$R$ in \eqref{eq:pseudo_pdf}). 

\subsection{Reconstruction Methods for Comparison}

We compared three \ac{DPS} methods (\ac{ODPS}, \ac{ImTDPS}, and \ac{ProjTDPS}) with two other two-step decomposition methods, as well as a one-step variational method, described in the following subsections. Each method was finely tuned with respect to the above-mentioned metrics. Both \ac{ODPS} and \ac{ProjTDPS} use a \ac{NN} trained on material images, while \ac{ImTDPS} uses a \ac{NN} trained on spectral images. Hybrid-\ac{ODPS} was compared with \ac{LBFGS} (see Section~\ref{sec:os} for description).

\subsubsection{One-step approaches}\label{sec:os}

We used the quasi-Newton \ac{LBFGS} \cite{lbfgs} to solve \eqref{eq:onestep}, where the regularizer (i.e., the negative $\log$-prior) is defined by a mix of Huber regularization \cite{huber2011robust} on image gradients and inner-product regularization between pairs of material images to promote material separability and mitigate crosstalks \cite{innerproductreg}. 

\Ac{LBFGS} was used in both Experiment~1 and~2.

\subsubsection{Two-step approaches}

We present here the two-step methods, which are also listed in Table~\ref{table:2steps}. These methods were used in Experiment~1 only.

\paragraph{FBP}

The \ac{FBP} approach consists in performing the analytical inversion of the projector $\boldcalA$, denoted $\boldcalA^\dag$, applied to the approximated sinograms $\boldb_k = [b_{1,k},\dots,b_{\Nb,k}]\in\mathbb{R}^\Nb$, where $b_{i,k} = \log (\bar{h}_{k}/y_{i,k})$ for all $(i,k)$, derived from the measurements $\boldy = \{\boldy_k\}_{k=1}^{\Nb}$. It is then followed by the application of $\boldcalF^\dag_{\mathrm{discr}}$ to decompose the multi-energy images into material images: 
\begin{align*}
	\boldxhat &= \left\{\boldcalA^\dag (\boldb_k)\right\}_{k=1,\dots,\Ne}, \\
	\boldzhat &= \boldcalF_{\mathrm{discr}}^\dag \left( \boldxhat \right) \, .
\end{align*}

\paragraph{U-Net-I and U-Net-P}

Inspired by \citeauthor{abascal}~\cite{abascal}, we implemented two additional \ac{DL}-based methods using the U-Net architecture. U-Net-I is a \ac{CNN} that performs material decomposition in the image domain (i.e., from the multi-energy images ${\boldx_k, k=1,\dots,\Ne}$), while U-Net-P is a \ac{CNN} that decomposes the material in the projection domain to provide the material sinograms for reconstruction. We used \ac{LBFGS} with a Huber regularizer as the negative $\log$-prior for the reconstruction steps of U-Net-I (i.e., solving~\eqref{eq:image_domain1}) and U-Net-P (i.e., solving~\eqref{eq:map_proj}).

\begin{table}
	\centering
	\begin{tabular}{|c|||c|c||c|c|}
		\hline
		Method &  Step 1 & Reg. & Step 2 & Reg. \\
		\hline
		&  \multicolumn{2}{|c||}{Recon.} & \multicolumn{2}{|c|}{Decomp.} \\
		\hline
		FBP & FBP & $\varnothing$ & $\boldcalF_{\mathrm{discr}}^\dagger$ & $\varnothing$ \\
		Im-TDPS & DPS & DM & $\boldcalF_{\mathrm{discr}}^\dagger$ & $\varnothing$ \\
		U-Net-I & L-BFGS & Huber & U-Net-I & NN \\
		\hline \hline
		& \multicolumn{2}{|c||}{Decomp.} & \multicolumn{2}{|c|}{Recon.} \\
		\hline 
		Proj-TDPS & L-BFGS & Huber & DPS & DM \\
		U-Net-P & U-Net-P & NN & FBP & $\varnothing$ \\
		\hline
	\end{tabular}
	\caption{Experiment~1---List of the two-step material decomposition methods used for comparison.}
	\label{table:2steps}
\end{table}

\subsection{Experiment 1---Results with no added materials ($\Nmtilde=0$)}\label{sec:no_iodine}

\subsubsection{Full-view Data}

\input{figures/ResultatsLowCount2}
\input{figures/TableLowCount}
\input{figures/Boxplots}

Figure~\ref{fig:ResultatsLowCount} shows the reconstructed material images on one slice of the test dataset using the proposed methods. The \ac{FBP}-reconstructed images suffer from noise amplification, resulting from the direct inversion of $\boldcalA$, which propagates into the material images. Noise amplification is partly mitigated in the \ac{LBFGS} reconstruction thanks to the regularizer and the use of the true forward model with accurate statistics. 

U-Net-I and U-Net-P produce sharp images that appear similar to the reference images. The \ac{ImTDPS}-reconstructed images suffer from crosstalk (i.e., bones visible in the soft tissue image), which can be attributed to the simplified forward model \eqref{eq:forward2} (in which the X-ray emission is assumed monochromatic for each bin), while \ac{ProjTDPS} produces accurate images. Note that \ac{ProjTDPS} uses the same diffusion prior as \ac{ODPS}, but its data-fidelity term is derived from the approximate model \eqref{eq:gaussian}, which does not utilize the true statistics of the measurement; this could explain the slight decrease in image quality. On the other hand, \ac{ODPS} uses the true forward model and therefore produces sharp and noise-free images. Table~\ref{tab:TableLowCount} presents the \ac{PSNR}, \ac{SSIM}, and \ac{LPIPS} metrics averaged over the test dataset for each of the methods presented, as well as the average computational time. The metrics were computed for each material separately. They corroborate the observations from Figure~\ref{fig:ResultatsLowCount}, with \ac{ODPS} outperforming all other methods, except for \ac{PSNR} and \ac{SSIM} for soft tissue, where U-Net-I performs slightly better.

\input{figures/ResultatsSparseView2}
\input{figures/TableSparseView}
\input{figures/BoxPlotsSparseView}

Figure~\ref{fig:boxplot_lc} presents boxplots computed over the testing dataset for all metrics and materials (we omitted \ac{FBP} as it is largely outperformed). Interestingly, the \ac{DL}-based methods show lower variability compared to \ac{LBFGS}. We believe this could be explained by the noisy \ac{FBP} initialization of \ac{LBFGS}, which increases its randomness. 

\subsubsection{Sparse-view Data}

Figure~\ref{fig:ResultatsSparseView} and Table~\ref{tab:TableSparseView} present the results for the sparse-view experiment. 

\Ac{FBP} images suffer from noise and streak artifacts. The \ac{LBFGS}-based \ac{MD} manages to control noise, but the material images still exhibit streak artifacts, as the handcrafted regularizer is primarily designed for Gaussian noise; this can be addressed with \ac{MD} combining the data-fidelity term with \ac{TV} regularizers, which can be optimized with a primal-dual algorithm \cite{chambolle2011first} or the \ac{ADMM} \cite{boyd2011distributed}.

U-Net-I produces oversmoothed images compared to the previous experiment. This is because the strength of the regularizer was increased to compensate for the sparsity of the measurement, which produces streak artifacts in the reconstruction step. For U-Net-P, the strength of the regularizer was reduced to improve the metrics, resulting in artifacts in the images. In both cases, a \ac{TV} regularizer in the reconstruction step could mitigate these effects.

While \ac{ImTDPS} suffers from the same crosstalks as in the full-view case and \ac{ProjTDPS} has minor artifacts (cf.~green magnified area), the \ac{ODPS} reconstruction remains accurate.

Table~\ref{tab:TableSparseView} shows the same metrics as Table~\ref{tab:TableLowCount} for the sparse-view data. The results are similar, except that this time \ac{ODPS} outperforms all methods. 

Figure~\ref{fig:boxplot_sv} presents the boxplots of the obtained metrics, which are similar to those of Figure~\ref{fig:boxplot_lc}.

\subsection{Experiment 2---Hybridization with added Iodine}\label{sec:iodine}

We used the same imaging setup as in the full-view experiment with two materials. 

We set the parameters and weights of the Huber regularizer on the iodine channel to be the same for both Hybrid-\ac{ODPS} and \ac{LBFGS}, so that the main difference between the two lies in the prior on the bone and soft tissue channels. For this experiment, the metrics were computed on a region that excludes the bed.

Figure~\ref{fig:resultats_hyb} shows the results on one slice. Rows~1 and~2 show bone and soft tissue material maps, which were used to train the \ac{DPS} model, while Row~3 shows the iodine map, which was only present to create the measurement $\boldy$ used for inference. 

Compared to \ac{LBFGS}, Hybrid-\ac{ODPS} shows reduced noise, particularly in the soft tissue and iodine distributions. Hybrid-\ac{ODPS} outperforms \ac{LBFGS} on soft tissue and iodine for all metrics and shows similar results on the bone distribution. Table~\ref{tab:TableHyb} and Figure~\ref{fig:boxplot_hyb} confirm these observations. 

When comparing line profiles, Hybrid-\ac{ODPS} closely matches the reference image, while \ac{LBFGS} shows noise amplification in the soft tissue and iodine images. Both Hybrid-\ac{ODPS} and \ac{LBFGS} have similar lower absolute values for all material images.

Some crosstalk in the iodine image is observed in both the Hybrid-\ac{ODPS} method and \ac{LBFGS}. This occurs due to the similar energy-dependent mass attenuation of the two materials and the absence of a trained prior on the iodine distribution in the Hybrid-\ac{ODPS} method. As a result, bone density is generally underestimated in regions where this crosstalk appears, as shown in the profile in Figure~\ref{fig:resultats_hyb}.

It should be noted that, in general, the metric results show lower values for \ac{PSNR} and \ac{SSIM}, and higher values for \ac{LPIPS}, compared to Experiment~1, as the addition of the third material makes the problem more complex.

This experiment shows that Hybrid-\ac{ODPS} is a viable alternative to \ac{LBFGS}, as it retains the advantages of \ac{ODPS} while incorporating a material that was not present during model training.

\input{figures/Resultats_hyb}
\input{figures/TableHybrid}
\input{figures/BoxPlotsHyb}

%% file: figures/ResultatsLowCount2.tex
\begin{figure*}[h!]
	\centering
	\settoheight{\tempdima}
	{\includegraphics[width=0.115\linewidth]{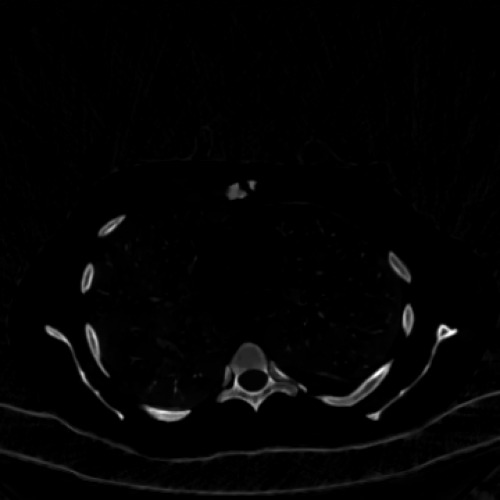}}%
	\footnotesize
	\begin{tabular}
		{@{}c@{\hspace{0.1cm}}c@{}c@{}c@{}c@{}c@{}c@{}c@{}c@{}}
		\vspace{-0.1cm}
		&   Reference & FBP  & L-BFGS & U-Net-I & U-Net-P &   \ac{ImTDPS}  &  \ac{ProjTDPS} &  \ac{ODPS}  \vspace{0.1cm} \\  
		
		\rowname{Bones      } 
		& 
		\begin{tikzpicture}
			\begin{scope}[spy using outlines={rectangle,yellow,magnification=2.5,size=6.6mm,connect spies}]
				\node[inner sep=0pt]{
					\begin{overpic}[height=\tempdima]{Images_lowcount/GT_0.png}
					\end{overpic}
					}; 
				\spy on (0.045,-0.55) in node [right] at (0.42,0.7);
			\end{scope}
		\end{tikzpicture} 
		&
		\begin{tikzpicture}
			\begin{scope}[spy using outlines={rectangle,yellow,magnification=2.5,size=6.6mm,connect spies}]
				\node[inner sep=0pt]{
					\begin{overpic}[height=\tempdima]{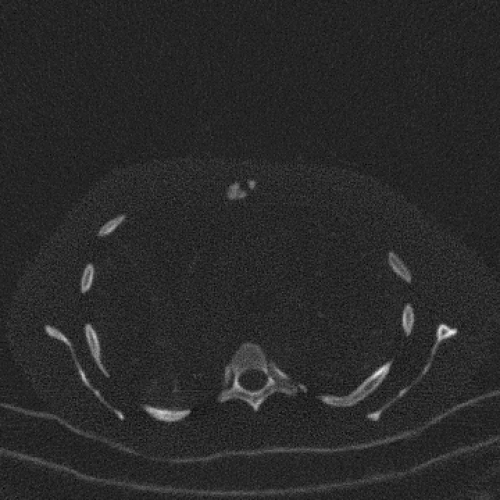}
						\put(3,90){\tiny \textbf{\color{white}{PSNR: 29.35}}}
						\put(3,83){\tiny \textbf{\color{white}{SSIM: 0.41}}}
						\put(3,76){\tiny \textbf{\color{white}{LPIPS: 0.54}}}
					\end{overpic}}
				; 
				\spy on (0.045,-0.55) in node [right] at (0.42,0.7);
			\end{scope}
		\end{tikzpicture} 
		&
		\begin{tikzpicture}
			\begin{scope}[spy using outlines={rectangle,yellow,magnification=2.5,size=6.6mm,connect spies}]
				\node[inner sep=0pt]{
					\begin{overpic}[height=\tempdima]{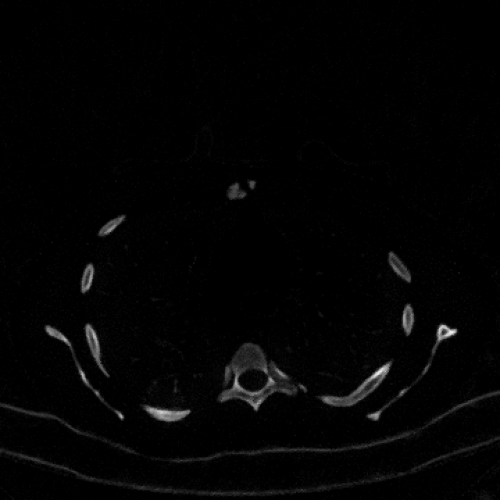}
						\put(3,90){\tiny \textbf{\color{white}{PSNR: 40.17}}}
						\put(3,83){\tiny \textbf{\color{white}{SSIM: 0.93}}}
						\put(3,76){\tiny \textbf{\color{white}{LPIPS: 0.05}}}
				\end{overpic}}
				; 
				\spy on (0.045,-0.55) in node [right] at (0.42,0.7);
			\end{scope}
		\end{tikzpicture} 

		&
		\begin{tikzpicture}
			\begin{scope}[spy using outlines={rectangle,yellow,magnification=2.5,size=6.6mm,connect spies}]
				\node[inner sep=0pt]{
					\begin{overpic}[height=\tempdima]{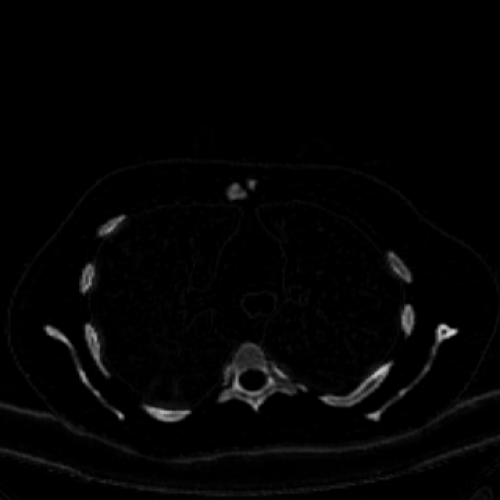}
						\put(3,90){\tiny\textbf{\color{white}{PSNR: 38.15}}}
						\put(3,83){\tiny \textbf{\color{white}{SSIM: 0.92}}}
						\put(3,76){\tiny \textbf{\color{white}{LPIPS: 0.13}}}
				\end{overpic}}
				; 
				\spy on (0.045,-0.55) in node [right] at (0.42,0.7);
			\end{scope}
		\end{tikzpicture} 
		&
		\begin{tikzpicture}
			\begin{scope}[spy using outlines={rectangle,yellow,magnification=2.5,size=6.6mm,connect spies}]
				\node[inner sep=0pt]{
					\begin{overpic}[height=\tempdima]{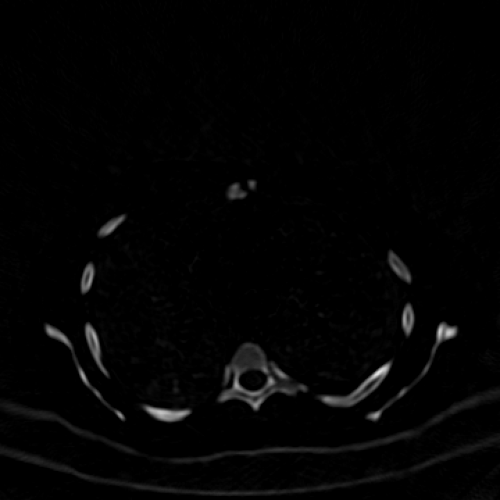}
						\put(3,90){\tiny\textbf{\color{white}{PSNR: 39.39}}}
						\put(3,83){\tiny \textbf{\color{white}{SSIM: 0.96}}}
						\put(3,76){\tiny \textbf{\color{white}{LPIPS: 0.08}}}
				\end{overpic}}
				; 
				\spy on (0.045,-0.55) in node [right] at (0.42,0.7);
			\end{scope}
		\end{tikzpicture} 
	
		&
		\begin{tikzpicture}
			\begin{scope}[spy using outlines={rectangle,yellow,magnification=2.5,size=6.6mm,connect spies}]
				\node[inner sep=0pt]{
					\begin{overpic}[height=\tempdima]{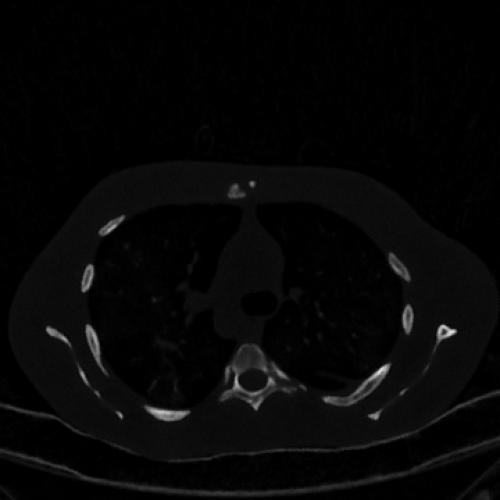}
						\put(3,90){\tiny\textbf{\color{white}{PSNR: 34.20}}}
						\put(3,83){\tiny \textbf{\color{white}{SSIM: 0.81}}}
						\put(3,76){\tiny \textbf{\color{white}{LPIPS: 0.07}}}
				\end{overpic}}
				; 
				\spy on (0.045,-0.55) in node [right] at (0.42,0.7);
			\end{scope}
		\end{tikzpicture} 
		
		&
		\begin{tikzpicture}
			\begin{scope}[spy using outlines={rectangle,yellow,magnification=2.5,size=6.6mm,connect spies}]
				\node[inner sep=0pt]{
					\begin{overpic}[height=\tempdima]{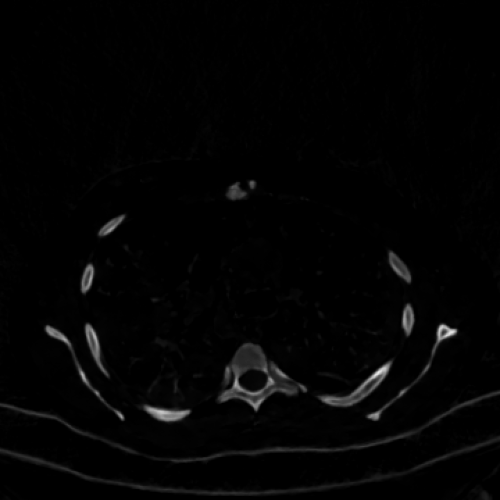}
						\put(3,90){\tiny\textbf{\color{white}{PSNR: 42.48}}}
						\put(3,83){\tiny \textbf{\color{white}{SSIM: 0.95}}}
						\put(3,76){\tiny \textbf{\color{white}{LPIPS: 0.03}}}
				\end{overpic}}
				; 
				\spy on (0.045,-0.55) in node [right] at (0.42,0.7);
			\end{scope}
		\end{tikzpicture} 
		&
		\begin{tikzpicture}
			\begin{scope}[spy using outlines={rectangle,yellow,magnification=2.5,size=6.6mm,connect spies}]
				\node[inner sep=0pt]{
					\begin{overpic}[height=\tempdima]{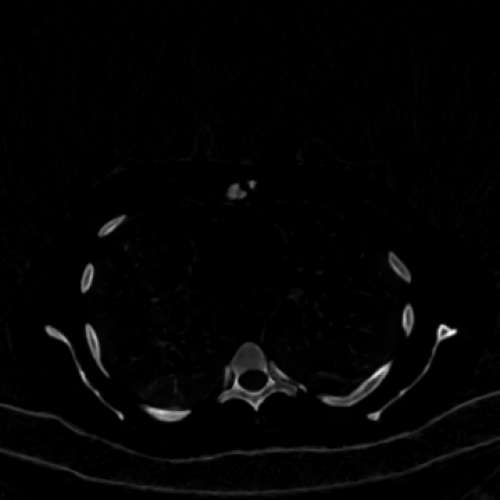}
						\put(3,90){\tiny\textbf{\color{white}{PSNR: 43.39}}}
						\put(3,83){\tiny \textbf{\color{white}{SSIM: 0.97}}}
						\put(3,76){\tiny \textbf{\color{white}{LPIPS: 0.01}}}
				\end{overpic}}
				; 
				\spy on (0.045,-0.55) in node [right] at (0.42,0.7);
			\end{scope}
		\end{tikzpicture} 
		\\
		\vspace{-0.6cm}

		\rowname{Soft Tissues      }  & 
		
		\begin{tikzpicture}
			\begin{scope}[spy using outlines={rectangle,yellow,magnification=2.5,size=6.6mm,connect spies, connect spies}]
				\node[inner sep=0pt]{
					\begin{overpic}[height=\tempdima]{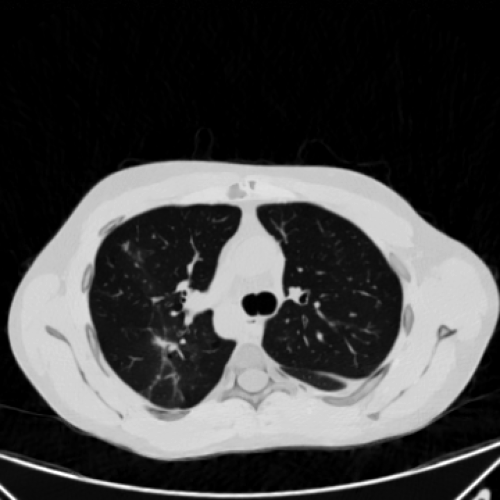}
					\end{overpic}	
					}; 
				\spy on (0.045,-0.55) in node [right] at (0.42,-0.7);
				\spy on (-0.2, -0.15) in node [right] at (0.42, 0.7);
			\end{scope}
		\end{tikzpicture} 
		&
		\begin{tikzpicture}
			\begin{scope}[spy using outlines={rectangle,yellow,magnification=2.5,size=6.6mm,connect spies}]
				\node[inner sep=0pt]{
					\begin{overpic}[height=\tempdima]{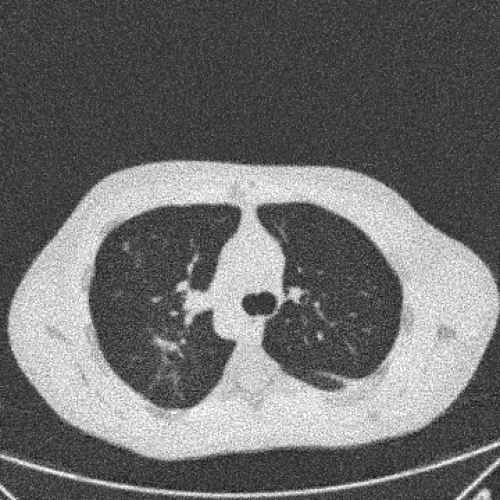}
						\put(3,90){\tiny\textbf{\color{white}{PSNR: 21.79}}}
						\put(3,83){\tiny \textbf{\color{white}{SSIM: 0.17}}}
						\put(3,76){\tiny \textbf{\color{white}{LPIPS: 0.63}}}
				\end{overpic}}
				; 
				\spy on (0.045,-0.55) in node [right] at (0.42,-0.7);
				\spy on (-0.2, -0.15) in node [right] at (0.42, 0.7);
			\end{scope}
		\end{tikzpicture} 
		&	
		\begin{tikzpicture}
			\begin{scope}[spy using outlines={rectangle,yellow,magnification=2.5,size=6.6mm,connect spies}]
				\node[inner sep=0pt]{
					\begin{overpic}[height=\tempdima]{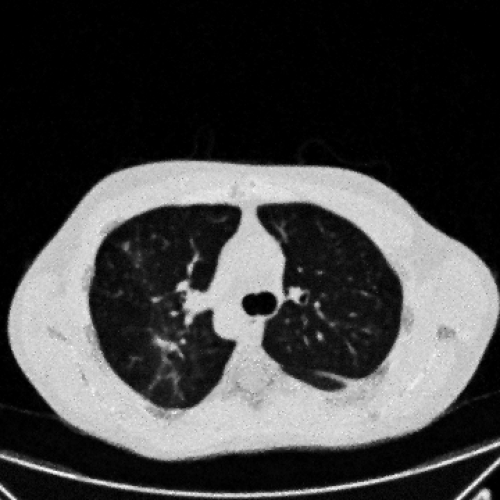}
						\put(3,90){\tiny\textbf{\color{white}{PSNR: 33.60}}}
						\put(3,83){\tiny \textbf{\color{white}{SSIM: 0.89}}}
						\put(3,76){\tiny \textbf{\color{white}{LPIPS: 0.09}}}
				\end{overpic}}
				;  
				\spy on (0.045,-0.55) in node [right] at (0.42,-0.7);
				\spy on (-0.2, -0.15) in node [right] at (0.42, 0.7);
			\end{scope}
		\end{tikzpicture} 
		&
		\begin{tikzpicture}
			\begin{scope}[spy using outlines={rectangle,yellow,magnification=2.5,size=6.6mm,connect spies}]
				\node[inner sep=0pt]{
					\begin{overpic}[height=\tempdima]{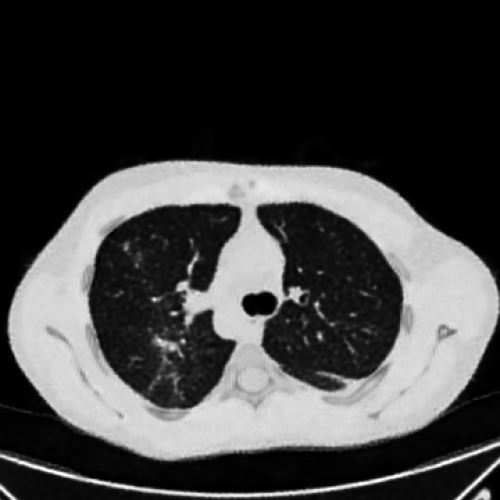}
						\put(3,90){\tiny\textbf{\color{white}{PSNR: 37.11}}}
						\put(3,83){\tiny \textbf{\color{white}{SSIM: 0.94}}}
						\put(3,76){\tiny \textbf{\color{white}{LPIPS: 0.06}}}
				\end{overpic}}
				; 
				\spy on (0.045,-0.55) in node [right] at (0.42,-0.7);
				\spy on (-0.2, -0.15) in node [right] at (0.42, 0.7);
			\end{scope}
		\end{tikzpicture} 
		
		&
		\begin{tikzpicture}
			\begin{scope}[spy using outlines={rectangle,yellow,magnification=2.5,size=6.6mm,connect spies}]
				\node[inner sep=0pt]{
					\begin{overpic}[height=\tempdima]{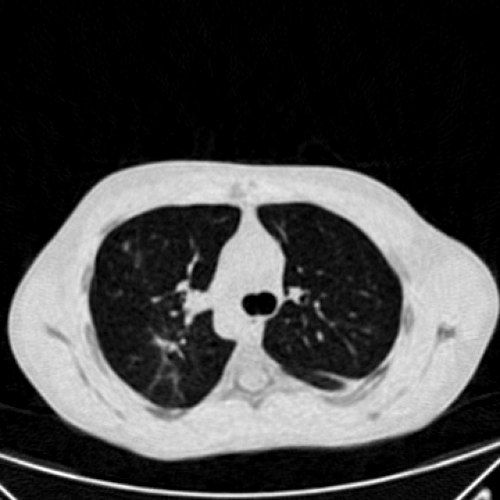}
						\put(3,90){\tiny\textbf{\color{white}{PSNR: 36.22}}}
						\put(3,83){\tiny \textbf{\color{white}{SSIM: 0.91}}}
						\put(3,76){\tiny \textbf{\color{white}{LPIPS: 0.08}}}
				\end{overpic}}
				; 
				\spy on (0.045,-0.55) in node [right] at (0.42,-0.7);
				\spy on (-0.2, -0.15) in node [right] at (0.42, 0.7);
			\end{scope}
		\end{tikzpicture} 

		&
		\begin{tikzpicture}
			\begin{scope}[spy using outlines={rectangle,yellow,magnification=2.5,size=6.6mm,connect spies}]
				\node[inner sep=0pt]{
					\begin{overpic}[height=\tempdima]{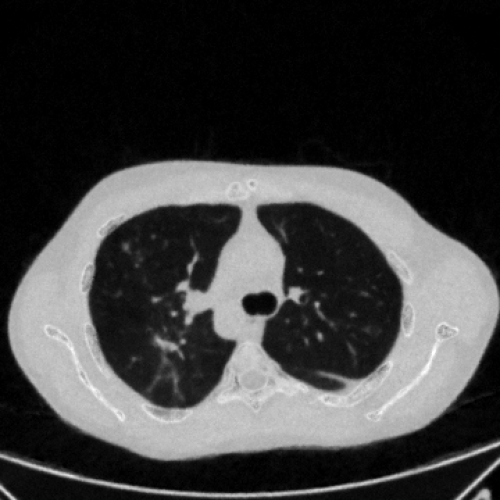}
						\put(3,90){\tiny\textbf{\color{white}{PSNR: 29.20}}}
						\put(3,83){\tiny \textbf{\color{white}{SSIM: 0.85}}}
						\put(3,76){\tiny \textbf{\color{white}{LPIPS: 0.05}}}
				\end{overpic}}
				; 
				\spy on (0.045,-0.55) in node [right] at (0.42,-0.7);
				\spy on (-0.2, -0.15) in node [right] at (0.42, 0.7);
			\end{scope}
		\end{tikzpicture} 
		&	
		\begin{tikzpicture}
			\begin{scope}[spy using outlines={rectangle,yellow,magnification=2.5,size=6.6mm,connect spies}]
				\node[inner sep=0pt]{
					\begin{overpic}[height=\tempdima]{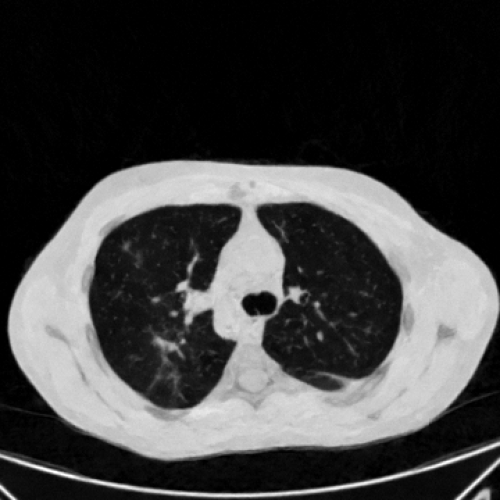}
						\put(3,90){\tiny\textbf{\color{white}{PSNR: 35.50}}}
						\put(3,83){\tiny \textbf{\color{white}{SSIM: 0.89}}}
						\put(3,76){\tiny \textbf{\color{white}{LPIPS: 0.09}}}
				\end{overpic}}
				;                                 
				\spy on (0.045,-0.55) in node [right] at (0.42,-0.7);
				\spy on (-0.2, -0.15) in node [right] at (0.42, 0.7);
			\end{scope}
		\end{tikzpicture} 
		&	
		\begin{tikzpicture}
			\begin{scope}[spy using outlines={rectangle,yellow,magnification=2.5,size=6.6mm,connect spies}]
				\node[inner sep=0pt]{
					\begin{overpic}[height=\tempdima]{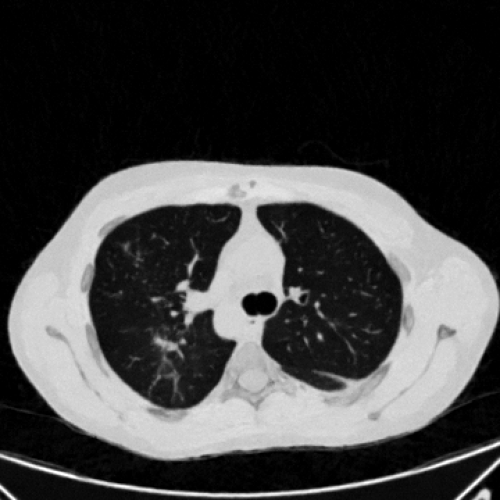}
						\put(3,90){\tiny\textbf{\color{white}{PSNR: 36.76}}}
						\put(3,83){\tiny \textbf{\color{white}{SSIM: 0.92}}}
						\put(3,76){\tiny \textbf{\color{white}{LPIPS: 0.05}}}
				\end{overpic}}
				; 
				\spy on (0.045,-0.55) in node [right] at (0.42,-0.7);
				\spy on (-0.2, -0.15) in node [right] at (0.42, 0.7);
			\end{scope}
		\end{tikzpicture} 

	\end{tabular}	
	\vspace{1cm}
	\caption{Experiment 1 (full-view data)---Material decomposition images obtained with the different methods.}
	\label{fig:ResultatsLowCount}
\end{figure*}	

%% file: figures/TableLowCount.tex
\begin{table*}
	\begin{center}
	    \begin{tabular}{|l|l|c|c|c|c|c|c|c|}
	    \hline  
	    \multicolumn{2}{|c|}{} & \ac{FBP} & \ac{LBFGS} & U-Net-I & U-Net-P & \ac{ImTDPS} & \ac{ProjTDPS} & \ac{ODPS} \\
	    \hline
	    \multicolumn{2}{|c|}{Avg. Time (in sec)} & 0.06 &
	28.59 &
	95.91 &
	88.26 &
	90.51 &
	89.3 &
	134.08\\
	    \hline 
	    \multirow{2}{*}{\scriptsize{PSNR} $\uparrow$} & \scriptsize{Bones} &  
	26.41 &
	37.97 &
	39.99 &
	\textbf{42.48} &
	30.12 &
	41.39 &
	\underline{42.13} \\
	    & \scriptsize{Tissues}  & 
	20.74 &
	31.31 &
	33.06 &
	33.05 &
	25.78 &
	\underline{34.52} &
	\textbf{35.86} \\
	    \hline 
	    \multirow{2}{*}{\scriptsize{SSIM} $\uparrow$} & \scriptsize{Bones} & 
	0.26 &
	0.89 &
	0.90 &
	\textbf{0.96} &
	0.64 &
	0.91 &
	\underline{0.95}\\
	    & \scriptsize{Tissues}  &
	0.12 &
	0.80 &
	\underline{0.86} &
	0.79 &
	0.56 &
	\textbf{0.87} &
	\textbf{0.87} \\
	    \hline 
	    \multirow{2}{*}{\scriptsize{LPIPS} $\downarrow$} & \scriptsize{Bones} &   
	0.64 &
	0.10 &
	0.07 &
	0.06 &
	0.09 &
	\underline{0.04} &
	\textbf{0.02}  \\
	    & \scriptsize{Tissues}  & 
	0.68 &
	0.17 &
	0.12 &
	0.21 &
	0.10 &
	\textbf{0.04} &
	\underline{0.06} \\
	    \hline 
	    \end{tabular}
	    \caption{Experiment 1 (full-view data)---Averaged performance metrics computed over the entire test dataset. Best and second best results are respectively in bold and underlined.}
	    \label{tab:TableLowCount}
	\end{center}
\end{table*}

%% file: figures/Boxplots.tex
\begin{figure*}
	\centering
	\begin{subfigure}[b]{0.30\textwidth}
		\begin{adjustbox}{width=\textwidth}    
			\begin{tikzpicture}
				\begin{axis}[
					height=0.8\textwidth,
					boxplot/draw direction = y,
					x axis line style = {opacity=1},
					axis x line* = bottom,
					axis y line = left,
					enlarge y limits,
					yticklabel style={font=\tiny},
					ymajorgrids,
					xtick = {1, 2, 3, 4, 5, 6, 7, 8, 9, 10, 11, 12},
					xticklabel style = {align=center, font=\tiny, rotate=90},
					xticklabels = {L-BFGS , U-Net-I , U-Net-P , im-TDPS , proj-TDPS , ODPS , L-BFGS , U-Net-I , U-Net-P , im-TDPS , proj-TDPS , ODPS},
					grid=major,
					name=border	
					]
                    \addplot+[
                        draw=RoyalBlue,
                        boxplot prepared={
                            lower whisker = 33.55, 
                            lower quartile = 36.255, 
                            median = 37.67, 
                            upper quartile = 39.52, 
                            upper whisker = 42.27,
                        },
                    ] coordinates {};
                    
                    \addplot+[
                        draw=RoyalBlue,
                        boxplot prepared={
                            lower whisker = 39.04, 
                            lower quartile = 39.32, 
                            median = 39.43, 
                            upper quartile = 41.23, 
                            upper whisker = 41.64,
                        },
                    ] coordinates {};
                    \addplot+[
                        draw=RoyalBlue,
                        boxplot prepared={
                            lower whisker = 41.13, 
                            lower quartile = 41.35, 
                            median = 41.64, 
                            upper quartile = 44.53, 
                            upper whisker = 44.72,
                        },
                    ] coordinates {};
                    
                    \addplot+[
                        draw=RoyalBlue,
                        boxplot prepared={
                            lower whisker = 27.48, 
                            lower quartile = 27.55, 
                            median = 30.17, 
                            upper quartile = 32.8, 
                            upper whisker = 32.9,
                        },
                    ] coordinates {};
                    
                    \addplot+[
                        draw=RoyalBlue,
                        boxplot prepared={
                            lower whisker = 39.41, 
                            lower quartile = 40.13, 
                            median = 41.05, 
                            upper quartile = 42.08, 
                            upper whisker = 44.83,
                        },
                    ] coordinates {};
                    
                    \addplot+[
                        draw=RoyalBlue,
                        boxplot prepared={
                            lower whisker = 40.43,
                            lower quartile = 40.81, 
                            median = 41.63, 
                            upper quartile = 44.02, 
                            upper whisker = 44.43,
                        },
                    ] coordinates {};
                    
                    
\addplot+[ 
 draw=Maroon, 
 boxplot prepared={
 lower whisker = 23.3 ,
 lower quartile = 29.405 ,
 median = 31.81 ,
 upper quartile = 33.48 ,
 upper whisker = 36.28 ,
 }, 
 ] coordinates {};
\addplot+[ 
 draw=Maroon, 
 boxplot prepared={
 lower whisker = 30.67 ,
 lower quartile = 31.05 ,
 median = 33.53 ,
 upper quartile = 34.495 ,
 upper whisker = 35.01 ,
 }, 
 ] coordinates {};
\addplot+[ 
 draw=Maroon, 
 boxplot prepared={
 lower whisker = 31.04 ,
 lower quartile = 31.53 ,
 median = 32.94 ,
 upper quartile = 34.64 ,
 upper whisker = 35.18 ,
 }, 
 ] coordinates {};
\addplot+[ 
 draw=Maroon, 
 boxplot prepared={
 lower whisker = 24.34 ,
 lower quartile = 24.47 ,
 median = 24.64 ,
 upper quartile = 28.3 ,
 upper whisker = 28.4 ,
 }, 
 ] coordinates {};
\addplot+[ 
 draw=Maroon, 
 boxplot prepared={
 lower whisker = 30.06 ,
 lower quartile = 32.045 ,
 median = 35.59 ,
 upper quartile = 36.28 ,
 upper whisker = 36.69 ,
 }, 
 ] coordinates {};
\addplot+[ 
 draw=Maroon, 
 boxplot prepared={
 lower whisker = 30.95 ,
 lower quartile = 33.645 ,
 median = 37.11 ,
 upper quartile = 37.46 ,
 upper whisker = 37.74 ,
 }, 
 ] coordinates {};
				\end{axis}
				\node[above left, xshift = -0.5cm] at (border.north) {\textcolor{RoyalBlue}{\tiny{Bones}}};
				\node[above right, xshift = 0.2cm] at (border.north) {\textcolor{Maroon}{\tiny{Soft Tissues}}};
			\end{tikzpicture}
		\end{adjustbox}
		\caption{PSNR}\label{subfig:psnr_lc}
	\end{subfigure}
	\hfill
	\begin{subfigure}[b]{0.30\textwidth}
		\begin{adjustbox}{width=\textwidth}    
			\begin{tikzpicture}
				\begin{axis}[
					height=0.8\textwidth,
					boxplot/draw direction = y,
					x axis line style = {opacity=1},
					axis x line* = bottom,
					axis y line = left,
					enlarge y limits,
					yticklabel style={font=\tiny},
					ymajorgrids,
					xtick = {1, 2, 3, 4, 5, 6, 7, 8, 9, 10, 11, 12},
					xticklabel style = {align=center, font=\tiny, rotate=90},
					xticklabels = {L-BFGS , U-Net-I , U-Net-P , im-TDPS , proj-TDPS , ODPS , L-BFGS , U-Net-I , U-Net-P , im-TDPS , proj-TDPS , ODPS},
					grid=major,
					name=border
					]
\addplot+[ 
 draw=RoyalBlue, 
 boxplot prepared={
 lower whisker = 0.8 ,
 lower quartile = 0.87 ,
 median = 0.89 ,
 upper quartile = 0.92 ,
 upper whisker = 0.96 ,
 }, 
 ] coordinates {};
\addplot+[ 
 draw=RoyalBlue, 
 boxplot prepared={
 lower whisker = 0.87 ,
 lower quartile = 0.88 ,
 median = 0.91 ,
 upper quartile = 0.92 ,
 upper whisker = 0.92 ,
 }, 
 ] coordinates {};
\addplot+[ 
 draw=RoyalBlue, 
 boxplot prepared={
 lower whisker = 0.94 ,
 lower quartile = 0.94 ,
 median = 0.95 ,
 upper quartile = 0.97 ,
 upper whisker = 0.97 ,
 }, 
 ] coordinates {};
\addplot+[ 
 draw=RoyalBlue, 
 boxplot prepared={
 lower whisker = 0.57 ,
 lower quartile = 0.57 ,
 median = 0.63 ,
 upper quartile = 0.72 ,
 upper whisker = 0.72 ,
 }, 
 ] coordinates {};
\addplot+[ 
 draw=RoyalBlue, 
 boxplot prepared={
 lower whisker = 0.88 ,
 lower quartile = 0.89 ,
 median = 0.9 ,
 upper quartile = 0.9 ,
 upper whisker = 0.91 ,
 }, 
 ] coordinates {};
\addplot+[ 
 draw=RoyalBlue, 
 boxplot prepared={
 lower whisker = 0.94 ,
 lower quartile = 0.94 ,
 median = 0.95 ,
 upper quartile = 0.97 ,
 upper whisker = 0.97 ,
 }, 
 ] coordinates {};
\addplot+[ 
 draw=Maroon, 
 boxplot prepared={
 lower whisker = 0.52 ,
 lower quartile = 0.73 ,
 median = 0.81 ,
 upper quartile = 0.87 ,
 upper whisker = 0.93 ,
 }, 
 ] coordinates {};
\addplot+[ 
 draw=Maroon, 
 boxplot prepared={
 lower whisker = 0.79 ,
 lower quartile = 0.8 ,
 median = 0.86 ,
 upper quartile = 0.91 ,
 upper whisker = 0.92 ,
 }, 
 ] coordinates {};
\addplot+[ 
 draw=Maroon, 
 boxplot prepared={
 lower whisker = 0.73 ,
 lower quartile = 0.75 ,
 median = 0.79 ,
 upper quartile = 0.84 ,
 upper whisker = 0.85 ,
 }, 
 ] coordinates {};
\addplot+[ 
 draw=Maroon, 
 boxplot prepared={
 lower whisker = 0.46 ,
 lower quartile = 0.46 ,
 median = 0.46 ,
 upper quartile = 0.74 ,
 upper whisker = 0.75 ,
 }, 
 ] coordinates {};
\addplot+[ 
 draw=Maroon, 
 boxplot prepared={
 lower whisker = 0.78 ,
 lower quartile = 0.84 ,
 median = 0.88 ,
 upper quartile = 0.9 ,
 upper whisker = 0.9 ,
 }, 
 ] coordinates {};
\addplot+[ 
 draw=Maroon, 
 boxplot prepared={
 lower whisker = 0.78 ,
 lower quartile = 0.81 ,
 median = 0.89 ,
 upper quartile = 0.91 ,
 upper whisker = 0.91 ,
 }, 
 ] coordinates {};
				\end{axis}
				\node[above left, xshift = -0.5cm] at (border.north) {\textcolor{RoyalBlue}{\tiny{Bones}}};
				\node[above right, xshift = 0.2cm] at (border.north) {\textcolor{Maroon}{\tiny{Soft Tissues}}};
			\end{tikzpicture}
		\end{adjustbox}
		\caption{SSIM}\label{subfig:ssim_lc}
	\end{subfigure}
	\hfill
	\begin{subfigure}[b]{0.30\textwidth}
		\begin{adjustbox}{width=\textwidth}    
			\begin{tikzpicture}
				\begin{axis}[
					height=0.8\textwidth,
					boxplot/draw direction = y,
					x axis line style = {opacity=1},
					axis x line* = bottom,
					axis y line = left,
					enlarge y limits,
					yticklabel style={font=\tiny},
					ymajorgrids,
					xtick = {1, 2, 3, 4, 5, 6, 7, 8, 9, 10, 11, 12},
					xticklabel style = {align=center, font=\tiny, rotate=90},
					xticklabels = {L-BFGS , U-Net-I , U-Net-P , im-TDPS , proj-TDPS , ODPS , L-BFGS , U-Net-I , U-Net-P , im-TDPS , proj-TDPS , ODPS},
					grid=major,
					name=border
					]
\addplot+[ 
 draw=RoyalBlue, 
 boxplot prepared={
 lower whisker = 0.03 ,
 lower quartile = 0.07 ,
 median = 0.09 ,
 upper quartile = 0.12 ,
 upper whisker = 0.19 ,
 }, 
 ] coordinates {};
\addplot+[ 
 draw=RoyalBlue, 
 boxplot prepared={
 lower whisker = 0.05 ,
 lower quartile = 0.06 ,
 median = 0.06 ,
 upper quartile = 0.08 ,
 upper whisker = 0.09 ,
 }, 
 ] coordinates {};
\addplot+[ 
 draw=RoyalBlue, 
 boxplot prepared={
 lower whisker = 0.04 ,
 lower quartile = 0.05 ,
 median = 0.06 ,
 upper quartile = 0.07 ,
 upper whisker = 0.07 ,
 }, 
 ] coordinates {};
\addplot+[ 
 draw=RoyalBlue, 
 boxplot prepared={
 lower whisker = 0.06 ,
 lower quartile = 0.07 ,
 median = 0.1 ,
 upper quartile = 0.11 ,
 upper whisker = 0.12 ,
 }, 
 ] coordinates {};
\addplot+[ 
 draw=RoyalBlue, 
 boxplot prepared={
 lower whisker = 0.02 ,
 lower quartile = 0.03 ,
 median = 0.04 ,
 upper quartile = 0.06 ,
 upper whisker = 0.09 ,
 }, 
 ] coordinates {};
\addplot+[ 
 draw=RoyalBlue, 
 boxplot prepared={
 lower whisker = 0.01 ,
 lower quartile = 0.02 ,
 median = 0.02 ,
 upper quartile = 0.03 ,
 upper whisker = 0.04 ,
 }, 
 ] coordinates {};
\addplot+[ 
 draw=Maroon, 
 boxplot prepared={
 lower whisker = 0.04 ,
 lower quartile = 0.1 ,
 median = 0.17 ,
 upper quartile = 0.23 ,
 upper whisker = 0.41 ,
 }, 
 ] coordinates {};
\addplot+[ 
 draw=Maroon, 
 boxplot prepared={
 lower whisker = 0.08 ,
 lower quartile = 0.09 ,
 median = 0.11 ,
 upper quartile = 0.15 ,
 upper whisker = 0.16 ,
 }, 
 ] coordinates {};
\addplot+[ 
 draw=Maroon, 
 boxplot prepared={
 lower whisker = 0.16 ,
 lower quartile = 0.18 ,
 median = 0.22 ,
 upper quartile = 0.23 ,
 upper whisker = 0.26 ,
 }, 
 ] coordinates {};
\addplot+[ 
 draw=Maroon, 
 boxplot prepared={
 lower whisker = 0.08 ,
 lower quartile = 0.08 ,
 median = 0.11 ,
 upper quartile = 0.12 ,
 upper whisker = 0.13 ,
 }, 
 ] coordinates {};
\addplot+[ 
 draw=Maroon, 
 boxplot prepared={
 lower whisker = 0.02 ,
 lower quartile = 0.03 ,
 median = 0.04 ,
 upper quartile = 0.05 ,
 upper whisker = 0.07 ,
 }, 
 ] coordinates {};
\addplot+[ 
 draw=Maroon, 
 boxplot prepared={
 lower whisker = 0.03 ,
 lower quartile = 0.03 ,
 median = 0.03 ,
 upper quartile = 0.12 ,
 upper whisker = 0.14 ,
 }, 
 ] coordinates {};
				\end{axis}
				\node[above left, xshift = -0.5cm] at (border.north) {\textcolor{RoyalBlue}{\tiny{Bones}}};
				\node[above right, xshift = 0.2cm] at (border.north) {\textcolor{Maroon}{\tiny{Soft Tissues}}};
			\end{tikzpicture}
		\end{adjustbox}
		\caption{LPIPS}\label{subfig:lpips_lc}
	\end{subfigure}
	\caption{Experiment 1 (full-view data)--Boxplots of \subref{subfig:psnr_lc} PSNR, \subref{subfig:ssim_lc} SSIM and \subref{subfig:lpips_lc} LPIPS, computed over the test dataset. }\label{fig:boxplot_lc}
\end{figure*}

%% file: figures/ResultatsSparseView2.tex
\definecolor{forestgreen4416044}{RGB}{44,160,44}
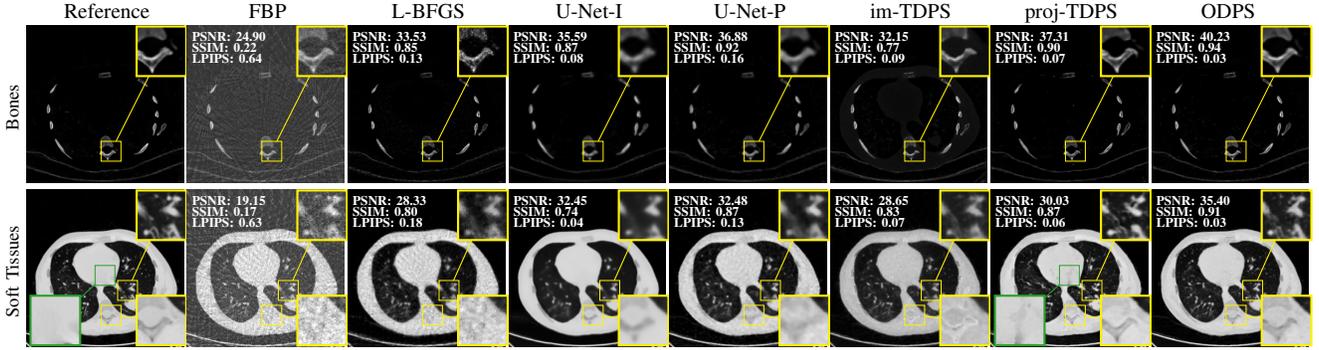
\begin{figure*}[h!]
	\centering
	\settoheight{\tempdima}
	{\includegraphics[width=0.115\linewidth]{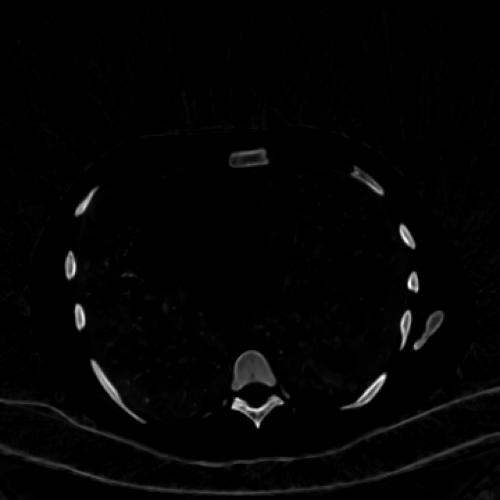}}%
	\footnotesize
	\begin{tabular}
		{@{}c@{\hspace{0.1cm}}c@{}c@{}c@{}c@{}c@{}c@{}c@{}c@{}}
		\vspace{-0.1cm}
		&   Reference & FBP  & L-BFGS & U-Net-I & U-Net-P &   \ac{ImTDPS}  &  \ac{ProjTDPS} &  \ac{ODPS}  \vspace{0.1cm} \\  
		
		\rowname{Bones      } 
		& 
		\begin{tikzpicture}
			\begin{scope}[spy using outlines={rectangle,yellow,magnification=2.5,size=6.6mm,connect spies}]
				\node[inner sep=0pt]{
					\begin{overpic}[height=\tempdima]{Images_sparseview/GT_0.png}
					\end{overpic}
					}; 
				\spy on (0.08,-0.63) in node [right] at (0.42,0.7);
			\end{scope}
		\end{tikzpicture} 
		&
		\begin{tikzpicture}
			\begin{scope}[spy using outlines={rectangle,yellow,magnification=2.5,size=6.6mm,connect spies}]
				\node[inner sep=0pt]{
					\begin{overpic}[height=\tempdima]{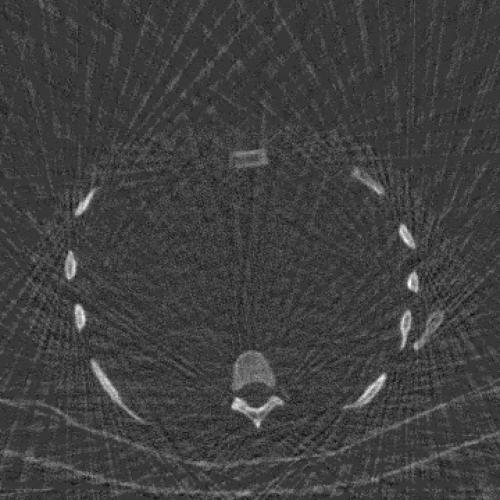}
						\put(3,90){\tiny \textbf{\color{white}{PSNR: 24.90}}}
						\put(3,83){\tiny \textbf{\color{white}{SSIM: 0.22}}}
						\put(3,76){\tiny \textbf{\color{white}{LPIPS: 0.64}}}
					\end{overpic}}
				; 
				\spy on (0.08,-0.63) in node [right] at (0.42,0.7);
			\end{scope}
		\end{tikzpicture} 
		&
		\begin{tikzpicture}
			\begin{scope}[spy using outlines={rectangle,yellow,magnification=2.5,size=6.6mm,connect spies}]
				\node[inner sep=0pt]{
					\begin{overpic}[height=\tempdima]{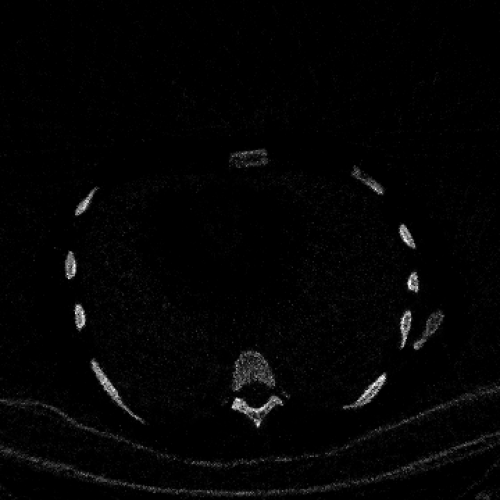}
						\put(3,90){\tiny \textbf{\color{white}{PSNR: 33.53}}}
						\put(3,83){\tiny \textbf{\color{white}{SSIM: 0.85}}}
						\put(3,76){\tiny \textbf{\color{white}{LPIPS: 0.13}}}
				\end{overpic}}
				; 
				\spy on (0.08,-0.63) in node [right] at (0.42,0.7);
			\end{scope}
		\end{tikzpicture} 

		&
		\begin{tikzpicture}
			\begin{scope}[spy using outlines={rectangle,yellow,magnification=2.5,size=6.6mm,connect spies}]
				\node[inner sep=0pt]{
					\begin{overpic}[height=\tempdima]{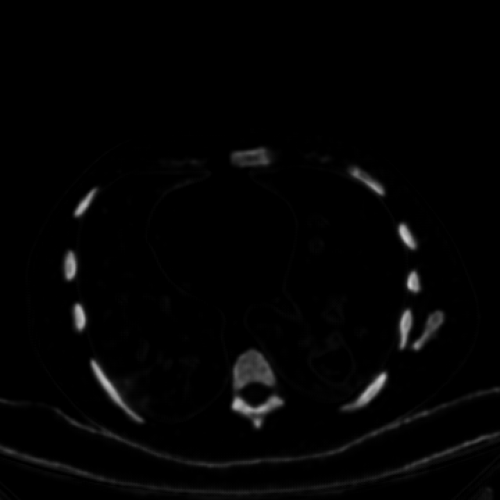}
						\put(3,90){\tiny\textbf{\color{white}{PSNR: 35.59}}}
						\put(3,83){\tiny \textbf{\color{white}{SSIM: 0.87}}}
						\put(3,76){\tiny \textbf{\color{white}{LPIPS: 0.08}}}
				\end{overpic}}
				; 
				\spy on (0.08,-0.63) in node [right] at (0.42,0.7);
			\end{scope}
		\end{tikzpicture} 
		&
		\begin{tikzpicture}
			\begin{scope}[spy using outlines={rectangle,yellow,magnification=2.5,size=6.6mm,connect spies}]
				\node[inner sep=0pt]{
					\begin{overpic}[height=\tempdima]{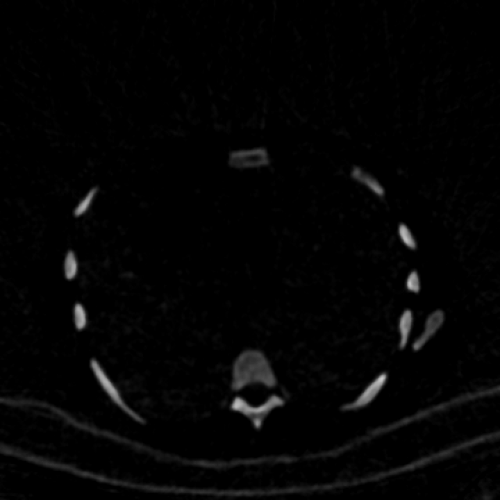}
						\put(3,90){\tiny\textbf{\color{white}{PSNR: 36.88}}}
						\put(3,83){\tiny \textbf{\color{white}{SSIM: 0.92}}}
						\put(3,76){\tiny \textbf{\color{white}{LPIPS: 0.16}}}
				\end{overpic}}
				; 
				\spy on (0.08,-0.63) in node [right] at (0.42,0.7);
			\end{scope}
		\end{tikzpicture} 
	
		&
		\begin{tikzpicture}
			\begin{scope}[spy using outlines={rectangle,yellow,magnification=2.5,size=6.6mm,connect spies}]
				\node[inner sep=0pt]{
					\begin{overpic}[height=\tempdima]{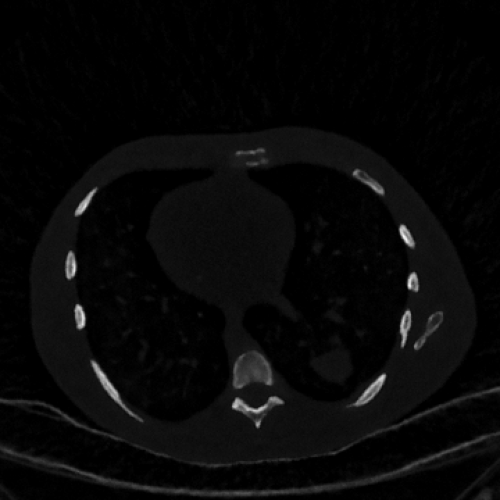}
						\put(3,90){\tiny\textbf{\color{white}{PSNR: 32.15}}}
						\put(3,83){\tiny \textbf{\color{white}{SSIM: 0.77}}}
						\put(3,76){\tiny \textbf{\color{white}{LPIPS: 0.09}}}
				\end{overpic}}
				; 
				\spy on (0.08,-0.63) in node [right] at (0.42,0.7);
			\end{scope}
		\end{tikzpicture} 
		
		&
		\begin{tikzpicture}
			\begin{scope}[spy using outlines={rectangle,yellow,magnification=2.5,size=6.6mm,connect spies}]
				\node[inner sep=0pt]{
					\begin{overpic}[height=\tempdima]{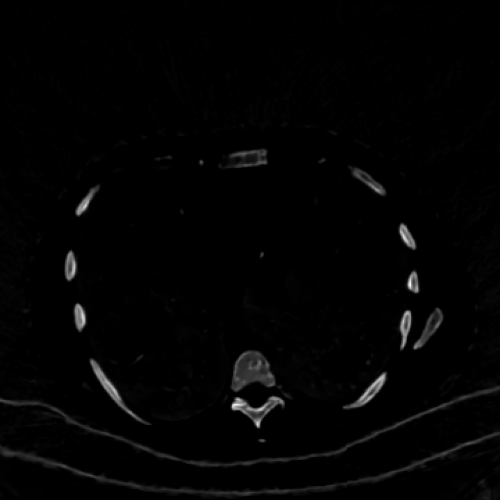}
						\put(3,90){\tiny\textbf{\color{white}{PSNR: 37.31}}}
						\put(3,83){\tiny \textbf{\color{white}{SSIM: 0.90}}}
						\put(3,76){\tiny \textbf{\color{white}{LPIPS: 0.07}}}
				\end{overpic}}
				; 
				\spy on (0.08,-0.63) in node [right] at (0.42,0.7);
			\end{scope}
		\end{tikzpicture} 
		&
		\begin{tikzpicture}
			\begin{scope}[spy using outlines={rectangle,yellow,magnification=2.5,size=6.6mm,connect spies}]
				\node[inner sep=0pt]{
					\begin{overpic}[height=\tempdima]{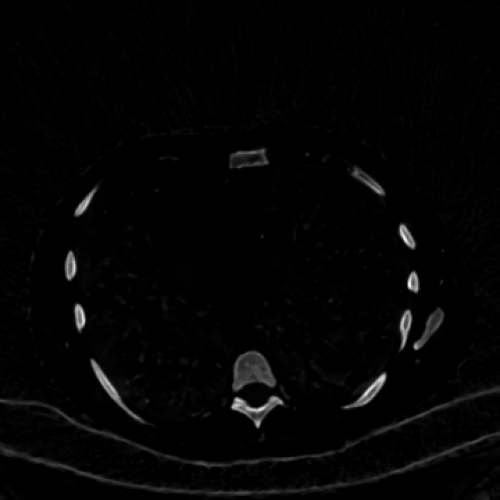}
						\put(3,90){\tiny\textbf{\color{white}{PSNR: 40.23}}}
						\put(3,83){\tiny \textbf{\color{white}{SSIM: 0.94}}}
						\put(3,76){\tiny \textbf{\color{white}{LPIPS: 0.03}}}
				\end{overpic}}
				; 
				\spy on (0.08,-0.63) in node [right] at (0.42,0.7);
			\end{scope}
		\end{tikzpicture} 
		\\
		\vspace{-0.6cm}

		\rowname{Soft Tissues      }  & 
		
		\begin{tikzpicture}
			\begin{scope}[spy using outlines={rectangle,yellow,magnification=2.5,size=6.6mm,connect spies, connect spies}]
				\node[inner sep=0pt]{
					\begin{overpic}[height=\tempdima]{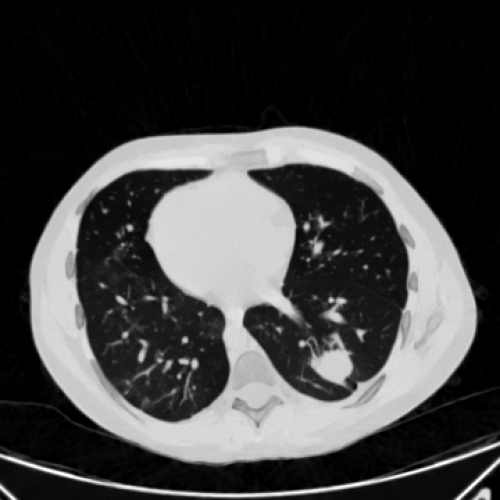}
					\end{overpic}	
					}; 
				\spy on (0.08,-0.63) in node [right] at (0.42,-0.7);
				\spy on (0.3,-0.3)in node [right] at (0.42, 0.7);
                \spy [forestgreen4416044] on (0,-0.1) in node [left] at (-0.33, -0.7);
			\end{scope}
		\end{tikzpicture} 
		&
		\begin{tikzpicture}
			\begin{scope}[spy using outlines={rectangle,yellow,magnification=2.5,size=6.6mm,connect spies}]
				\node[inner sep=0pt]{
					\begin{overpic}[height=\tempdima]{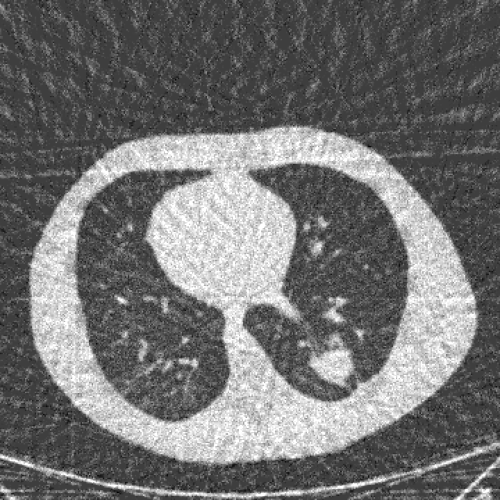}
						\put(3,90){\tiny\textbf{\color{white}{PSNR: 19.15}}}
						\put(3,83){\tiny \textbf{\color{white}{SSIM: 0.17}}}
						\put(3,76){\tiny \textbf{\color{white}{LPIPS: 0.63}}}
				\end{overpic}}
				; 
				\spy on (0.08,-0.63) in node [right] at (0.42,-0.7);
				\spy on (0.3,-0.3)in node [right] at (0.42, 0.7);
			\end{scope}
		\end{tikzpicture} 
		&	
		\begin{tikzpicture}
			\begin{scope}[spy using outlines={rectangle,yellow,magnification=2.5,size=6.6mm,connect spies}]
				\node[inner sep=0pt]{
					\begin{overpic}[height=\tempdima]{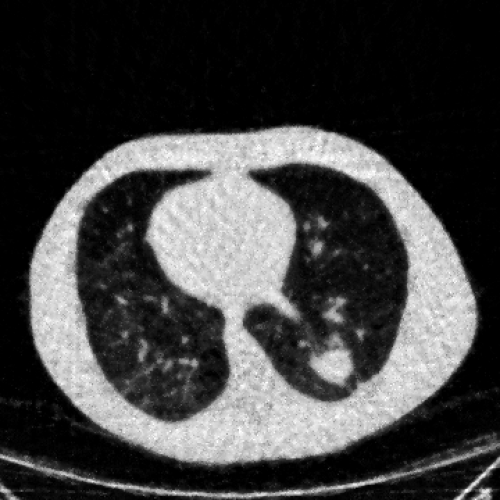}
						\put(3,90){\tiny\textbf{\color{white}{PSNR: 28.33}}}
						\put(3,83){\tiny \textbf{\color{white}{SSIM: 0.80}}}
						\put(3,76){\tiny \textbf{\color{white}{LPIPS: 0.18}}}
				\end{overpic}}
				;  
				\spy on (0.08,-0.63) in node [right] at (0.42,-0.7);
				\spy on (0.3,-0.3)in node [right] at (0.42, 0.7);
			\end{scope}
		\end{tikzpicture} 
		&
		\begin{tikzpicture}
			\begin{scope}[spy using outlines={rectangle,yellow,magnification=2.5,size=6.6mm,connect spies}]
				\node[inner sep=0pt]{
					\begin{overpic}[height=\tempdima]{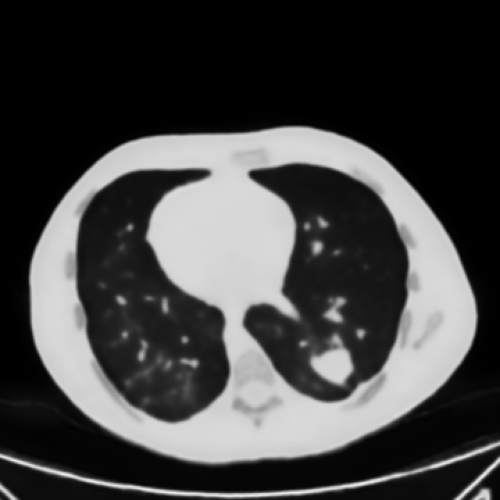}
						\put(3,90){\tiny\textbf{\color{white}{PSNR: 32.45}}}
						\put(3,83){\tiny \textbf{\color{white}{SSIM: 0.74}}}
						\put(3,76){\tiny \textbf{\color{white}{LPIPS: 0.04}}}
				\end{overpic}}
				; 
				\spy on (0.08,-0.63) in node [right] at (0.42,-0.7);
				\spy on (0.3,-0.3)in node [right] at (0.42, 0.7);
			\end{scope}
		\end{tikzpicture} 
		
		&
		\begin{tikzpicture}
			\begin{scope}[spy using outlines={rectangle,yellow,magnification=2.5,size=6.6mm,connect spies}]
				\node[inner sep=0pt]{
					\begin{overpic}[height=\tempdima]{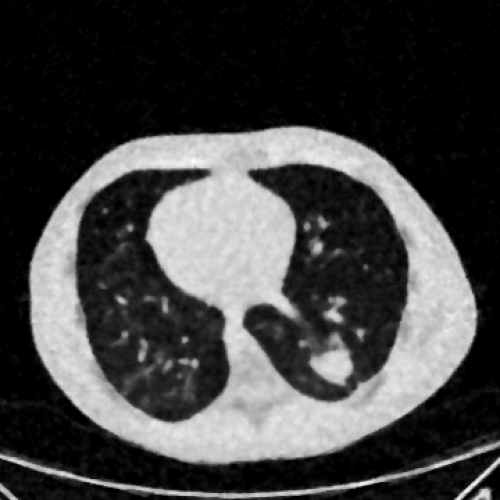}
						\put(3,90){\tiny\textbf{\color{white}{PSNR: 32.48}}}
						\put(3,83){\tiny \textbf{\color{white}{SSIM: 0.87}}}
						\put(3,76){\tiny \textbf{\color{white}{LPIPS: 0.13}}}
				\end{overpic}}
				; 
				\spy on (0.08,-0.63) in node [right] at (0.42,-0.7);
				\spy on (0.3,-0.3)in node [right] at (0.42, 0.7);
			\end{scope}
		\end{tikzpicture} 

		&
		\begin{tikzpicture}
			\begin{scope}[spy using outlines={rectangle,yellow,magnification=2.5,size=6.6mm,connect spies}]
				\node[inner sep=0pt]{
					\begin{overpic}[height=\tempdima]{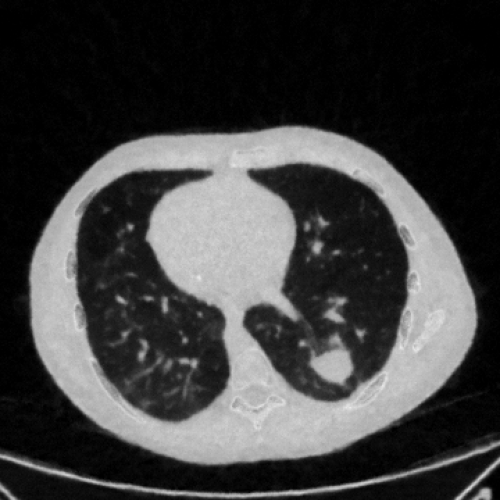}
						\put(3,90){\tiny\textbf{\color{white}{PSNR: 28.65}}}
						\put(3,83){\tiny \textbf{\color{white}{SSIM: 0.83}}}
						\put(3,76){\tiny \textbf{\color{white}{LPIPS: 0.07}}}
				\end{overpic}}
				; 
				\spy on (0.08,-0.63) in node [right] at (0.42,-0.7);
				\spy on (0.3,-0.3)in node [right] at (0.42, 0.7);
			\end{scope}
		\end{tikzpicture} 
		&	
		\begin{tikzpicture}
			\begin{scope}[spy using outlines={rectangle,yellow,magnification=2.5,size=6.6mm,connect spies}]
				\node[inner sep=0pt]{
					\begin{overpic}[height=\tempdima]{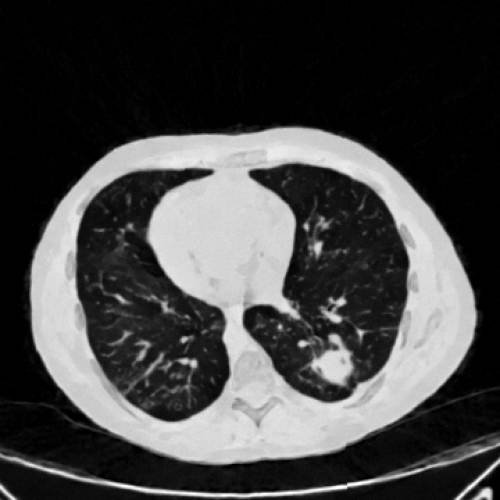}
						\put(3,90){\tiny\textbf{\color{white}{PSNR: 30.03}}}
						\put(3,83){\tiny \textbf{\color{white}{SSIM: 0.87}}}
						\put(3,76){\tiny \textbf{\color{white}{LPIPS: 0.06}}}
				\end{overpic}}
				;                                 
				\spy on (0.08,-0.63) in node [right] at (0.42,-0.7);
				\spy on (0.3,-0.3)in node [right] at (0.42, 0.7);
                \spy [forestgreen4416044] on (0,-0.1) in node [left] at (-0.33, -0.7);
			\end{scope}
		\end{tikzpicture} 
		&	
		\begin{tikzpicture}
			\begin{scope}[spy using outlines={rectangle,yellow,magnification=2.5,size=6.6mm,connect spies}]
				\node[inner sep=0pt]{
					\begin{overpic}[height=\tempdima]{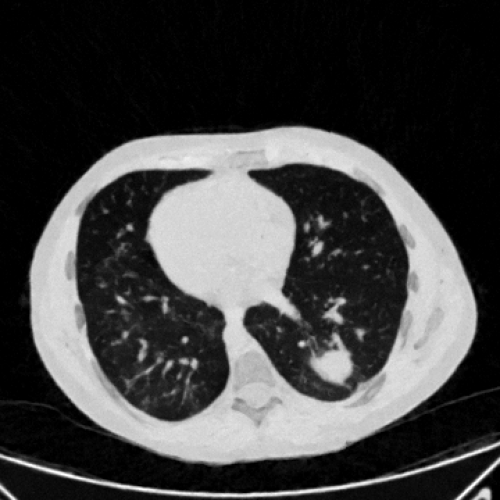}
						\put(3,90){\tiny\textbf{\color{white}{PSNR: 35.40}}}
						\put(3,83){\tiny \textbf{\color{white}{SSIM: 0.91}}}
						\put(3,76){\tiny \textbf{\color{white}{LPIPS: 0.03}}}
				\end{overpic}}
				; 
				\spy on (0.08,-0.63) in node [right] at (0.42,-0.7);
				\spy on (0.3,-0.3)in node [right] at (0.42, 0.7);
			\end{scope}
		\end{tikzpicture} 

	\end{tabular}	
	\vspace{1cm}
	\caption{Experiment 1 (sparse-view data)---Material decomposition images obtained with the different methods.}
	\label{fig:ResultatsSparseView}
\end{figure*}	

%% file: figures/TableSparseView.tex
\begin{table*}
	\begin{center}
	    \begin{tabular}{|l|l|c|c|c|c|c|c|c|}
	    \hline  
	    \multicolumn{2}{|c|}{} & FBP & \ac{LBFGS} & U-Net-I & U-Net-P  & \ac{ImTDPS} & \ac{ProjTDPS} & \ac{ODPS} \\
	    \hline
	    \multicolumn{2}{|c|}{Avg. Time (in sec)} & 0.05 &
23.67 &
104.38 &
105.35 &
90.38 &
82.97 &
88.89 \\
	    \hline 
	    \multirow{2}{*}{\scriptsize{PSNR} $\uparrow$} & \scriptsize{Bones} &  
25.22 &
35.09 &
37.52 &
\textbf{41.29} &
30.09 &
37.69 &
\underline{40.58} \\
	    & \scriptsize{Tissues}  & 
19.13 &
29.57 &
\underline{33.35} &
32.46 &
25.55 &
31.48 &
\textbf{35.05}  \\
	    \hline 
	    \multirow{2}{*}{\scriptsize{SSIM} $\uparrow$} & \scriptsize{Bones} & 
0.22 &
0.87 &
\underline{0.93} &
\textbf{0.94} &
0.64 &
0.90 &
\textbf{0.94} \\
	    & \scriptsize{Tissues}  &
0.11 &
0.78 &
\textbf{0.90} &
0.83 &
0.55 &
0.82 &
\underline{0.86} \\
	    \hline 
	    \multirow{2}{*}{\scriptsize{LPIPS} $\downarrow$} & \scriptsize{Bones} & 
0.69 &
0.11 &
0.08 &
0.08 &
0.09 &
\underline{0.07} &
\textbf{0.04}\\
	    & \scriptsize{Tissues}  & 
0.67 &
0.18 &
0.08 &
0.16 &
0.11 &
\textbf{0.06} &
\underline{0.07} \\
	    \hline 
	    \end{tabular}
	    \caption{Experiment 1 (sparse-view data)---Averaged performance metrics computed over the entire test dataset. Best and second best results are respectively in bold and underlined.}
	    \label{tab:TableSparseView}
	\end{center}
\end{table*}

%% file: figures/BoxPlotsSparseView.tex
\begin{figure*}
	\centering
	\begin{subfigure}[b]{0.30\textwidth}
		\begin{adjustbox}{width=\textwidth}    
			\begin{tikzpicture}
				\begin{axis}[
					height=0.8\textwidth,
					boxplot/draw direction = y,
					x axis line style = {opacity=1},
					axis x line* = bottom,
					axis y line = left,
					enlarge y limits,
					yticklabel style={font=\tiny},
					ymajorgrids,
					xtick = {1, 2, 3, 4, 5, 6, 7, 8, 9, 10, 11, 12},
					xticklabel style = {align=center, font=\tiny, rotate=90},
					xticklabels = {L-BFGS ,  U-Net-I ,  U-Net-P ,  im-TDPS ,  proj-TDPS ,  ODPS , L-BFGS ,U-Net-I ,U-Net-P ,im-TDPS ,proj-TDPS ,ODPS },
					grid=major,
                    name=border
					]
\addplot+[ 
 draw=RoyalBlue, 
 boxplot prepared={
 lower whisker = 29.76 ,
 lower quartile = 32.87 ,
 median = 34.79 ,
 upper quartile = 37.03 ,
 upper whisker = 41.26 ,
 }, 
 ] coordinates {};
\addplot+[ 
 draw=RoyalBlue, 
 boxplot prepared={
 lower whisker = 36.82 ,
 lower quartile = 36.95 ,
 median = 37.4 ,
 upper quartile = 38.23 ,
 upper whisker = 38.38 ,
 }, 
 ] coordinates {};
\addplot+[ 
 draw=RoyalBlue, 
 boxplot prepared={
 lower whisker = 40.15 ,
 lower quartile = 40.32 ,
 median = 40.37 ,
 upper quartile = 43.24 ,
 upper whisker = 43.43 ,
 }, 
 ] coordinates {};
\addplot+[ 
 draw=RoyalBlue, 
 boxplot prepared={
 lower whisker = 27.46 ,
 lower quartile = 27.55 ,
 median = 30.14 ,
 upper quartile = 32.73 ,
 upper whisker = 32.91 ,
 }, 
 ] coordinates {};
\addplot+[ 
 draw=RoyalBlue, 
 boxplot prepared={
 lower whisker = 33.4 ,
 lower quartile = 36.8 ,
 median = 37.4 ,
 upper quartile = 39.355 ,
 upper whisker = 40.31 ,
 }, 
 ] coordinates {};
\addplot+[ 
 draw=RoyalBlue, 
 boxplot prepared={
 lower whisker = 39.07 ,
 lower quartile = 39.6 ,
 median = 40.08 ,
 upper quartile = 42.13 ,
 upper whisker = 42.52 ,
 }, 
 ] coordinates {};
\addplot+[ 
 draw=Maroon, 
 boxplot prepared={
 lower whisker = 25.9 ,
 lower quartile = 28.655 ,
 median = 29.89 ,
 upper quartile = 30.54 ,
 upper whisker = 32.44 ,
 }, 
 ] coordinates {};
\addplot+[ 
 draw=Maroon, 
 boxplot prepared={
 lower whisker = 31.61 ,
 lower quartile = 31.89 ,
 median = 33.75 ,
 upper quartile = 34.35 ,
 upper whisker = 34.62 ,
 }, 
 ] coordinates {};
\addplot+[ 
 draw=Maroon, 
 boxplot prepared={
 lower whisker = 31.16 ,
 lower quartile = 31.37 ,
 median = 32.74 ,
 upper quartile = 33.24 ,
 upper whisker = 33.4 ,
 }, 
 ] coordinates {};
\addplot+[ 
 draw=Maroon, 
 boxplot prepared={
 lower whisker = 24.16 ,
 lower quartile = 24.375 ,
 median = 24.5 ,
 upper quartile = 27.83 ,
 upper whisker = 27.94 ,
 }, 
 ] coordinates {};
\addplot+[ 
 draw=Maroon, 
 boxplot prepared={
 lower whisker = 29.2 ,
 lower quartile = 30.22 ,
 median = 32.38 ,
 upper quartile = 32.78 ,
 upper whisker = 33.86 ,
 }, 
 ] coordinates {};
\addplot+[ 
 draw=Maroon, 
 boxplot prepared={
 lower whisker = 30.71 ,
 lower quartile = 32.89 ,
 median = 36.29 ,
 upper quartile = 36.41 ,
 upper whisker = 36.73 ,
 }, 
 ] coordinates {};
				\end{axis}
                \node[above left, xshift = -0.5cm] at (border.north) {\textcolor{RoyalBlue}{\tiny{Bones}}};
				\node[above right, xshift = 0.2cm] at (border.north) {\textcolor{Maroon}{\tiny{Soft Tissues}}};
			\end{tikzpicture}
		\end{adjustbox}
		\caption{PSNR}\label{subfig:psnr_sv}
	\end{subfigure}
	\hfill
	\begin{subfigure}[b]{0.30\textwidth}
		\begin{adjustbox}{width=\textwidth}    
			\begin{tikzpicture}
				\begin{axis}[
					height=0.8\textwidth,
					boxplot/draw direction = y,
					x axis line style = {opacity=1},
					axis x line* = bottom,
					axis y line = left,
					enlarge y limits,
					yticklabel style={font=\tiny},
					ymajorgrids,
					xtick = {1, 2, 3, 4, 5, 6, 7, 8, 9, 10, 11, 12},
					xticklabel style = {align=center, font=\tiny, rotate=90},
					xticklabels = {L-BFGS ,  U-Net-I ,  U-Net-P ,  im-TDPS ,  proj-TDPS ,  ODPS , L-BFGS ,U-Net-I ,U-Net-P ,im-TDPS ,proj-TDPS ,ODPS },
					grid=major,
                    name=border
					]
\addplot+[ 
 draw=RoyalBlue, 
 boxplot prepared={
 lower whisker = 0.74 ,
 lower quartile = 0.84 ,
 median = 0.87 ,
 upper quartile = 0.91 ,
 upper whisker = 0.95 ,
 }, 
 ] coordinates {};
\addplot+[ 
 draw=RoyalBlue, 
 boxplot prepared={
 lower whisker = 0.9 ,
 lower quartile = 0.91 ,
 median = 0.93 ,
 upper quartile = 0.94 ,
 upper whisker = 0.94 ,
 }, 
 ] coordinates {};
\addplot+[ 
 draw=RoyalBlue, 
 boxplot prepared={
 lower whisker = 0.93 ,
 lower quartile = 0.93 ,
 median = 0.94 ,
 upper quartile = 0.96 ,
 upper whisker = 0.97 ,
 }, 
 ] coordinates {};
\addplot+[ 
 draw=RoyalBlue, 
 boxplot prepared={
 lower whisker = 0.57 ,
 lower quartile = 0.57 ,
 median = 0.63 ,
 upper quartile = 0.71 ,
 upper whisker = 0.72 ,
 }, 
 ] coordinates {};
\addplot+[ 
 draw=RoyalBlue, 
 boxplot prepared={
 lower whisker = 0.89 ,
 lower quartile = 0.89 ,
 median = 0.9 ,
 upper quartile = 0.93 ,
 upper whisker = 0.94 ,
 }, 
 ] coordinates {};
\addplot+[ 
 draw=RoyalBlue, 
 boxplot prepared={
 lower whisker = 0.93 ,
 lower quartile = 0.93 ,
 median = 0.93 ,
 upper quartile = 0.96 ,
 upper whisker = 0.96 ,
 }, 
 ] coordinates {};
\addplot+[ 
 draw=Maroon, 
 boxplot prepared={
 lower whisker = 0.6 ,
 lower quartile = 0.73 ,
 median = 0.79 ,
 upper quartile = 0.83 ,
 upper whisker = 0.89 ,
 }, 
 ] coordinates {};
\addplot+[ 
 draw=Maroon, 
 boxplot prepared={
 lower whisker = 0.87 ,
 lower quartile = 0.88 ,
 median = 0.9 ,
 upper quartile = 0.92 ,
 upper whisker = 0.92 ,
 }, 
 ] coordinates {};
\addplot+[ 
 draw=Maroon, 
 boxplot prepared={
 lower whisker = 0.8 ,
 lower quartile = 0.81 ,
 median = 0.84 ,
 upper quartile = 0.85 ,
 upper whisker = 0.85 ,
 }, 
 ] coordinates {};
\addplot+[ 
 draw=Maroon, 
 boxplot prepared={
 lower whisker = 0.45 ,
 lower quartile = 0.46 ,
 median = 0.46 ,
 upper quartile = 0.73 ,
 upper whisker = 0.74 ,
 }, 
 ] coordinates {};
\addplot+[ 
 draw=Maroon, 
 boxplot prepared={
 lower whisker = 0.81 ,
 lower quartile = 0.82 ,
 median = 0.85 ,
 upper quartile = 0.88 ,
 upper whisker = 0.89 ,
 }, 
 ] coordinates {};
\addplot+[ 
 draw=Maroon, 
 boxplot prepared={
 lower whisker = 0.77 ,
 lower quartile = 0.79 ,
 median = 0.88 ,
 upper quartile = 0.9 ,
 upper whisker = 0.9 ,
 }, 
 ] coordinates {};
				\end{axis}
				\node[above left, xshift = -0.5cm] at (border.north) {\textcolor{RoyalBlue}{\tiny{Bones}}};
				\node[above right, xshift = 0.2cm] at (border.north) {\textcolor{Maroon}{\tiny{Soft Tissues}}};
			\end{tikzpicture}
		\end{adjustbox}
		\caption{SSIM}\label{subfig:ssim_sv}
	\end{subfigure}
	\hfill
	\begin{subfigure}[b]{0.30\textwidth}
		\begin{adjustbox}{width=\textwidth}    
			\begin{tikzpicture}
				\pgfplotstableread[col sep=comma]{csvfiles/LPIPS_SparseView_0.csv}\csvdata
				\begin{axis}[
					height=0.8\textwidth,
					boxplot/draw direction = y,
					x axis line style = {opacity=1},
					axis x line* = bottom,
					axis y line = left,
					enlarge y limits,
					yticklabel style={font=\tiny},
					ymajorgrids,
					xtick = {1, 2, 3, 4, 5, 6, 7, 8, 9, 10, 11, 12},
					xticklabel style = {align=center, font=\tiny, rotate=90},
					xticklabels = {L-BFGS ,  U-Net-I ,  U-Net-P ,  im-TDPS ,  proj-TDPS ,  ODPS , L-BFGS ,U-Net-I ,U-Net-P ,im-TDPS ,proj-TDPS ,ODPS },
					grid=major,
                    name=border
					]
\addplot+[ 
 draw=RoyalBlue, 
 boxplot prepared={
 lower whisker = 0.04 ,
 lower quartile = 0.08 ,
 median = 0.1 ,
 upper quartile = 0.14 ,
 upper whisker = 0.23 ,
 }, 
 ] coordinates {};
\addplot+[ 
 draw=RoyalBlue, 
 boxplot prepared={
 lower whisker = 0.07 ,
 lower quartile = 0.07 ,
 median = 0.07 ,
 upper quartile = 0.1 ,
 upper whisker = 0.1 ,
 }, 
 ] coordinates {};
\addplot+[ 
 draw=RoyalBlue, 
 boxplot prepared={
 lower whisker = 0.06 ,
 lower quartile = 0.07 ,
 median = 0.08 ,
 upper quartile = 0.09 ,
 upper whisker = 0.1 ,
 }, 
 ] coordinates {};
\addplot+[ 
 draw=RoyalBlue, 
 boxplot prepared={
 lower whisker = 0.06 ,
 lower quartile = 0.07 ,
 median = 0.1 ,
 upper quartile = 0.11 ,
 upper whisker = 0.13 ,
 }, 
 ] coordinates {};
\addplot+[ 
 draw=RoyalBlue, 
 boxplot prepared={
 lower whisker = 0.05 ,
 lower quartile = 0.06 ,
 median = 0.07 ,
 upper quartile = 0.07 ,
 upper whisker = 0.08 ,
 }, 
 ] coordinates {};
\addplot+[ 
 draw=RoyalBlue, 
 boxplot prepared={
 lower whisker = 0.02 ,
 lower quartile = 0.03 ,
 median = 0.04 ,
 upper quartile = 0.04 ,
 upper whisker = 0.05 ,
 }, 
 ] coordinates {};
\addplot+[ 
 draw=Maroon, 
 boxplot prepared={
 lower whisker = 0.08 ,
 lower quartile = 0.135 ,
 median = 0.17 ,
 upper quartile = 0.215 ,
 upper whisker = 0.33 ,
 }, 
 ] coordinates {};
\addplot+[ 
 draw=Maroon, 
 boxplot prepared={
 lower whisker = 0.06 ,
 lower quartile = 0.07 ,
 median = 0.07 ,
 upper quartile = 0.1 ,
 upper whisker = 0.1 ,
 }, 
 ] coordinates {};
\addplot+[ 
 draw=Maroon, 
 boxplot prepared={
 lower whisker = 0.14 ,
 lower quartile = 0.15 ,
 median = 0.15 ,
 upper quartile = 0.17 ,
 upper whisker = 0.18 ,
 }, 
 ] coordinates {};
\addplot+[ 
 draw=Maroon, 
 boxplot prepared={
 lower whisker = 0.08 ,
 lower quartile = 0.09 ,
 median = 0.12 ,
 upper quartile = 0.13 ,
 upper whisker = 0.14 ,
 }, 
 ] coordinates {};
\addplot+[ 
 draw=Maroon, 
 boxplot prepared={
 lower whisker = 0.03 ,
 lower quartile = 0.04 ,
 median = 0.06 ,
 upper quartile = 0.08 ,
 upper whisker = 0.14 ,
 }, 
 ] coordinates {};
\addplot+[ 
 draw=Maroon, 
 boxplot prepared={
 lower whisker = 0.04 ,
 lower quartile = 0.04 ,
 median = 0.04 ,
 upper quartile = 0.12 ,
 upper whisker = 0.14 ,
 }, 
 ] coordinates {};
				\end{axis}
				\node[above left, xshift = -0.5cm] at (border.north) {\textcolor{RoyalBlue}{\tiny{Bones}}};
				\node[above right, xshift = 0.2cm] at (border.north) {\textcolor{Maroon}{\tiny{Soft Tissues}}};
			\end{tikzpicture}
		\end{adjustbox}
		\caption{LPIPS}\label{subfig:lpips_sv}
	\end{subfigure}
	\caption{Experiment 1 (sparse-view data)---Boxplots of \subref{subfig:psnr_sv} PSNR, \subref{subfig:ssim_sv} SSIM and \subref{subfig:lpips_sv} LPIPS, computed over the test dataset. }\label{fig:boxplot_sv}
\end{figure*}

%% file: figures/Resultats_hyb.tex
\begin{figure*}
		\centering
		\settoheight{\tempdima}
        {\includegraphics[width=0.14\linewidth]{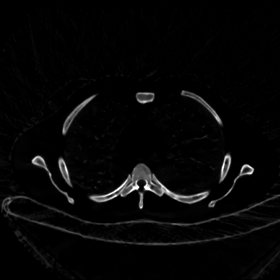}}%
		\begin{tabular}{@{}c@{}c@{\hspace{-0.1cm}}c@{\hspace{-0.1cm}}c@{}c@{}}
            \vspace{-0.1cm}
			&  Reference &  \ac{LBFGS} &  Hybrid-\ac{ODPS} &   \\
            
			\rowname{Bones      } & 
			\begin{tikzpicture}
	  	    \begin{scope}[spy using outlines={rectangle,yellow,magnification=2.5,size=8mm,connect spies}]
	  	    \node{
                \begin{overpic}[height=\tempdima]{Images_hyb/GT_Bones.png}
                \end{overpic}
                };
                \spy on (-0.7,-0.35) in node [right] at (0.4,0.8);
            \end{scope}
            \draw[red] (-1.05,-0.39) -- (1.3,-0.39);
			\end{tikzpicture} 
                
			&

			\begin{tikzpicture}
	  	    \begin{scope}[spy using outlines={rectangle,yellow,magnification=2.5,size=8mm,connect spies}]
                \node{
					\begin{overpic}[height=\tempdima]{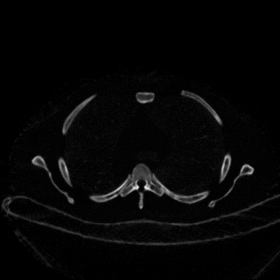}
						\put(3,90){\tiny\textbf{\color{white}{PSNR: 30.92}}}
						\put(3,83){\tiny \textbf{\color{white}{SSIM: 0.65}}}
                        \put(3,76){\tiny \textbf{\color{white}{LPIPS: 0.21}}}
					\end{overpic}}
                ; 
                \spy on (-0.7,-0.35) in node [right] at (0.4,0.8);
            \end{scope}
			\end{tikzpicture} 
            
			&
			\begin{tikzpicture}
	  	    \begin{scope}[spy using outlines={rectangle,yellow,magnification=2.5,size=8mm,connect spies}]
                \node{
					\begin{overpic}[height=\tempdima]{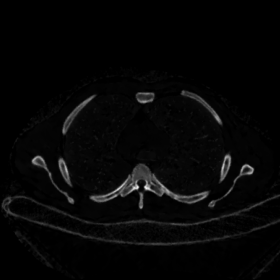}
						\put(3,90){\tiny\textbf{\color{white}{PSNR: 30.98}}}
						\put(3,83){\tiny \textbf{\color{white}{SSIM: 0.66}}}
                        \put(3,76){\tiny \textbf{\color{white}{LPIPS: 0.09}}}
					\end{overpic}}
                ; 
                \spy on (-0.7,-0.35) in node [right] at (0.4,0.8);
            \end{scope}
			\end{tikzpicture} 	
	        
	        & 
			\scalebox{0.4}{
			\input{figures/profiles/bone}
			}
            \\
            \vspace{-0.6cm}			
			  \\
              
			\rowname{Soft Tissues      }  & 
			
			\begin{tikzpicture}
	  	    \begin{scope}[spy using outlines={rectangle,yellow,magnification=2.5,size=8mm,connect spies}]
	  	    \node{
                \begin{overpic}[height=\tempdima]{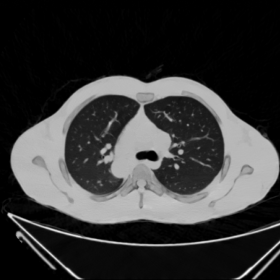}
                \end{overpic}
                };
                \spy on (-0.7,-0.35) in node [right] at (0.4,0.8);
            \end{scope}
            \draw[red] (-1.05,-0.39) -- (1.3,-0.39);
			\end{tikzpicture} 
			&
		
			\begin{tikzpicture}
	  	    \begin{scope}[spy using outlines={rectangle,yellow,magnification=2.5,size=8mm,connect spies}]
                \node{
					\begin{overpic}[height=\tempdima]{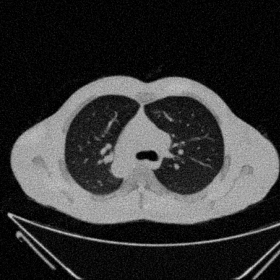}
						\put(3,90){\tiny\textbf{\color{white}{PSNR: 20.92}}}
						\put(3,83){\tiny \textbf{\color{white}{SSIM: 0.39}}}
                        \put(3,76){\tiny \textbf{\color{white}{LPIPS: 0.59}}}
					\end{overpic}}
                ;  
                \spy on (-0.7,-0.35) in node [right] at (0.4,0.8);
            \end{scope}
			\end{tikzpicture} 
			&	
			\begin{tikzpicture}
	  	    \begin{scope}[spy using outlines={rectangle,yellow,magnification=2.5,size=8mm,connect spies}]
                \node{
					\begin{overpic}[height=\tempdima]{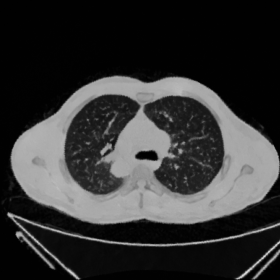}
						\put(3,90){\tiny\textbf{\color{white}{PSNR: 23.95}}}
						\put(3,83){\tiny \textbf{\color{white}{SSIM: 0.74}}}
                        \put(3,76){\tiny \textbf{\color{white}{LPIPS: 0.14}}}
					\end{overpic}}
                ; 
                \spy on (-0.7,-0.35) in node [right] at (0.4,0.8);
            \end{scope}
			\end{tikzpicture} 
			&
			\scalebox{0.4}{
			\input{figures/profiles/st}
			}
			            \\
            \vspace{-0.6cm}			
			  \\
              
			\rowname{Iodine     }  & 
			
			\begin{tikzpicture}
	  	    \begin{scope}[spy using outlines={rectangle,yellow,magnification=2.5,size=8mm,connect spies}]
	  	    	\node{
                \begin{overpic}[height=\tempdima]{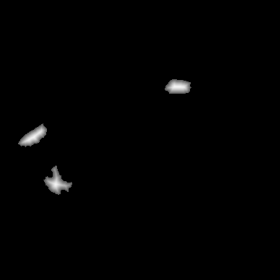}
                \end{overpic}
                };
                \spy on (-0.7,-0.35) in node [right] at (0.4,-0.8);
            \end{scope}
            \draw[red] (-0.8,-0.39) -- (-0.55,-0.39);
			\end{tikzpicture} 
			&
			
			\begin{tikzpicture}
	  	    \begin{scope}[spy using outlines={rectangle,yellow,magnification=2.5,size=8mm,connect spies}]
                \node{
					\begin{overpic}[height=\tempdima]{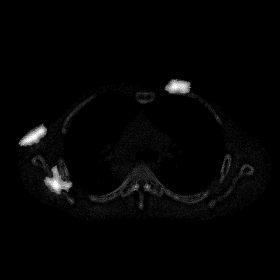}
						\put(3,90){\tiny\textbf{\color{white}{PSNR: 26.12}}}
						\put(3,83){\tiny \textbf{\color{white}{SSIM: 0.73}}}
                        \put(3,76){\tiny \textbf{\color{white}{LPIPS: 0.45}}}
					\end{overpic}}
                ;  
                \spy on (-0.7,-0.35) in node [right] at (0.4,-0.8);
            \end{scope}
			\end{tikzpicture} 
			&	
			\begin{tikzpicture}
	  	    \begin{scope}[spy using outlines={rectangle,yellow,magnification=2.5,size=8mm,connect spies}]
                \node{
					\begin{overpic}[height=\tempdima]{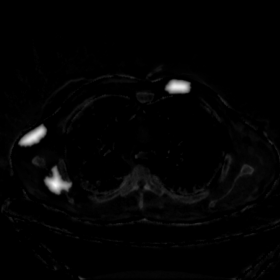}
						\put(3,90){\tiny\textbf{\color{white}{PSNR: 31.13}}}
						\put(3,83){\tiny \textbf{\color{white}{SSIM: 0.81}}}
                        \put(3,76){\tiny \textbf{\color{white}{LPIPS: 0.35}}}
					\end{overpic}}
                ; 
                \spy on (-0.7,-0.35) in node [right] at (0.4,-0.8);
            \end{scope}
			\end{tikzpicture} 
			
			&
			\scalebox{0.4}{
			\input{figures/profiles/iod}
			}

		\end{tabular}	

        \caption{Experiment 2---Hybrid-\ac{ODPS} and \ac{LBFGS} material images as well as their associated metrics and density profiles. The red line on the reference image indicates the location of the density profile.}
        \label{fig:resultats_hyb}
\end{figure*}
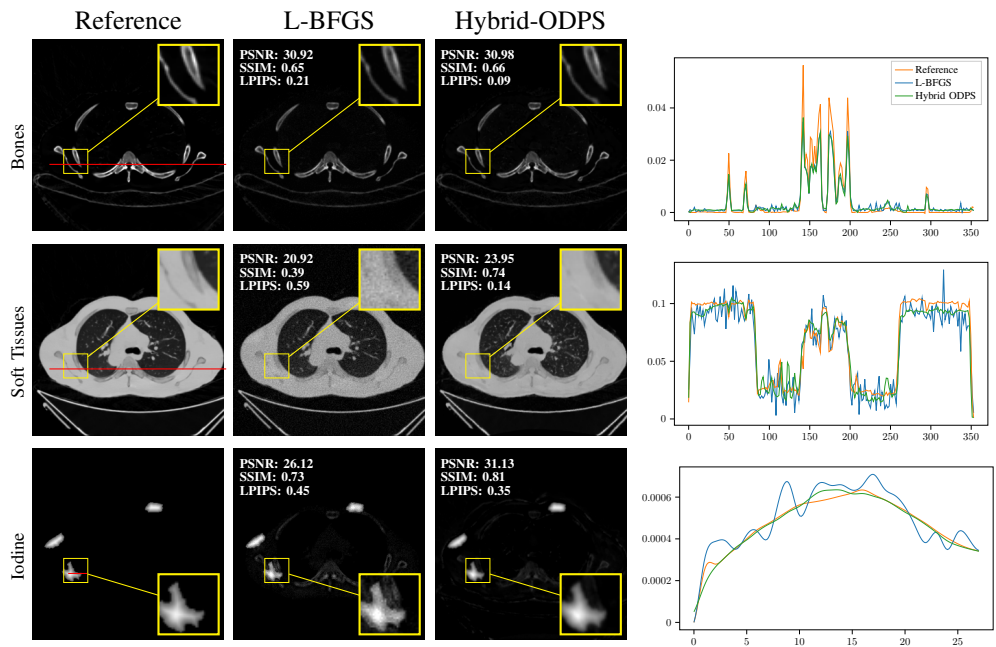	

%% file: figures/profiles/bone.tex
\begin{tikzpicture}

\definecolor{darkgrey176}{RGB}{176,176,176}
\definecolor{darkorange25512714}{RGB}{255,127,14}
\definecolor{forestgreen4416044}{RGB}{44,160,44}
\definecolor{lightgrey204}{RGB}{204,204,204}
\definecolor{steelblue31119180}{RGB}{31,119,180}

\begin{axis}[
legend cell align={left},
legend style={fill opacity=0.8, draw opacity=1, text opacity=1, draw=lightgrey204},
tick align=outside,
tick pos=left,
x grid style={darkgrey176},
xmin=-17.65, xmax=370.65,
xtick style={color=black},
y grid style={darkgrey176},
ymin=-0.00316011660906952, ymax=0.0591900830011582,
ytick style={color=black},
yticklabel style={
        /pgf/number format/fixed,
        /pgf/number format/precision=5
},
scaled y ticks=false,
		width=12cm,
		height=7cm,
]
\addplot [semithick, darkorange25512714]
table {%
0 0.000548926473129541
1.77386934673367 -2.38403754337924e-05
3.54773869346734 -3.30470265907934e-06
5.321608040201 1.35709584014876e-07
7.09547738693467 1.49284687722684e-06
8.86934673366834 -1.97360914171441e-05
10.643216080402 0.000327435962390155
12.4170854271357 0.000261452078120783
14.1909547738693 0.000188522099051625
15.964824120603 0.000202223702217452
17.7386934673367 0.000214094819966704
19.5125628140704 0.00021808544988744
21.286432160804 0.000209366509807296
23.0603015075377 5.45193870493677e-05
24.8341708542714 0.000155506873852573
26.608040201005 0.000115978182293475
28.3819095477387 0.000131519787828438
30.1557788944724 9.85386868705973e-05
31.929648241206 -1.27532848637202e-06
33.7035175879397 -4.43329838617501e-07
35.4773869346734 -5.09426634209831e-08
37.251256281407 -2.50182585759262e-09
39.0251256281407 1.07160880169488e-09
40.7989949748744 -8.09146243341274e-08
42.572864321608 -1.38427992624202e-06
44.3467336683417 -1.35493119159946e-05
46.1206030150754 -5.90670060773846e-05
47.894472361809 0.00772883230820298
49.6683417085427 0.0226332265883684
51.4422110552764 0.00802210718393326
53.21608040201 0.000410633452702314
54.9899497487437 0.00014585662574973
56.7638190954774 4.00208045903128e-05
58.5376884422111 -7.21017443083838e-07
60.3115577889447 -4.58901695310487e-07
62.0854271356784 -1.37270080813323e-06
63.8592964824121 6.4781641412992e-05
65.6331658291457 -2.22656017285772e-05
67.4070351758794 -9.33338887989521e-05
69.1809045226131 0.0107541233301163
70.9547738693467 0.0158564765006304
72.7286432160804 0.00620406586676836
74.5025125628141 0.000150327061419375
76.2763819095477 2.32619913731469e-06
78.0502512562814 -8.2810423407409e-08
79.8241206030151 4.26764972871752e-06
81.5979899497488 0.000197700021089986
83.3718592964824 0.00265018618665636
85.1457286432161 0.00177330290898681
86.9195979899498 0.00176247803028673
88.6934673366834 0.000772169034462422
90.4673366834171 0.000685770995914936
92.2412060301507 0.000853164703585207
94.0150753768844 3.50341033481527e-05
95.7889447236181 7.35945423002704e-06
97.5628140703518 1.90372393262805e-05
99.3366834170854 0.000195645348867401
101.110552763819 0.000266851362539455
102.884422110553 -2.76471582765225e-05
104.658291457286 0.000158539420226589
106.43216080402 7.12546898284927e-06
108.206030150754 1.28259089251515e-05
109.979899497487 0.000186127945198677
111.753768844221 -9.64111677603796e-06
113.527638190955 1.06381621662877e-05
115.301507537688 0.000116540744784288
117.075376884422 0.000266205024672672
118.849246231156 0.00021387888409663
120.623115577889 0.00020286776998546
122.396984924623 0.00016443467757199
124.170854271357 0.000226005460717715
125.94472361809 0.00125518988352269
127.718592964824 0.000678858836181462
129.492462311558 0.00127574056386948
131.266331658291 0.00048558201524429
133.040201005025 0.000451524014351889
134.814070351759 0.000609563488978893
136.587939698492 0.000636092852801085
138.361809045226 0.00482356874272227
140.13567839196 0.0304486658424139
141.909547738693 0.0563559830188751
143.683417085427 0.01704465970397
145.457286432161 0.0222453996539116
147.231155778894 0.020012304186821
149.005025125628 0.00732386205345392
150.778894472362 0.0287592280656099
152.552763819095 0.0278719048947096
154.326633165829 0.0172044020146132
156.100502512563 0.0254252478480339
157.874371859296 0.0187148097902536
159.64824120603 0.0295674111694098
161.422110552764 0.0376745201647282
163.195979899498 0.041477657854557
164.969849246231 0.00594110134989023
166.743718592965 0.00114995241165161
168.517587939699 0.000181276351213455
170.291457286432 -7.67497476772405e-05
172.065326633166 0.00796425342559814
173.8391959799 0.0439013689756393
175.613065326633 0.0391351357102394
177.386934673367 0.0348547473549843
179.1608040201 0.030694393441081
180.934673366834 0.0176116600632668
182.708542713568 0.00436642905697227
184.482412060301 0.00635869894176722
186.256281407035 0.0155862607061863
188.030150753769 0.0188654363155365
189.804020100503 0.0172421224415302
191.577889447236 0.0105814142152667
193.35175879397 0.00885576009750366
195.125628140704 0.0187954399734735
196.899497487437 0.0438822880387306
198.673366834171 0.0304956529289484
200.447236180905 0.00290483655408025
202.221105527638 9.37677214096766e-06
203.994974874372 6.69106059447699e-09
205.768844221106 1.25499836940435e-05
207.542713567839 0.000276009523076937
209.316582914573 0.000229146447964013
211.090452261307 3.78057166017243e-06
212.86432160804 -5.68013490465091e-07
214.638190954774 -1.18412394556344e-07
216.412060301508 -3.34556737868752e-08
218.185929648241 -1.36168537778758e-07
219.959798994975 3.09737828274592e-07
221.733668341709 1.81200593942776e-05
223.507537688442 0.000362788938218728
225.281407035176 0.00091453839559108
227.05527638191 3.4935128496727e-06
228.829145728643 0.000163470715051517
230.603015075377 0.000236158302868716
232.376884422111 0.000336830475134775
234.150753768844 0.00047805585199967
235.924623115578 0.000461229879874736
237.698492462312 0.000512058031745255
239.472361809045 0.000571239506825805
241.246231155779 0.000972909387201071
243.020100502513 0.00140473514329642
244.793969849246 0.0015546316280961
246.56783919598 0.00151292909868062
248.341708542714 0.00179922732058913
250.115577889447 0.00119949073996395
251.889447236181 0.00118388305418193
253.663316582915 0.00107053224928677
255.437185929648 0.000964939885307103
257.211055276382 0.000893320713657886
258.984924623116 0.000434344838140532
260.758793969849 0.000379245117073879
262.532663316583 0.000496248598210514
264.306532663317 0.000114945127279498
266.08040201005 1.71285728356452e-06
267.854271356784 -3.07607564309365e-07
269.628140703518 -5.97480749320312e-08
271.402010050251 -5.76792347217747e-09
273.175879396985 -3.08257364078912e-10
274.949748743719 9.50802457599931e-12
276.723618090452 4.19489529918105e-12
278.497487437186 -2.52417119324388e-12
280.27135678392 -2.48436514416195e-11
282.045226130653 -4.88332881076747e-11
283.819095477387 1.80226655821514e-09
285.592964824121 3.30858753727625e-08
287.366834170854 3.3889932637976e-07
289.140703517588 1.69895338331116e-06
290.914572864322 -1.03545771708013e-05
292.688442211055 -0.000326016626786441
294.462311557789 0.00961440056562424
296.236180904523 0.00893766619265079
298.010050251256 1.00332781585166e-05
299.78391959799 -1.96490855159936e-05
301.557788944724 -2.85230748886534e-06
303.331658291457 -2.35519806324191e-07
305.105527638191 -8.77492567497029e-09
306.879396984925 1.01990871304025e-09
308.653266331658 2.29062838141836e-10
310.427135678392 2.34025247097813e-11
312.201005025126 1.36842143519389e-12
313.974874371859 -1.87757761541032e-14
315.748743718593 -1.63390300773761e-14
317.522613065327 -2.1256334162994e-15
319.29648241206 -1.61797441022768e-16
321.070351758794 -4.40935739274659e-18
322.844221105528 8.89542482716436e-19
324.618090452261 -1.20435367197644e-18
326.391959798995 -1.49433585641828e-17
328.165829145729 -8.48590429086064e-17
329.939698492462 3.18499899458464e-16
331.713567839196 1.32298101442638e-14
333.48743718593 1.78066301937882e-13
335.261306532663 1.40844053000272e-12
337.035175879397 2.20585572377474e-12
338.809045226131 -1.09758084565481e-10
340.582914572864 -1.94282190335571e-09
342.356783919598 -1.94556317723027e-08
344.130653266332 -9.21601781556092e-08
345.904522613065 6.68979964757455e-07
347.678391959799 1.9349905414856e-05
349.452261306533 0.000327834277413785
351.226130653266 0.0021393911447376
353 0.00190574792213738
};
\addlegendentry{Reference}
\addplot [semithick, steelblue31119180]
table {%
0 5.89429061250871e-19
1.77386934673367 0.00106045561000029
3.54773869346734 0.000693668662404386
5.321608040201 0.000700838511019043
7.09547738693467 0.00196706494176399
8.86934673366834 0.000420874697006613
10.643216080402 0.000479822983000486
12.4170854271357 0.000977728781980676
14.1909547738693 0.000352204153806283
15.964824120603 0.000740607429781046
17.7386934673367 0.000453566122531712
19.5125628140704 0.00143648334346969
21.286432160804 0.00113234856323492
23.0603015075377 0.000495027062043921
24.8341708542714 0.000608934415791676
26.608040201005 0.000572441330206606
28.3819095477387 0.000576214156345116
30.1557788944724 0.000569687328297286
31.929648241206 0.000569247028029625
33.7035175879397 0.000811470808931984
35.4773869346734 0.000940001449870324
37.251256281407 0.000611335975161366
39.0251256281407 0.000616601037589094
40.7989949748744 0.00107617345410022
42.572864321608 0.00057171261300945
44.3467336683417 0.000589827661882932
46.1206030150754 0.00199444235270376
47.894472361809 0.00623911436718781
49.6683417085427 0.0141475150498297
51.4422110552764 0.00565669797733472
53.21608040201 0.00075283933731728
54.9899497487437 0.000649269125227674
56.7638190954774 0.000666510304539038
58.5376884422111 0.000667687078137118
60.3115577889447 0.000675629307030711
62.0854271356784 0.00069365289796872
63.8592964824121 0.000699278016460075
65.6331658291457 0.000728554532617086
67.4070351758794 0.00110241564577341
69.1809045226131 0.00793775053711696
70.9547738693467 0.0104995587058728
72.7286432160804 0.00331314692353784
74.5025125628141 0.000704477305785698
76.2763819095477 0.000656332821254892
78.0502512562814 0.000615321751199124
79.8241206030151 0.000639143410313808
81.5979899497488 0.00114468503151467
83.3718592964824 0.00266869008935346
85.1457286432161 0.00094996858931214
86.9195979899498 0.00208523775977767
88.6934673366834 0.00202951586352567
90.4673366834171 0.00160896858464175
92.2412060301507 0.00185797975797497
94.0150753768844 0.00095765824696343
95.7889447236181 0.000680936614309578
97.5628140703518 0.00144581978923211
99.3366834170854 0.00038625847912486
101.110552763819 0.000569294406095013
102.884422110553 0.00291320740233223
104.658291457286 0.00115413179155772
106.43216080402 0.00041934456665111
108.206030150754 0.00242083949849365
109.979899497487 0.000623842473480747
111.753768844221 0.00166145746079953
113.527638190955 0.000854998041217724
115.301507537688 0.00196462170931691
117.075376884422 0.00207680930270388
118.849246231156 0.00113753007538828
120.623115577889 0.00222490737654103
122.396984924623 0.000450227054094493
124.170854271357 0.000872753214157963
125.94472361809 0.00121538229328059
127.718592964824 0.00146491873234977
129.492462311558 0.000656733394613023
131.266331658291 0.00326123082674177
133.040201005025 0.00270142718564058
134.814070351759 0.000927415827119069
136.587939698492 0.00255540633985494
138.361809045226 0.00663829626486609
140.13567839196 0.0235688148692985
141.909547738693 0.0347205284251503
143.683417085427 0.0173289909719958
145.457286432161 0.0162787411452117
147.231155778894 0.0145001562481812
149.005025125628 0.0072284147253014
150.778894472362 0.0176406102387722
152.552763819095 0.0173700063858328
154.326633165829 0.0156776921101875
156.100502512563 0.0181752372739237
157.874371859296 0.0163815639351567
159.64824120603 0.02156027230202
161.422110552764 0.0283586705156262
163.195979899498 0.0304164146888947
164.969849246231 0.00656671351686328
166.743718592965 0.000766527402453261
168.517587939699 0.000864061163871487
170.291457286432 0.000714195082387904
172.065326633166 0.00740835615641527
173.8391959799 0.0291118239843357
175.613065326633 0.0308964600246367
177.386934673367 0.0267746994018264
179.1608040201 0.0211626364490561
180.934673366834 0.0112796279306366
182.708542713568 0.00414340594003257
184.482412060301 0.00572145860716439
186.256281407035 0.0132596707820528
188.030150753769 0.0132840383317842
189.804020100503 0.0105042658021091
191.577889447236 0.00911944238175478
193.35175879397 0.00777550594569771
195.125628140704 0.0147598655410316
196.899497487437 0.0310876271443064
198.673366834171 0.0228230025366485
200.447236180905 0.0059612575137261
202.221105527638 0.00278912124992414
203.994974874372 0.000734435682377262
205.768844221106 0.000664722467691778
207.542713567839 0.000929737624523812
209.316582914573 0.000502194788919765
211.090452261307 0.00268325598118852
212.86432160804 0.000574622219656775
214.638190954774 0.00104463911033691
216.412060301508 0.00152636774415068
218.185929648241 0.000648544706464209
219.959798994975 0.00117287233937314
221.733668341709 0.00127372703438881
223.507537688442 0.00194337852278913
225.281407035176 0.00316889705774645
227.05527638191 0.000843958219845462
228.829145728643 0.000933575948636204
230.603015075377 0.000621853850549559
232.376884422111 0.00366578708324949
234.150753768844 0.000692387747885092
235.924623115578 0.000617230006296649
237.698492462312 0.00261767388715284
239.472361809045 0.000680934698723573
241.246231155779 0.000784999996231456
243.020100502513 0.00320410604759887
244.793969849246 0.00321419330265399
246.56783919598 0.00374250664189565
248.341708542714 0.00408014167583782
250.115577889447 0.000780668920357565
251.889447236181 0.00157949162433286
253.663316582915 0.00072015548666946
255.437185929648 0.00244726944537928
257.211055276382 0.00356993499400215
258.984924623116 0.00125150767442905
260.758793969849 0.00340121021707294
262.532663316583 0.00133504820338757
264.306532663317 0.000924587467781999
266.08040201005 0.000676105251183418
267.854271356784 0.00067895863346495
269.628140703518 0.00069219976288469
271.402010050251 0.0008059290322679
273.175879396985 0.00094519659019303
274.949748743719 0.000749564902003553
276.723618090452 0.00113188961607871
278.497487437186 0.000612024199010243
280.27135678392 0.00157039120322167
282.045226130653 0.00077322515516947
283.819095477387 0.000734740714754826
285.592964824121 0.00103843444073296
287.366834170854 0.000774972270849799
289.140703517588 0.000751744812426602
290.914572864322 0.000740232279944428
292.688442211055 0.00106347068514787
294.462311557789 0.00704612424253818
296.236180904523 0.00672649655435424
298.010050251256 0.00134847331586956
299.78391959799 0.000680296841831097
301.557788944724 0.000544286794769539
303.331658291457 0.00172751566853553
305.105527638191 0.000702046623033541
306.879396984925 0.000675534836111238
308.653266331658 0.000692633167960006
310.427135678392 0.000632954172289002
312.201005025126 0.00119379408040656
313.974874371859 0.000795031752317348
315.748743718593 0.000467335077956841
317.522613065327 0.00113143553958779
319.29648241206 0.000848114412824297
321.070351758794 0.000553797406743837
322.844221105528 0.000985000051084766
324.618090452261 0.000518230681622039
326.391959798995 0.000629835017670432
328.165829145729 0.000904011572281597
329.939698492462 0.000616069772773878
331.713567839196 0.00037185263846168
333.48743718593 0.00116875104036817
335.261306532663 0.000544575659134148
337.035175879397 0.00356539214683574
338.809045226131 8.36603785550586e-05
340.582914572864 0.00156803270356973
342.356783919598 0.000898701467644117
344.130653266332 0.00168598912101335
345.904522613065 0.00108789489213992
347.678391959799 0.000893874892201927
349.452261306533 0.00130532414213947
351.226130653266 0.0012195346548077
353 0.000721111765004016
};
\addlegendentry{L-BFGS}

\addplot [semithick, forestgreen4416044]
table {%
0 0.000376508280169219
1.77386934673367 7.29711473468342e-06
3.54773869346734 0.000557559949811548
5.321608040201 0.000902442669030279
7.09547738693467 0.000995663809590042
8.86934673366834 0.00100736925378442
10.643216080402 0.0009646873222664
12.4170854271357 0.000897378253284842
14.1909547738693 0.000884991721250117
15.964824120603 0.000867830298375338
17.7386934673367 0.000891213538125157
19.5125628140704 0.00086616980843246
21.286432160804 0.000794868567027152
23.0603015075377 0.00101309572346509
24.8341708542714 0.00104300712700933
26.608040201005 0.00101458665449172
28.3819095477387 0.000941151927690953
30.1557788944724 0.000846796261612326
31.929648241206 0.000840081425849348
33.7035175879397 0.000788351986557245
35.4773869346734 0.000802903261501342
37.251256281407 0.000924295338336378
39.0251256281407 0.00101801252458245
40.7989949748744 0.00106376048643142
42.572864321608 0.000984014361165464
44.3467336683417 0.00101517967414111
46.1206030150754 0.00121862394735217
47.894472361809 0.00657686172053218
49.6683417085427 0.0146077666431665
51.4422110552764 0.00524762365967035
53.21608040201 0.000742703210562468
54.9899497487437 0.000842968875076622
56.7638190954774 0.000884983048308641
58.5376884422111 0.00090424029622227
60.3115577889447 0.000947572058066726
62.0854271356784 0.00100018677767366
63.8592964824121 0.000948528060689569
65.6331658291457 0.00054417276987806
67.4070351758794 0.00114934158045799
69.1809045226131 0.00825223606079817
70.9547738693467 0.0107815470546484
72.7286432160804 0.00250758766196668
74.5025125628141 0.000706041173543781
76.2763819095477 0.000202842216822319
78.0502512562814 0.000118810989079066
79.8241206030151 0.000147556216688827
81.5979899497488 0.000553620746359229
83.3718592964824 0.000881935586221516
85.1457286432161 0.0018615365261212
86.9195979899498 0.00158647482749075
88.6934673366834 0.00146547541953623
90.4673366834171 0.00113969238009304
92.2412060301507 0.00103026058059186
94.0150753768844 0.00158192252274603
95.7889447236181 0.00124351482372731
97.5628140703518 0.00111026002559811
99.3366834170854 0.000209978490602225
101.110552763819 0.00159311352763325
102.884422110553 0.00175464025232941
104.658291457286 0.00189652945846319
106.43216080402 0.00148561701644212
108.206030150754 0.00145401933696121
109.979899497487 0.00144223135430366
111.753768844221 0.0013953591696918
113.527638190955 0.00128328800201416
115.301507537688 0.00318467034958303
117.075376884422 0.000995973707176745
118.849246231156 0.00121069489978254
120.623115577889 0.00158949079923332
122.396984924623 0.000171958628925495
124.170854271357 0.000364144536433741
125.94472361809 0.00174745079129934
127.718592964824 0.00138438655994833
129.492462311558 0.00185903266537935
131.266331658291 0.00356369209475815
133.040201005025 0.00177427870221436
134.814070351759 0.00177739222999662
136.587939698492 0.00248150154948235
138.361809045226 0.00566681195050478
140.13567839196 0.0229810029268265
141.909547738693 0.0363133512437344
143.683417085427 0.0177286826074123
145.457286432161 0.0165360756218433
147.231155778894 0.0136459898203611
149.005025125628 0.00744134094566107
150.778894472362 0.0166112016886473
152.552763819095 0.0185869876295328
154.326633165829 0.0152165088802576
156.100502512563 0.0175833851099014
157.874371859296 0.0155613524839282
159.64824120603 0.0207438077777624
161.422110552764 0.0278345085680485
163.195979899498 0.030668705701828
164.969849246231 0.0049825725145638
166.743718592965 0.00166933599393815
168.517587939699 0.00155479635577649
170.291457286432 0.00135261635296047
172.065326633166 0.00750148063525558
173.8391959799 0.0296139009296894
175.613065326633 0.0292322486639023
177.386934673367 0.0276054311543703
179.1608040201 0.022234695032239
180.934673366834 0.011757499538362
182.708542713568 0.00330943218432367
184.482412060301 0.00415320182219148
186.256281407035 0.0135303186252713
188.030150753769 0.0147281158715487
189.804020100503 0.0108904307708144
191.577889447236 0.0080171125009656
193.35175879397 0.00653634779155254
195.125628140704 0.0175686366856098
196.899497487437 0.0292771272361279
198.673366834171 0.0206467360258102
200.447236180905 0.00529926596209407
202.221105527638 0.00182502612005919
203.994974874372 0.00138203869573772
205.768844221106 0.000904474698472768
207.542713567839 0.000609687878750265
209.316582914573 3.11531766783446e-05
211.090452261307 0.000927794608287513
212.86432160804 0.00107981206383556
214.638190954774 0.00110829493496567
216.412060301508 0.00135748414322734
218.185929648241 0.00126718950923532
219.959798994975 0.00145109568256885
221.733668341709 0.00126147957053035
223.507537688442 0.00133477000053972
225.281407035176 0.000830748409498483
227.05527638191 0.000882309395819902
228.829145728643 0.000702314951922745
230.603015075377 0.0010170271852985
232.376884422111 0.00135888298973441
234.150753768844 0.0012312171747908
235.924623115578 0.00136597245000303
237.698492462312 0.00144389586057514
239.472361809045 0.0016409921227023
241.246231155779 0.00189856078941375
243.020100502513 0.00232933345250785
244.793969849246 0.0035931202583015
246.56783919598 0.00465857004746795
248.341708542714 0.0030845336150378
250.115577889447 0.00227937242016196
251.889447236181 0.000963085214607418
253.663316582915 0.00109053298365325
255.437185929648 0.000918715842999518
257.211055276382 0.00173157756216824
258.984924623116 0.00189486006274819
260.758793969849 0.00154287915211171
262.532663316583 0.000696015369612724
264.306532663317 0.000459754810435697
266.08040201005 0.000135757320094854
267.854271356784 8.47871488076635e-05
269.628140703518 0.000152235064888373
271.402010050251 0.00245405989699066
273.175879396985 0.000677530711982399
274.949748743719 0.000766162411309779
276.723618090452 0.00103067827876657
278.497487437186 0.00110768643207848
280.27135678392 0.00115516886580735
282.045226130653 0.00114700349513441
283.819095477387 0.00109886040445417
285.592964824121 0.00102652853820473
287.366834170854 0.00104484229814261
289.140703517588 0.00110459967982024
290.914572864322 0.000998736126348376
292.688442211055 0.00134919409174472
294.462311557789 0.00669559184461832
296.236180904523 0.0055724261328578
298.010050251256 0.00144591450225562
299.78391959799 0.00116224924568087
301.557788944724 0.00110707653220743
303.331658291457 0.00107878854032606
305.105527638191 0.0010911354329437
306.879396984925 0.00098903407342732
308.653266331658 0.000922941835597157
310.427135678392 0.000668164808303118
312.201005025126 0.000839308020658791
313.974874371859 0.000696763396263123
315.748743718593 0.000717964372597635
317.522613065327 0.000755764311179519
319.29648241206 0.000900524493772537
321.070351758794 0.000951730762608349
322.844221105528 0.000963489525020123
324.618090452261 0.00114567996934056
326.391959798995 0.00114233826752752
328.165829145729 0.000749383994843811
329.939698492462 0.000994449481368065
331.713567839196 0.00111329078208655
333.48743718593 0.00107784301508218
335.261306532663 0.00108521117363125
337.035175879397 0.00107643764931709
338.809045226131 0.0010708844056353
340.582914572864 0.000955989817157388
342.356783919598 0.000899529550224543
344.130653266332 0.000914083328098059
345.904522613065 0.000881779415067285
347.678391959799 0.000759251648560166
349.452261306533 0.00173617142718285
351.226130653266 0.0010953233577311
353 0.000711591506842524
};
\addlegendentry{Hybrid ODPS}
\end{axis}

\end{tikzpicture}

%% file: figures/profiles/st.tex
\begin{tikzpicture}

\definecolor{darkgrey176}{RGB}{176,176,176}
\definecolor{darkorange25512714}{RGB}{255,127,14}
\definecolor{forestgreen4416044}{RGB}{44,160,44}
\definecolor{lightgrey204}{RGB}{204,204,204}
\definecolor{steelblue31119180}{RGB}{31,119,180}

\begin{axis}[
tick align=outside,
tick pos=left,
x grid style={darkgrey176},
xmin=-17.65, xmax=370.65,
xtick style={color=black},
y grid style={darkgrey176},
ymin=-0.00546332036353996, ymax=0.135740481798648,
ytick style={color=black},
yticklabel style={
        /pgf/number format/fixed,
        /pgf/number format/precision=5,
},
scaled y ticks=false,
		width=12cm,
		height=7cm,
]
\addplot [semithick, steelblue31119180]
table {%
0 0.0252672904243296
1.77386934673367 0.0553282657098247
3.54773869346734 0.0964895146413367
5.321608040201 0.0843829081590024
7.09547738693467 0.0735225489085544
8.86934673366834 0.0918224094519699
10.643216080402 0.0984115740431569
12.4170854271357 0.0896694504283078
14.1909547738693 0.0973469525481053
15.964824120603 0.0874856153096036
17.7386934673367 0.0940108733602161
19.5125628140704 0.0810242014384132
21.286432160804 0.0876012538818164
23.0603015075377 0.0939169835040464
24.8341708542714 0.0917605056467976
26.608040201005 0.0978923396923951
28.3819095477387 0.0932056477216908
30.1557788944724 0.100433020121495
31.929648241206 0.104180840771329
33.7035175879397 0.0862218336094659
35.4773869346734 0.0845105364115755
37.251256281407 0.0934155904906934
39.0251256281407 0.106083084051532
40.7989949748744 0.08846299536436
42.572864321608 0.0982165759939631
44.3467336683417 0.112844951028294
46.1206030150754 0.0780021780687118
47.894472361809 0.100369087358993
49.6683417085427 0.0962056605546705
51.4422110552764 0.102081117838807
53.21608040201 0.0933197343188202
54.9899497487437 0.115654704359387
56.7638190954774 0.0862633884988428
58.5376884422111 0.110667414896881
60.3115577889447 0.102983413170052
62.0854271356784 0.0841560664688606
63.8592964824121 0.0884892309168603
65.6331658291457 0.0954257292595586
67.4070351758794 0.0898702161805627
69.1809045226131 0.085090389385894
70.9547738693467 0.089145497774526
72.7286432160804 0.0954683697608953
74.5025125628141 0.108320940714299
76.2763819095477 0.0909030188152179
78.0502512562814 0.103350910357854
79.8241206030151 0.0872594907379783
81.5979899497488 0.0732409476839782
83.3718592964824 0.0463893335061201
85.1457286432161 0.0265352849990306
86.9195979899498 0.0202324078557547
88.6934673366834 0.0178862753562305
90.4673366834171 0.0199905734340865
92.2412060301507 0.018028788957337
94.0150753768844 0.0199944488684939
95.7889447236181 0.0250327756034402
97.5628140703518 0.021896675477728
99.3366834170854 0.0516152505158663
101.110552763819 0.0372390994922032
102.884422110553 0.0144755940707222
104.658291457286 0.028886795634981
106.43216080402 0.0340927893153411
108.206030150754 0.00300657430499864
109.979899497487 0.0395209789265625
111.753768844221 0.0273643798564596
113.527638190955 0.0445573081461418
115.301507537688 0.0203289184166327
117.075376884422 0.0139072412916183
118.849246231156 0.0243109701644654
120.623115577889 0.0115734973666971
122.396984924623 0.029557940231227
124.170854271357 0.0323291486123072
125.94472361809 0.0238698086704676
127.718592964824 0.0231900794349275
129.492462311558 0.0361682515943775
131.266331658291 0.00527138424429764
133.040201005025 0.0203764845719583
134.814070351759 0.0259117022317411
136.587939698492 0.0159354063721091
138.361809045226 0.0306344624566296
140.13567839196 0.0647712917175883
141.909547738693 0.0672981148384777
143.683417085427 0.0664664027956253
145.457286432161 0.0809661179395771
147.231155778894 0.0786119257070156
149.005025125628 0.0839960455263943
150.778894472362 0.0755832737380936
152.552763819095 0.0955907654498629
154.326633165829 0.0638062302412086
156.100502512563 0.073975318700526
157.874371859296 0.064342549398248
159.64824120603 0.0726736462456844
161.422110552764 0.0768555363396497
163.195979899498 0.0619493554602727
164.969849246231 0.087199952515963
166.743718592965 0.107204955626156
168.517587939699 0.0975389276272023
170.291457286432 0.0974241611748554
172.065326633166 0.0849084679385302
173.8391959799 0.0782422819358961
175.613065326633 0.0603296752511369
177.386934673367 0.0623969325722577
179.1608040201 0.0775854911515226
180.934673366834 0.0871177025645107
182.708542713568 0.0779279617638501
184.482412060301 0.0786670936296942
186.256281407035 0.0707537429874062
188.030150753769 0.0787094326150132
189.804020100503 0.0893289464746241
191.577889447236 0.076979087602951
193.35175879397 0.0752803310551263
195.125628140704 0.085358566278399
196.899497487437 0.0598519586491214
198.673366834171 0.0520954767531188
200.447236180905 0.0223040469980447
202.221105527638 0.0118635945106198
203.994974874372 0.0293648150565297
205.768844221106 0.0304865030813245
207.542713567839 0.0348103824079835
209.316582914573 0.0397134682675865
211.090452261307 0.00385287000377481
212.86432160804 0.029780870442948
214.638190954774 0.0159437136116623
216.412060301508 0.0173422462205378
218.185929648241 0.0344394965698182
219.959798994975 0.0160104045416393
221.733668341709 0.016803261738049
223.507537688442 0.0127848664515316
225.281407035176 0.00949079730482056
227.05527638191 0.0208235418035684
228.829145728643 0.0205883198125824
230.603015075377 0.0245686276450686
232.376884422111 0.00727664321014174
234.150753768844 0.025716547513522
235.924623115578 0.0370510269455436
237.698492462312 0.0111373411069421
239.472361809045 0.0233766768389645
241.246231155779 0.0191437906952318
243.020100502513 0.00887830531544009
244.793969849246 0.0122835480274464
246.56783919598 0.0096752680047333
248.341708542714 0.0111101134778953
250.115577889447 0.0302353511082493
251.889447236181 0.0208438725739422
253.663316582915 0.0341578811820165
255.437185929648 0.0205039950996145
257.211055276382 0.013116288771673
258.984924623116 0.0436681684343712
260.758793969849 0.0554223857929042
262.532663316583 0.0858519773074709
264.306532663317 0.0839323413691253
266.08040201005 0.102525231097873
267.854271356784 0.106186028009361
269.628140703518 0.0933995017784272
271.402010050251 0.093070880391939
273.175879396985 0.0932435859662067
274.949748743719 0.100749058714998
276.723618090452 0.0955026859882087
278.497487437186 0.0924857082988537
280.27135678392 0.0891102830353979
282.045226130653 0.0953261271909486
283.819095477387 0.0914213374007085
285.592964824121 0.0907593764966892
287.366834170854 0.0951592464452225
289.140703517588 0.104581496693978
290.914572864322 0.0913739613158841
292.688442211055 0.0887466694139565
294.462311557789 0.0872429478560745
296.236180904523 0.0912928265117643
298.010050251256 0.0908348114804337
299.78391959799 0.0866983031902752
301.557788944724 0.0981384485999255
303.331658291457 0.0753982151304292
305.105527638191 0.0949904837415546
306.879396984925 0.0997082768669001
308.653266331658 0.0927148959036528
310.427135678392 0.0924214018201318
312.201005025126 0.086181021098787
313.974874371859 0.0811940685592744
315.748743718593 0.129322127154912
317.522613065327 0.0859762582272698
319.29648241206 0.090941139974306
321.070351758794 0.0947599515155006
322.844221105528 0.0824238212759183
324.618090452261 0.0975555210963134
326.391959798995 0.0932288009529978
328.165829145729 0.0886056479279037
329.939698492462 0.0927480882052284
331.713567839196 0.0947761884477005
333.48743718593 0.080314472535701
335.261306532663 0.0929254660798864
337.035175879397 0.0585411112244648
338.809045226131 0.0930232874877175
340.582914572864 0.0830225288378846
342.356783919598 0.0911180826828249
344.130653266332 0.0773880473370501
345.904522613065 0.0807357273118162
347.678391959799 0.0806357772583499
349.452261306533 0.0588400048068167
351.226130653266 0.0218284714505043
353 0.0052222264174595
};
\addplot [semithick, darkorange25512714]
table {%
0 0.0144259529188275
1.77386934673367 0.0837511047720909
3.54773869346734 0.10124633461237
5.321608040201 0.0998178347945213
7.09547738693467 0.1002007573843
8.86934673366834 0.0986308380961418
10.643216080402 0.0994263887405396
12.4170854271357 0.0997193530201912
14.1909547738693 0.0994973704218864
15.964824120603 0.0987595319747925
17.7386934673367 0.0988104417920113
19.5125628140704 0.100483268499374
21.286432160804 0.099287211894989
23.0603015075377 0.100369334220886
24.8341708542714 0.100258812308311
26.608040201005 0.100502073764801
28.3819095477387 0.101329505443573
30.1557788944724 0.100540943443775
31.929648241206 0.0992640554904938
33.7035175879397 0.100932329893112
35.4773869346734 0.0991204902529716
37.251256281407 0.0990833938121796
39.0251256281407 0.0999778732657433
40.7989949748744 0.0997036024928093
42.572864321608 0.100202858448029
44.3467336683417 0.100147992372513
46.1206030150754 0.102131567895412
47.894472361809 0.0966833829879761
49.6683417085427 0.0883326157927513
51.4422110552764 0.0963595733046532
53.21608040201 0.10019426047802
54.9899497487437 0.0998191982507706
56.7638190954774 0.0992551669478416
58.5376884422111 0.100191600620747
60.3115577889447 0.0998933091759682
62.0854271356784 0.101124100387096
63.8592964824121 0.102787211537361
65.6331658291457 0.101739853620529
67.4070351758794 0.100079506635666
69.1809045226131 0.0892803892493248
70.9547738693467 0.0862966924905777
72.7286432160804 0.0975647047162056
74.5025125628141 0.102825202047825
76.2763819095477 0.100647926330566
78.0502512562814 0.100201934576035
79.8241206030151 0.101220294833183
81.5979899497488 0.0940979495644569
83.3718592964824 0.0564373917877674
85.1457286432161 0.0221403520554304
86.9195979899498 0.0249907299876213
88.6934673366834 0.0255835670977831
90.4673366834171 0.0258732941001654
92.2412060301507 0.0266328062862158
94.0150753768844 0.0260121114552021
95.7889447236181 0.0273212399333715
97.5628140703518 0.0253281649202108
99.3366834170854 0.0286447517573833
101.110552763819 0.0315718837082386
102.884422110553 0.0330998674035072
104.658291457286 0.0333558432757854
106.43216080402 0.0272783972322941
108.206030150754 0.0279312524944544
109.979899497487 0.0279408693313599
111.753768844221 0.0466082580387592
113.527638190955 0.0506482385098934
115.301507537688 0.0319802165031433
117.075376884422 0.0223573036491871
118.849246231156 0.0248015150427818
120.623115577889 0.0241495575755835
122.396984924623 0.0228680670261383
124.170854271357 0.0246471762657166
125.94472361809 0.0269938819110394
127.718592964824 0.0246908143162727
129.492462311558 0.0275611039251089
131.266331658291 0.0246424414217472
133.040201005025 0.0251248143613338
134.814070351759 0.0217231083661318
136.587939698492 0.0205752681940794
138.361809045226 0.0379373803734779
140.13567839196 0.0547613762319088
141.909547738693 0.0433984994888306
143.683417085427 0.0774803012609482
145.457286432161 0.0753434523940086
147.231155778894 0.0772146359086037
149.005025125628 0.0837021321058273
150.778894472362 0.0700838565826416
152.552763819095 0.0697722807526588
154.326633165829 0.0758200734853745
156.100502512563 0.0720693841576576
157.874371859296 0.0744026303291321
159.64824120603 0.0680216327309608
161.422110552764 0.0629535540938377
163.195979899498 0.0575234368443489
164.969849246231 0.0925478264689445
166.743718592965 0.0965173467993736
168.517587939699 0.09608343988657
170.291457286432 0.0955935567617416
172.065326633166 0.0896496623754501
173.8391959799 0.058302991092205
175.613065326633 0.0591799467802048
177.386934673367 0.0600635185837746
179.1608040201 0.0683307945728302
180.934673366834 0.0805849507451057
182.708542713568 0.086064338684082
184.482412060301 0.0881703943014145
186.256281407035 0.081134982407093
188.030150753769 0.0791432783007622
189.804020100503 0.0848533362150192
191.577889447236 0.0827278792858124
193.35175879397 0.0812079012393951
195.125628140704 0.0801981464028358
196.899497487437 0.065204881131649
198.673366834171 0.0523243360221386
200.447236180905 0.0280568171292543
202.221105527638 0.0243471805006266
203.994974874372 0.0238807182759047
205.768844221106 0.0228989459574223
207.542713567839 0.037334993481636
209.316582914573 0.0311070214956999
211.090452261307 0.0247912593185902
212.86432160804 0.0230315178632736
214.638190954774 0.0236428435891867
216.412060301508 0.0237053483724594
218.185929648241 0.0243535246700048
219.959798994975 0.0250019989907742
221.733668341709 0.0244106519967318
223.507537688442 0.024206543341279
225.281407035176 0.0247700158506632
227.05527638191 0.0213213656097651
228.829145728643 0.0216799899935722
230.603015075377 0.0207491926848888
232.376884422111 0.0218337830156088
234.150753768844 0.0221025664359331
235.924623115578 0.021591380238533
237.698492462312 0.0206604860723019
239.472361809045 0.0201198700815439
241.246231155779 0.0198216438293457
243.020100502513 0.020725579932332
244.793969849246 0.0231056585907936
246.56783919598 0.0204654298722744
248.341708542714 0.0226284898817539
250.115577889447 0.0213529355823994
251.889447236181 0.0219791159033775
253.663316582915 0.0222180392593145
255.437185929648 0.0232826098799706
257.211055276382 0.0278521850705147
258.984924623116 0.0458544120192528
260.758793969849 0.0855441093444824
262.532663316583 0.0992322489619255
264.306532663317 0.100592695176601
266.08040201005 0.101887807250023
267.854271356784 0.10136216878891
269.628140703518 0.101030021905899
271.402010050251 0.101605758070946
273.175879396985 0.101419053971767
274.949748743719 0.101869374513626
276.723618090452 0.104001998901367
278.497487437186 0.103737875819206
280.27135678392 0.103219866752625
282.045226130653 0.101832948625088
283.819095477387 0.103004179894924
285.592964824121 0.104230105876923
287.366834170854 0.103726111352444
289.140703517588 0.103025034070015
290.914572864322 0.102870985865593
292.688442211055 0.102395989000797
294.462311557789 0.0936373099684715
296.236180904523 0.0954014211893082
298.010050251256 0.100538052618504
299.78391959799 0.100376255810261
301.557788944724 0.100216500461102
303.331658291457 0.0997431874275208
305.105527638191 0.100568428635597
306.879396984925 0.100346274673939
308.653266331658 0.101974911987782
310.427135678392 0.100911103188992
312.201005025126 0.101304098963737
313.974874371859 0.100814551115036
315.748743718593 0.100687682628632
317.522613065327 0.100672401487827
319.29648241206 0.100634217262268
321.070351758794 0.100274011492729
322.844221105528 0.100906983017921
324.618090452261 0.100975371897221
326.391959798995 0.101897396147251
328.165829145729 0.100392714142799
329.939698492462 0.0998132526874542
331.713567839196 0.100425511598587
333.48743718593 0.101477175951004
335.261306532663 0.093846283853054
337.035175879397 0.0950823053717613
338.809045226131 0.100134640932083
340.582914572864 0.10027813911438
342.356783919598 0.0993484407663345
344.130653266332 0.0978519916534424
345.904522613065 0.0968887731432915
347.678391959799 0.102260805666447
349.452261306533 0.0728944838047028
351.226130653266 0.01397212035954
353 0.000955034280195832
};
\addplot [semithick, forestgreen4416044]
table {%
0 0.0183153990656137
1.77386934673367 0.0833855122327805
3.54773869346734 0.0913695767521858
5.321608040201 0.0928246304392815
7.09547738693467 0.0926549658179283
8.86934673366834 0.090868666768074
10.643216080402 0.090571254491806
12.4170854271357 0.0897963121533394
14.1909547738693 0.0893096029758453
15.964824120603 0.0890611857175827
17.7386934673367 0.0888539478182793
19.5125628140704 0.089959591627121
21.286432160804 0.0949179828166962
23.0603015075377 0.0939888656139374
24.8341708542714 0.095175676047802
26.608040201005 0.0965084210038185
28.3819095477387 0.0958170369267464
30.1557788944724 0.097447007894516
31.929648241206 0.0977403968572617
33.7035175879397 0.0980697274208069
35.4773869346734 0.0980925038456917
37.251256281407 0.0988302007317543
39.0251256281407 0.099525548517704
40.7989949748744 0.0997715294361115
42.572864321608 0.0978135466575623
44.3467336683417 0.0987199693918228
46.1206030150754 0.0992444008588791
47.894472361809 0.0996298789978027
49.6683417085427 0.092445395886898
51.4422110552764 0.0988718345761299
53.21608040201 0.104933179914951
54.9899497487437 0.10256365686655
56.7638190954774 0.10288006067276
58.5376884422111 0.101388469338417
60.3115577889447 0.100028306245804
62.0854271356784 0.100560337305069
63.8592964824121 0.100343279540539
65.6331658291457 0.0996236577630043
67.4070351758794 0.100899338722229
69.1809045226131 0.09546859562397
70.9547738693467 0.0870554223656654
72.7286432160804 0.0937067940831184
74.5025125628141 0.094757154583931
76.2763819095477 0.0948523208498955
78.0502512562814 0.0944729670882225
79.8241206030151 0.0973938629031181
81.5979899497488 0.0881210491061211
83.3718592964824 0.061591237783432
85.1457286432161 0.0275831259787083
86.9195979899498 0.0204629879444838
88.6934673366834 0.0211840625852346
90.4673366834171 0.0329636260867119
92.2412060301507 0.0364319644868374
94.0150753768844 0.0267538484185934
95.7889447236181 0.024347823113203
97.5628140703518 0.0234809704124928
99.3366834170854 0.0222801342606544
101.110552763819 0.0287953391671181
102.884422110553 0.0356605462729931
104.658291457286 0.0258011631667614
106.43216080402 0.0221237447112799
108.206030150754 0.0198301617056131
109.979899497487 0.0192956831306219
111.753768844221 0.0203565582633018
113.527638190955 0.0353659205138683
115.301507537688 0.0494083315134048
117.075376884422 0.0217389985918999
118.849246231156 0.0180662497878075
120.623115577889 0.0161939784884453
122.396984924623 0.0489027090370655
124.170854271357 0.0458939075469971
125.94472361809 0.0189298633486032
127.718592964824 0.0179077181965113
129.492462311558 0.0230406429618597
131.266331658291 0.0398302897810936
133.040201005025 0.0286549981683493
134.814070351759 0.0229458026587963
136.587939698492 0.0263074077665806
138.361809045226 0.0396419242024422
140.13567839196 0.0603163726627827
141.909547738693 0.0656680092215538
143.683417085427 0.0742502957582474
145.457286432161 0.0752419158816338
147.231155778894 0.0792483538389206
149.005025125628 0.0855780020356178
150.778894472362 0.0838931724429131
152.552763819095 0.0826960802078247
154.326633165829 0.0825530216097832
156.100502512563 0.0770080760121346
157.874371859296 0.0785501301288605
159.64824120603 0.0761375352740288
161.422110552764 0.073751300573349
163.195979899498 0.0739270895719528
164.969849246231 0.0893260762095451
166.743718592965 0.0903443396091461
168.517587939699 0.0918876603245735
170.291457286432 0.0924119129776955
172.065326633166 0.0880680233240128
173.8391959799 0.0692313835024834
175.613065326633 0.0682794153690338
177.386934673367 0.0728893652558327
179.1608040201 0.0740348100662231
180.934673366834 0.0799498483538628
182.708542713568 0.0868628621101379
184.482412060301 0.0869027674198151
186.256281407035 0.083798311650753
188.030150753769 0.0836497396230698
189.804020100503 0.0838153958320618
191.577889447236 0.0815290510654449
193.35175879397 0.08365498483181
195.125628140704 0.0748589187860489
196.899497487437 0.0641944259405136
198.673366834171 0.0572599843144417
200.447236180905 0.0453211143612862
202.221105527638 0.0286594871431589
203.994974874372 0.0254907757043839
205.768844221106 0.0184601694345474
207.542713567839 0.0230924151837826
209.316582914573 0.021621061488986
211.090452261307 0.0269100163131952
212.86432160804 0.0222358126193285
214.638190954774 0.0207821205258369
216.412060301508 0.0211368482559919
218.185929648241 0.0201281197369099
219.959798994975 0.022614061832428
221.733668341709 0.0165406130254269
223.507537688442 0.0154670430347323
225.281407035176 0.0160481072962284
227.05527638191 0.0183473657816648
228.829145728643 0.015242800116539
230.603015075377 0.0187014471739531
232.376884422111 0.0161531381309032
234.150753768844 0.0172102507203817
235.924623115578 0.0184918660670519
237.698492462312 0.0147500783205032
239.472361809045 0.015037976205349
241.246231155779 0.016447726637125
243.020100502513 0.0283033233135939
244.793969849246 0.0353944897651672
246.56783919598 0.03515949472785
248.341708542714 0.026990732178092
250.115577889447 0.0235968250781298
251.889447236181 0.019140213727951
253.663316582915 0.021137023344636
255.437185929648 0.0209272094070911
257.211055276382 0.0223217252641916
258.984924623116 0.0304244197905064
260.758793969849 0.0687246918678284
262.532663316583 0.0887333750724792
264.306532663317 0.0896144509315491
266.08040201005 0.0884614661335945
267.854271356784 0.0899779424071312
269.628140703518 0.0941231623291969
271.402010050251 0.0932549461722374
273.175879396985 0.0946923047304153
274.949748743719 0.0941211208701134
276.723618090452 0.0936884060502052
278.497487437186 0.0911916494369507
280.27135678392 0.0936716198921204
282.045226130653 0.0938766896724701
283.819095477387 0.0936088114976883
285.592964824121 0.0930105000734329
287.366834170854 0.0929457917809486
289.140703517588 0.0936585441231728
290.914572864322 0.0924912467598915
292.688442211055 0.0927004441618919
294.462311557789 0.0923867225646973
296.236180904523 0.0912148877978325
298.010050251256 0.0945170000195503
299.78391959799 0.0937090963125229
301.557788944724 0.0938513576984406
303.331658291457 0.0943838804960251
305.105527638191 0.0953582227230072
306.879396984925 0.0950284376740456
308.653266331658 0.0970252156257629
310.427135678392 0.093299426138401
312.201005025126 0.094956710934639
313.974874371859 0.0938789322972298
315.748743718593 0.0916665866971016
317.522613065327 0.0941782295703888
319.29648241206 0.0930795222520828
321.070351758794 0.0918150395154953
322.844221105528 0.0935559570789337
324.618090452261 0.0910676121711731
326.391959798995 0.0900265946984291
328.165829145729 0.0910028219223022
329.939698492462 0.0925835520029068
331.713567839196 0.0930346250534058
333.48743718593 0.0939126163721085
335.261306532663 0.0943205058574677
337.035175879397 0.0939085856080055
338.809045226131 0.0936714559793472
340.582914572864 0.0934347212314606
342.356783919598 0.0931629911065102
344.130653266332 0.0934744030237198
345.904522613065 0.0939496755599976
347.678391959799 0.0874186530709267
349.452261306533 0.03190878033638
351.226130653266 0.00161153625231236
353 0.00114052300341427
};
\end{axis}

\end{tikzpicture}

%% file: figures/profiles/iod.tex
\begin{tikzpicture}

	\definecolor{darkgrey176}{RGB}{176,176,176}
	\definecolor{darkorange25512714}{RGB}{255,127,14}
	\definecolor{forestgreen4416044}{RGB}{44,160,44}
	\definecolor{lightgrey204}{RGB}{204,204,204}
	\definecolor{steelblue31119180}{RGB}{31,119,180}
	
	\begin{axis}[
		tick align=outside,
		tick pos=left,
		x grid style={darkgrey176},
		xmin=-1.35, xmax=28.35,
		xtick style={color=black},
		y grid style={darkgrey176},
		ymin=-3.54597941525159e-05, ymax=0.000744655677202835,
		ytick style={color=black},
        yticklabel style={
        /pgf/number format/fixed,
        /pgf/number format/precision=5
},
scaled y ticks=false,
		width=12cm,
		height=7cm,
	]
		\addplot [semithick, darkorange25512714]
		table {%
			0 1.91999331516255e-20
			0.135678391959799 2.56615530815907e-05
			0.271356783919598 5.8642886870075e-05
			0.407035175879397 9.63321144809015e-05
			0.542713567839196 0.000136117349029519
			0.678391959798995 0.000175386710907333
			0.814070351758794 0.000211528313229792
			0.949748743718593 0.000241930261836387
			1.08542713567839 0.000264157046331093
			1.22110552763819 0.000278125400654972
			1.35678391959799 0.000285426765913144
			1.49246231155779 0.000287688366370276
			1.62814070351759 0.000286537542706355
			1.76381909547739 0.000283601577393711
			1.89949748743719 0.000280507723800838
			2.03517587939699 0.000278877327218652
			2.17085427135678 0.000279667496215552
			2.30653266331658 0.000282578024780378
			2.44221105527638 0.000287168513750657
			2.57788944723618 0.000292998709483072
			2.71356783919598 0.000299628212815151
			2.84924623115578 0.000306616711895913
			2.98492462311558 0.000313523894874379
			3.12060301507538 0.000319974002195522
			3.25628140703518 0.00032595288939774
			3.39195979899498 0.000331571995047852
			3.52763819095477 0.000336942990543321
			3.66331658291457 0.000342177518177778
			3.79899497487437 0.000347387132933363
			3.93467336683417 0.000352683506207541
			4.07035175879397 0.000358174904249609
			4.20603015075377 0.000363898172508925
			4.34170854271357 0.000369822228094563
			4.47738693467337 0.000375913281459361
			4.61306532663317 0.00038213754305616
			4.74874371859296 0.000388461194233969
			4.88442211055276 0.000394850445445627
			5.02010050251256 0.000401271507143974
			5.15577889447236 0.000407683663070202
			5.29145728643216 0.000414028618251905
			5.42713567839196 0.000420245167333633
			5.56281407035176 0.000426272163167596
			5.69849246231156 0.000432048429502174
			5.83417085427136 0.000437512819189578
			5.96984924623116 0.000442604126874357
			6.10552763819095 0.000447272788733244
			6.24120603015075 0.000451560801593587
			6.37688442211055 0.000455554021755233
			6.51256281407035 0.000459338538348675
			6.64824120603015 0.000463000382296741
			6.78391959798995 0.000466625642729923
			6.91959798994975 0.000470300379674882
			7.05527638190955 0.000474109401693568
			7.19095477386935 0.000478089263197035
			7.32663316582915 0.000482215167721733
			7.46231155778895 0.000486458331579342
			7.59798994974874 0.00049078994197771
			7.73366834170854 0.000495181244332343
			7.86934673366834 0.000499603396747261
			8.00502512562814 0.000504027586430311
			8.14070351758794 0.00050843064673245
			8.27638190954774 0.000512809434439987
			8.41206030150754 0.000517165113706142
			8.54773869346734 0.000521499023307115
			8.68341708542714 0.000525812443811446
			8.81909547738693 0.000530106655787677
			8.95477386934673 0.000534382939804345
			9.09045226130653 0.000538639607839286
			9.22613065326633 0.000542840745765716
			9.36180904522613 0.000546928378753364
			9.49748743718593 0.000550844124518335
			9.63316582914573 0.000554529600776732
			9.76884422110553 0.00055792648345232
			9.90452261306533 0.000560976332053542
			10.0402010050251 0.000563621404580772
			10.1758793969849 0.000565846392419189
			10.3115577889447 0.000567708921153098
			10.4472361809045 0.000569273717701435
			10.5829145728643 0.000570605567190796
			10.7185929648241 0.000571769254747778
			10.8542713567839 0.000572829507291317
			10.9899497487437 0.00057385116815567
			11.1256281407035 0.000574890640564263
			11.2613065326633 0.000575963349547237
			11.3969849246231 0.000577071739826351
			11.5326633165829 0.000578218256123364
			11.6683417085427 0.000579405401367694
			11.8040201005025 0.000580635678488761
			11.9396984924623 0.000581911474000663
			12.0753768844221 0.000583235290832818
			12.2110552763819 0.000584608351346105
			12.3467336683417 0.000586031062994152
			12.4824120603015 0.000587503658607602
			12.6180904522613 0.00058902648743242
			12.7537688442211 0.00059059978229925
			12.8894472361809 0.000592223950661719
			13.0251256281407 0.000593899167142808
			13.1608040201005 0.000595621531829238
			13.2964824120603 0.000597377249505371
			13.4321608040201 0.000599151069764048
			13.5678391959799 0.000600927800405771
			13.7035175879397 0.00060269224923104
			13.8391959798995 0.000604429165832698
			13.9748743718593 0.000606123358011246
			14.1105527638191 0.000607765279710293
			14.2462311557789 0.000609385839197785
			14.3819095477387 0.000611033290624619
			14.5175879396985 0.000612755713518709
			14.6532663316583 0.000614601536653936
			14.7889447236181 0.000616618955973536
			14.9246231155779 0.000618856225628406
			15.0603015075377 0.000621358107309788
			15.1959798994975 0.000624063890427351
			15.3316582914573 0.000626790861133486
			15.4673366834171 0.00062934955349192
			15.6030150753769 0.000631550443358719
			15.7386934673367 0.000633204006589949
			15.8743718592965 0.000634120835456997
			16.0100502512563 0.000634111347608268
			16.1457286432161 0.000633062503766268
			16.2814070351759 0.000631106027867645
			16.4170854271357 0.000628422771114856
			16.5527638190955 0.000625193351879716
			16.6884422110553 0.000621598563157022
			16.8241206030151 0.000617819256149232
			16.9597989949749 0.000614036049228162
			17.0954773869347 0.000610416464041919
			17.2311557788945 0.000606992223765701
			17.3668341708543 0.000603715830948204
			17.5025125628141 0.000600538740400225
			17.6381909547739 0.000597412523347884
			17.7738693467337 0.000594288692809641
			17.9095477386935 0.000591118645388633
			18.0452261306533 0.000587854417972267
			18.1809045226131 0.000584474648348987
			18.3165829145729 0.000580999301746488
			18.4522613065327 0.00057745206868276
			18.5879396984925 0.000573856465052813
			18.7236180904523 0.000570236123166978
			18.8592964824121 0.000566614675335586
			18.9949748743719 0.000563015579245985
			19.1306532663317 0.000559454667381942
			19.2663316582915 0.000555912090931088
			19.4020100502513 0.000552358163986355
			19.5376884422111 0.000548763258848339
			19.6733668341709 0.000545097689609975
			19.8090452261307 0.000541331770364195
			19.9447236180905 0.000537435931619257
			20.0804020100503 0.000533382699359208
			20.21608040201 0.000529180571902543
			20.3517587939699 0.000524866103660315
			20.4874371859297 0.000520476547535509
			20.6231155778894 0.00051604921463877
			20.7587939698492 0.000511621299665421
			20.894472361809 0.000507230113726109
			21.0301507537688 0.000502912851516157
			21.1658291457286 0.000498684355989099
			21.3015075376884 0.0004945132532157
			21.4371859296482 0.000490362057462335
			21.572864321608 0.000486193341203034
			21.7085427135678 0.000481969735119492
			21.8442211055276 0.000477653898997232
			21.9798994974874 0.000473208347102627
			22.1155778894472 0.000468603509943932
			22.251256281407 0.000463858275907114
			22.3869346733668 0.000459010421764106
			22.5226130653266 0.000454097724286839
			22.6582914572864 0.000449157931143418
			22.7939698492462 0.000444228877313435
			22.929648241206 0.000439348339568824
			23.0653266331658 0.000434553279774264
			23.2010050251256 0.000429861480370164
			23.3366834170854 0.000425270263804123
			23.4723618090452 0.000420776137616485
			23.608040201005 0.000416375522036105
			23.7437185929648 0.000412064895499498
			23.8793969849246 0.000407840678235516
			24.0150753768844 0.000403699348680675
			24.1507537688442 0.000399639480747283
			24.286432160804 0.000395665818359703
			24.4221105527638 0.000391784124076366
			24.5577889447236 0.000388000247767195
			24.6934673366834 0.000384320039302111
			24.8291457286432 0.000380749232135713
			24.964824120603 0.000377293734345585
			25.1005025125628 0.000373959250282496
			25.2361809045226 0.000370751222362742
			25.3718592964824 0.000367674831068143
			25.5075376884422 0.000364735285984352
			25.643216080402 0.000361937796697021
			25.7788944723618 0.000359287514584139
			25.9145728643216 0.00035678967833519
			26.0502512562814 0.000354449352016672
			26.1859296482412 0.000352266652043909
			26.321608040201 0.000350234593497589
			26.4572864321608 0.000348345754900947
			26.5929648241206 0.000346592598361894
			26.7286432160804 0.000344967702403665
			26.8643216080402 0.000343463558238
			27 0.000342072715284303
		};
		\addplot [semithick, steelblue31119180]
		table {%
			0 2.79857899546685e-20
			0.135678391959799 2.90658693165327e-05
			0.271356783919598 6.66399901953319e-05
			0.407035175879397 0.000110008155984111
			0.542713567839196 0.000156456160030583
			0.678391959798995 0.00020326979568246
			0.814070351758794 0.000247734856287457
			0.949748743718593 0.000287137135193283
			1.08542713567839 0.000318927351914502
			1.22110552763819 0.000342756091946848
			1.35678391959799 0.0003598400511996
			1.49246231155779 0.000371429494921918
			1.62814070351759 0.000378774688362963
			1.76381909547739 0.000383125896771897
			1.89949748743719 0.00038573338539788
			2.03517587939699 0.000387842719835629
			2.17085427135678 0.000390179735964547
			2.30653266331658 0.000392486178465498
			2.44221105527638 0.000394394179087716
			2.57788944723618 0.000395535869580437
			2.71356783919598 0.000395543381692895
			2.84924623115578 0.000394048847174325
			2.98492462311558 0.000390684397773963
			3.12060301507538 0.00038522486629175
			3.25628140703518 0.000378243598243014
			3.39195979899498 0.000370591537280787
			3.52763819095477 0.000363119905771089
			3.66331658291457 0.000356679926079942
			3.79899497487437 0.000352122820573368
			3.93467336683417 0.000350299811617387
			4.07035175879397 0.000352024223162397
			4.20603015075377 0.000357309078978997
			4.34170854271357 0.00036540484914848
			4.47738693467337 0.00037553166015916
			4.61306532663317 0.000386909638499355
			4.74874371859296 0.000398758910657379
			4.88442211055276 0.000410299603121548
			5.02010050251256 0.000420752194073334
			5.15577889447236 0.000429499462590232
			5.29145728643216 0.000436343647780058
			5.42713567839196 0.000441153848916596
			5.56281407035176 0.00044379916527363
			5.69849246231156 0.000444148696124943
			5.83417085427136 0.000442071540744321
			5.96984924623116 0.000437436798405548
			6.10552763819095 0.000430226622778748
			6.24120603015075 0.000421321010611682
			6.37688442211055 0.000412028115097726
			6.51256281407035 0.000403658726267489
			6.64824120603015 0.000397523634151578
			6.78391959798995 0.0003949336287806
			6.91959798994975 0.000397199500185165
			7.05527638190955 0.000405603749555188
			7.19095477386935 0.000420375793573787
			7.32663316582915 0.000440403443746821
			7.46231155778895 0.000464487455914132
			7.59798994974874 0.00049142858591556
			7.73366834170854 0.000520027589590948
			7.86934673366834 0.000549085222780137
			8.00502512562814 0.000577402243693427
			8.14070351758794 0.000603831437383237
			8.27638190954774 0.000627411843005574
			8.41206030150754 0.000647224162910225
			8.54773869346734 0.000662349099446977
			8.68341708542714 0.000671867354965615
			8.81909547738693 0.00067485963181593
			8.95477386934673 0.000670406632347705
			9.09045226130653 0.000657755595169986
			9.22613065326633 0.000638089742905702
			9.36180904522613 0.000613841320122222
			9.49748743718593 0.000587463388419324
			9.63316582914573 0.000561409009396784
			9.76884422110553 0.000538131244654381
			9.90452261306533 0.000520083155791891
			10.0402010050251 0.00050970031150669
			10.1758793969849 0.000508023392358761
			10.3115577889447 0.000513704887740382
			10.4472361809045 0.000525162943650139
			10.5829145728643 0.000540815706086616
			10.7185929648241 0.000559081321048398
			10.8542713567839 0.00057837793453407
			10.9899497487437 0.000597123692542216
			11.1256281407035 0.000613948383862404
			11.2613065326633 0.000628539780971764
			11.3969849246231 0.000640915575288866
			11.5326633165829 0.000651093566593387
			11.6683417085427 0.000659091554665005
			11.8040201005025 0.000664927339283398
			11.9396984924623 0.000668618720228245
			12.0753768844221 0.000670196433730629
			12.2110552763819 0.000669923451197265
			12.3467336683417 0.000668263621252293
			12.4824120603015 0.000665687415982972
			12.6180904522613 0.000662665307476563
			12.7537688442211 0.000659667767820325
			12.8894472361809 0.000657165269101517
			13.0251256281407 0.00065562738295527
			13.1608040201005 0.000655291234702254
			13.2964824120603 0.000655853267779399
			13.4321608040201 0.000656933221509458
			13.5678391959799 0.000658150835215181
			13.7035175879397 0.000659125848219319
			13.8391959798995 0.000659477999844623
			13.9748743718593 0.000658827029413843
			14.1105527638191 0.000656861780584131
			14.2462311557789 0.000653758208585418
			14.3819095477387 0.000649901684374758
			14.5175879396985 0.000645678390145287
			14.6532663316583 0.000641474508090139
			14.7889447236181 0.000637676220402451
			14.9246231155779 0.000634669709275359
			15.0603015075377 0.000632835522947571
			15.1959798994975 0.000632383342383732
			15.3316582914573 0.000633325308021432
			15.4673366834171 0.000635662556481026
			15.6030150753769 0.000639396224382869
			15.7386934673367 0.000644527448347316
			15.8743718592965 0.000651057364994719
			16.0100502512563 0.000658987059750289
			16.1457286432161 0.000668161748019563
			16.2814070351759 0.000677926801405618
			16.4170854271357 0.000687527600993103
			16.5527638190955 0.000696209527866667
			16.6884422110553 0.000703217963110958
			16.8241206030151 0.000707798287810626
			16.9597989949749 0.000709195883050319
			17.0954773869347 0.000706750484619518
			17.2311557788945 0.000700763396155476
			17.3668341708543 0.000692097551202749
			17.5025125628141 0.000681622926549647
			17.6381909547739 0.000670209498984477
			17.7738693467337 0.000658727245295547
			17.9095477386935 0.000648046142271163
			18.0452261306533 0.000639028042166594
			18.1809045226131 0.000632047325254631
			18.3165829145729 0.00062672279023523
			18.4522613065327 0.000622608239544018
			18.5879396984925 0.000619257475616619
			18.7236180904523 0.000616224300888662
			18.8592964824121 0.000613062517795771
			18.9949748743719 0.000609325928773574
			19.1306532663317 0.00060464099783399
			19.2663316582915 0.000598959020465616
			19.4020100502513 0.000592322028176421
			19.5376884422111 0.000584772056608509
			19.6733668341709 0.000576351141403986
			19.8090452261307 0.000567101318204957
			19.9447236180905 0.000557064622653526
			20.0804020100503 0.000546282721579151
			20.21608040201 0.000534791598080622
			20.3517587939699 0.000522622773848258
			20.4874371859297 0.000509807650726275
			20.6231155778894 0.00049637763055889
			20.7587939698492 0.000482364115190319
			20.894472361809 0.000467798506464777
			21.0301507537688 0.000452713564900028
			21.1658291457286 0.000437362665630851
			21.3015075376884 0.000422461812134407
			21.4371859296482 0.000408785261016137
			21.572864321608 0.000397107268881485
			21.7085427135678 0.000388202092335892
			21.8442211055276 0.000382843987984801
			21.9798994974874 0.000381807212433654
			22.1155778894472 0.000385618780973517
			22.251256281407 0.000393254593254053
			22.3869346733668 0.000403090378610138
			22.5226130653266 0.000413500565855221
			22.6582914572864 0.000422859583802754
			22.7939698492462 0.00042954186126619
			22.929648241206 0.000431921827058978
			23.0653266331658 0.000428439984301212
			23.2010050251256 0.000419197325242791
			23.3366834170854 0.0004060375256558
			23.4723618090452 0.000390886786527354
			23.608040201005 0.000375671308844568
			23.7437185929648 0.000362317293594555
			23.8793969849246 0.00035275094176443
			24.0150753768844 0.000348897634632768
			24.1507537688442 0.000351866323772518
			24.286432160804 0.000360417495790149
			24.4221105527638 0.000372891946519762
			24.5577889447236 0.000387630471795455
			24.6934673366834 0.00040297386745133
			24.8291457286432 0.000417262929321485
			24.964824120603 0.00042883845324002
			25.1005025125628 0.000436194725576185
			25.2361809045226 0.000439204051537477
			25.3718592964824 0.000438466511703807
			25.5075376884422 0.000434588767561776
			25.643216080402 0.000428177480597985
			25.7788944723618 0.000419839312299037
			25.9145728643216 0.000410180924151535
			26.0502512562814 0.000399807244538611
			26.1859296482412 0.000389242347367241
			26.321608040201 0.000378896732807843
			26.4572864321608 0.00036917238629349
			26.5929648241206 0.000360471293257255
			26.7286432160804 0.000353195439132213
			26.8643216080402 0.000347746809351436
			27 0.000344527389347998
		};
		\addplot [semithick, forestgreen4416044]
		table {%
			0 5.04745112266392e-05
			0.135678391959799 6.51297814329155e-05
			0.271356783919598 8.15624662209302e-05
			0.407035175879397 9.92436616797931e-05
			0.542713567839196 0.000117644485726487
			0.678391959798995 0.000136236049002036
			0.814070351758794 0.000154489447595552
			0.949748743718593 0.000171875799424015
			1.08542713567839 0.000187893791007809
			1.22110552763819 0.000202409661142156
			1.35678391959799 0.000215551364817657
			1.49246231155779 0.000227452488616109
			1.62814070351759 0.000238246590015478
			1.76381909547739 0.000248067255597562
			1.89949748743719 0.000257047999184579
			2.03517587939699 0.00026532236370258
			2.17085427135678 0.000273014506092295
			2.30653266331658 0.000280231033684686
			2.44221105527638 0.000287076574750245
			2.57788944723618 0.000293655757559463
			2.71356783919598 0.000300073239486665
			2.84924623115578 0.00030643367790617
			2.98492462311558 0.000312841642880812
			3.12060301507538 0.000319379847496748
			3.25628140703518 0.000326008041156456
			3.39195979899498 0.000332643219735473
			3.52763819095477 0.000339202379109338
			3.66331658291457 0.000345602515153587
			3.79899497487437 0.000351760565536097
			3.93467336683417 0.000357593526132405
			4.07035175879397 0.000363022700184956
			4.20603015075377 0.000368060515029356
			4.34170854271357 0.000372806243831292
			4.47738693467337 0.000377362623112276
			4.61306532663317 0.000381832447601482
			4.74874371859296 0.000386318395612761
			4.88442211055276 0.000390923232771456
			5.02010050251256 0.000395749608287588
			5.15577889447236 0.000400856370106339
			5.29145728643216 0.000406189094064757
			5.42713567839196 0.000411675340728834
			5.56281407035176 0.000417242699768394
			5.69849246231156 0.0004228187026456
			5.83417085427136 0.000428330939030275
			5.96984924623116 0.000433706940384582
			6.10552763819095 0.000438880437286571
			6.24120603015075 0.000443834200268611
			6.37688442211055 0.00044857437023893
			6.51256281407035 0.000453107233624905
			6.64824120603015 0.000457439076853916
			6.78391959798995 0.000461576215457171
			6.91959798994975 0.000465524906758219
			7.05527638190955 0.00046929222298786
			7.19095477386935 0.00047291288501583
			7.32663316582915 0.000476456974865869
			7.46231155778895 0.000479996931971982
			7.59798994974874 0.000483605137560517
			7.73366834170854 0.000487353943753988
			7.86934673366834 0.000491315731778741
			8.00502512562814 0.000495562911964953
			8.14070351758794 0.000500134017784148
			8.27638190954774 0.000504946045111865
			8.41206030150754 0.000509888981468976
			8.54773869346734 0.000514852814376354
			8.68341708542714 0.000519727473147213
			8.81909547738693 0.000524402887094766
			8.95477386934673 0.000528769043739885
			9.09045226130653 0.000532727688550949
			9.22613065326633 0.00053631782066077
			9.36180904522613 0.00053966703126207
			9.49748743718593 0.000542904308531433
			9.63316582914573 0.000546158698853105
			9.76884422110553 0.000549559248611331
			9.90452261306533 0.000553235004190356
			10.0402010050251 0.000557314022444189
			10.1758793969849 0.000561850785743445
			10.3115577889447 0.000566773640457541
			10.4472361809045 0.000571998651139438
			10.5829145728643 0.000577441824134439
			10.7185929648241 0.000583019049372524
			10.8542713567839 0.000588646449614316
			10.9899497487437 0.000594240031205118
			11.1256281407035 0.000599716557189822
			11.2613065326633 0.000604997447226197
			11.3969849246231 0.000610005401540548
			11.5326633165829 0.000614663120359182
			11.6683417085427 0.000618893303908408
			11.8040201005025 0.000622618652414531
			11.9396984924623 0.000625761866103858
			12.0753768844221 0.0006282510003075
			12.2110552763819 0.00063010910525918
			12.3467336683417 0.000631441595032811
			12.4824120603015 0.000632356503047049
			12.6180904522613 0.000632961979135871
			12.7537688442211 0.00063336594030261
			12.8894472361809 0.000633676594588906
			13.0251256281407 0.000634001742582768
			13.1608040201005 0.000634392083156854
			13.2964824120603 0.000634765310678631
			13.4321608040201 0.00063502014381811
			13.5678391959799 0.000635055534075946
			13.7035175879397 0.000634770200122148
			13.8391959798995 0.000634062977042049
			13.9748743718593 0.000632832699920982
			14.1105527638191 0.000631004921160638
			14.2462311557789 0.000628693669568747
			14.3819095477387 0.000626094057224691
			14.5175879396985 0.000623401487246156
			14.6532663316583 0.000620811362750828
			14.7889447236181 0.000618519028648734
			14.9246231155779 0.000616720004472882
			15.0603015075377 0.000615603988990188
			15.1959798994975 0.000615189492236823
			15.3316582914573 0.000615297118201852
			15.4673366834171 0.000615736236795783
			15.6030150753769 0.000616316508967429
			15.7386934673367 0.00061684736283496
			15.8743718592965 0.000617138342931867
			16.0100502512563 0.000616998935583979
			16.1457286432161 0.000616298755630851
			16.2814070351759 0.000615099736023694
			16.4170854271357 0.000613502401392907
			16.5527638190955 0.000611607218161225
			16.6884422110553 0.000609514594543725
			16.8241206030151 0.000607325113378465
			16.9597989949749 0.000605139241088182
			17.0954773869347 0.000603047898039222
			17.2311557788945 0.000601044914219528
			17.3668341708543 0.000599067250732332
			17.5025125628141 0.000597051403019577
			17.6381909547739 0.000594933517277241
			17.7738693467337 0.000592650088947266
			17.9095477386935 0.000590137380640954
			18.0452261306533 0.000587332353461534
			18.1809045226131 0.000584207766223699
			18.3165829145729 0.000580792140681297
			18.4522613065327 0.000577118538785726
			18.5879396984925 0.000573220197111368
			18.7236180904523 0.000569130410440266
			18.8592964824121 0.000564882182516158
			18.9949748743719 0.000560508924536407
			19.1306532663317 0.000556044629774988
			19.2663316582915 0.000551528006326407
			19.4020100502513 0.00054699886823073
			19.5376884422111 0.000542497145943344
			19.6733668341709 0.00053806253708899
			19.8090452261307 0.000533735030330718
			19.9447236180905 0.00052955438150093
			20.0804020100503 0.000525557203218341
			20.21608040201 0.000521729583851993
			20.3517587939699 0.000518018030561507
			20.4874371859297 0.000514368002768606
			20.6231155778894 0.000510724843479693
			20.7587939698492 0.000507034012116492
			20.894472361809 0.000503240909893066
			21.0301507537688 0.000499291170854121
			21.1658291457286 0.000495153304655105
			21.3015075376884 0.00049084413331002
			21.4371859296482 0.000486386503325775
			21.572864321608 0.00048180329031311
			21.7085427135678 0.000477117369882762
			21.8442211055276 0.000472351588541642
			21.9798994974874 0.000467528821900487
			22.1155778894472 0.000462668045656756
			22.251256281407 0.000457763846497983
			22.3869346733668 0.000452801352366805
			22.5226130653266 0.000447765691205859
			22.6582914572864 0.000442641961853951
			22.7939698492462 0.000437415292253718
			22.929648241206 0.000432070781243965
			23.0653266331658 0.000426595390308648
			23.2010050251256 0.000421021686634049
			23.3366834170854 0.000415430084103718
			23.4723618090452 0.000409903324907646
			23.608040201005 0.000404524151235819
			23.7437185929648 0.000399375305278227
			23.8793969849246 0.000394539441913366
			24.0150753768844 0.000390099332435057
			24.1507537688442 0.00038610611227341
			24.286432160804 0.00038252052036114
			24.4221105527638 0.000379286968382075
			24.5577889447236 0.000376349984435365
			24.6934673366834 0.000373654038412496
			24.8291457286432 0.00037114360020496
			24.964824120603 0.000368763168808073
			25.1005025125628 0.000366461084922776
			25.2361809045226 0.000364220817573369
			25.3718592964824 0.000362044287612662
			25.5075376884422 0.000359933619620278
			25.643216080402 0.000357890879968181
			25.7788944723618 0.000355918193235993
			25.9145728643216 0.000354017654899508
			26.0502512562814 0.000352191214915365
			26.1859296482412 0.00035043410025537
			26.321608040201 0.000348731904523447
			26.4572864321608 0.000347069668350741
			26.5929648241206 0.000345432286849245
			26.7286432160804 0.000343804713338614
			26.8643216080402 0.00034217193024233
			27 0.000340518861776218
		};
	\end{axis}

\end{tikzpicture}

%% file: figures/TableHybrid.tex
\begin{table}
\begin{center}
    \begin{tabular}{|l|l|c|c|}
    \hline  
    \multicolumn{2}{|c|}{} & \ac{LBFGS} & Hybrid \ac{ODPS} \\
    \hline
    \multirow{3}{*}{\scriptsize{PSNR} $\uparrow$} & \scriptsize{Bones} &  
 32.00 &
 \textbf{32.16} \\
    & \scriptsize{Tissues}  & 
 21.53 &
 \textbf{25.39} \\
 & \scriptsize{Iodine} & 25.54 & \textbf{29.79} \\
    \hline 
    \multirow{3}{*}{\scriptsize{SSIM} $\uparrow$} & \scriptsize{Bones} & 
  0.72 &
 \textbf{0.74} \\
    & \scriptsize{Tissues}  &
  0.37 &
 \textbf{0.76} \\
 & \scriptsize{Iodine} & 
 0.56 & 
 \textbf{0.65} \\
    \hline 
    \multirow{3}{*}{\scriptsize{LPIPS} $\downarrow$} & \scriptsize{Bones} &   
     0.21  &
 \textbf{0.09}   \\
    & \scriptsize{Tissues}  & 
 0.60 &
 \textbf{0.13}\\
 & \scriptsize{Iodine} &
 0.53 &
 \textbf{0.44}\\
    \hline 
    \end{tabular}
    \caption{Experiment 2 (Hybridization with iodine)---Averaged metrics computed over the test dataset.}
    \label{tab:TableHyb}
\end{center}
\end{table}

%% file: figures/BoxPlotsHyb.tex
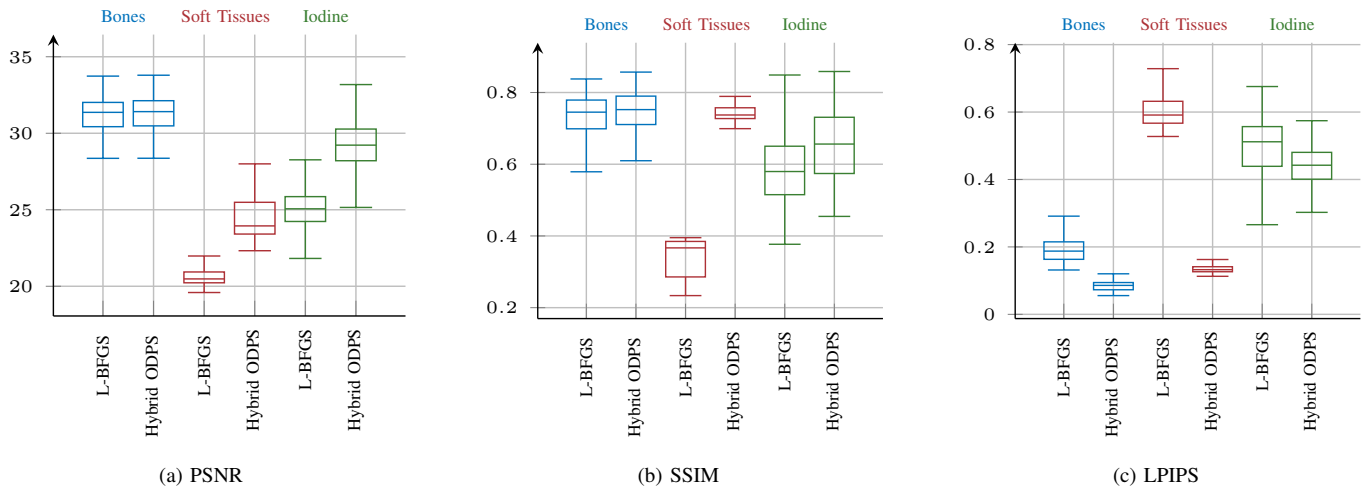
\begin{figure*}
	\centering
	\begin{subfigure}[b]{0.30\textwidth}
		\begin{adjustbox}{width=\textwidth}    
			\begin{tikzpicture}
				\pgfplotstableread[col sep=comma]{csvfiles/PSNR_Hybrid_0.csv}\csvdata
				\pgfplotstabletranspose\datatransposed{\csvdata}
				\pgfplotstableread[col sep=comma]{csvfiles/PSNR_Hybrid_1.csv}\csvdatabis
				\pgfplotstabletranspose\datatransposedbis{\csvdatabis}
				\pgfplotstableread[col sep=comma]{csvfiles/PSNR_Hybrid_2.csv}\csvdataiod
				\pgfplotstabletranspose\datatransposediod{\csvdataiod}
				
				\begin{axis}[
					height=0.8\textwidth,
					boxplot/draw direction = y,
					x axis line style = {opacity=1},
					axis x line* = bottom,
					axis y line = left,
					enlarge y limits,
					yticklabel style={font=\tiny},
					ymajorgrids,
					xtick = {1, 2, 3, 4, 5, 6},
					xticklabel style = {align=center, font=\tiny, rotate=90},
					xticklabels = {L-BFGS , Hybrid ODPS , L-BFGS , Hybrid ODPS, L-BFGS, Hybrid ODPS},
					grid=major,
					name=border	
					]
					\foreach \n in {1,...,2} {
						\addplot+[boxplot, draw=RoyalBlue] table[y index=\n] {\datatransposed};
					}
					\foreach \n in {1,...,2} {        
						\addplot+[boxplot, draw=Maroon] table[y index=\n] {\datatransposedbis};
					}
					\foreach \n in {1,...,2} {        
						\addplot+[boxplot, draw=OliveGreen] table[y index=\n] {\datatransposediod};
					}
				\end{axis}
				\node[above left, xshift = -0.7cm] at (border.north) {\textcolor{RoyalBlue}{\tiny{Bones}}};
				\node[above right, xshift = -0.6cm] at (border.north) {\textcolor{Maroon}{\tiny{Soft Tissues}}};
				\node[above right, xshift = 0.6cm] at (border.north) {\textcolor{OliveGreen}{\tiny{Iodine}}};
			\end{tikzpicture}
		\end{adjustbox}
		\caption{PSNR}\label{subfig:psnr_hyb}
	\end{subfigure}
	\hfill
	\begin{subfigure}[b]{0.30\textwidth}
		\begin{adjustbox}{width=\textwidth}    
			\begin{tikzpicture}
				\pgfplotstableread[col sep=comma]{csvfiles/SSIM_Hybrid_0.csv}\csvdata
				\pgfplotstabletranspose\datatransposed{\csvdata} 
				\pgfplotstableread[col sep=comma]{csvfiles/SSIM_Hybrid_1.csv}\csvdatabis
				\pgfplotstabletranspose\datatransposedbis{\csvdatabis} 
				\pgfplotstableread[col sep=comma]{csvfiles/SSIM_Hybrid_2.csv}\csvdataiod
				\pgfplotstabletranspose\datatransposediod{\csvdataiod} 
				\begin{axis}[
					height=0.8\textwidth,
					boxplot/draw direction = y,
					x axis line style = {opacity=1},
					axis x line* = bottom,
					axis y line = left,
					enlarge y limits,
					yticklabel style={font=\tiny},
					ymajorgrids,
					xtick = {1, 2, 3, 4, 5, 6},
					xticklabel style = {align=center, font=\tiny, rotate=90},
					xticklabels = {L-BFGS , Hybrid ODPS , L-BFGS , Hybrid ODPS, L-BFGS, Hybrid ODPS},
					grid=major,
					name=border
					]
					\foreach \n in {1,...,2} {
						\addplot+[boxplot, draw=RoyalBlue] table[y index=\n] {\datatransposed};
					}
					\foreach \n in {1,...,2} {        
						\addplot+[boxplot, draw=Maroon] table[y index=\n] {\datatransposedbis};
					}
					\foreach \n in {1,...,2} {        
						\addplot+[boxplot, draw=OliveGreen] table[y index=\n] {\datatransposediod};
					}
				\end{axis}
				\node[above left, xshift = -0.7cm] at (border.north) {\textcolor{RoyalBlue}{\tiny{Bones}}};
				\node[above right, xshift = -0.6cm] at (border.north) {\textcolor{Maroon}{\tiny{Soft Tissues}}};
				\node[above right, xshift = 0.6cm] at (border.north) {\textcolor{OliveGreen}{\tiny{Iodine}}};
			\end{tikzpicture}
		\end{adjustbox}
		\caption{SSIM}\label{subfig:ssim_hyb}
	\end{subfigure}
	\hfill
	\begin{subfigure}[b]{0.30\textwidth}
		\begin{adjustbox}{width=\textwidth}    
			\begin{tikzpicture}
				\pgfplotstableread[col sep=comma]{csvfiles/LPIPS_Hybrid_0.csv}\csvdata
				\pgfplotstabletranspose\datatransposed{\csvdata} 
				\pgfplotstableread[col sep=comma]{csvfiles/LPIPS_Hybrid_1.csv}\csvdatabis
				\pgfplotstabletranspose\datatransposedbis{\csvdatabis} 
				\pgfplotstableread[col sep=comma]{csvfiles/LPIPS_Hybrid_2.csv}\csvdataiod
				\pgfplotstabletranspose\datatransposediod{\csvdataiod} 
				\begin{axis}[
					height=0.8\textwidth,
					boxplot/draw direction = y,
					x axis line style = {opacity=1},
					axis x line* = bottom,
					axis y line = left,
					enlarge y limits,
					yticklabel style={font=\tiny},
					ymajorgrids,
					xtick = {1, 2, 3, 4, 5, 6},
					xticklabel style = {align=center, font=\tiny, rotate=90},
					xticklabels = {L-BFGS , Hybrid ODPS , L-BFGS , Hybrid ODPS, L-BFGS, Hybrid ODPS},
					grid=major,
					name=border
					]
					\foreach \n in {1,...,2} {
						\addplot+[boxplot, draw=RoyalBlue] table[y index=\n] {\datatransposed};
					}
					\foreach \n in {1,...,2} {        
						\addplot+[boxplot, draw=Maroon] table[y index=\n] {\datatransposedbis};
					}
					\foreach \n in {1,...,2} {        
						\addplot+[boxplot, draw=OliveGreen] table[y index=\n] {\datatransposediod};
					}
				\end{axis}
				\node[above left, xshift = -0.7cm] at (border.north) {\textcolor{RoyalBlue}{\tiny{Bones}}};
				\node[above right, xshift = -0.6cm] at (border.north) {\textcolor{Maroon}{\tiny{Soft Tissues}}};
				\node[above right, xshift = 0.7cm] at (border.north) {\textcolor{OliveGreen}{\tiny{Iodine}}};
			\end{tikzpicture}
		\end{adjustbox}
		\caption{LPIPS}\label{subfig:lpips_hyb}
	\end{subfigure}
	\caption{Experiment 2---Boxplots of \subref{subfig:psnr_hyb} PSNR, \subref{subfig:ssim_hyb} SSIM 
		and \subref{subfig:lpips_hyb} LPIPS 
		computed over the test dataset. 
	 }\label{fig:boxplot_hyb}
\end{figure*}

%% file: content/discussion.tex
\section{Discussion}\label{sec:discussion}

The one-step methods generally give better results than the two-step methods, as already mentioned in previous works on this topic \cite{mechlem2017joint, Mory2018}. We believe that combining a strong prior learned using \ac{DM} with the one-step forward model \eqref{eq:onestep} is crucial to obtain satisfactory results. As we showed, relying only on \ac{DM} is not sufficient, as \ac{ImTDPS} and \ac{ProjTDPS} suffer from either crosstalk (for the former) or slight over-smoothing (for the latter). These results contrast with our previous work \cite{vazia2024material}, where the performance of \ac{ImTDPS} was comparable to that of \ac{ODPS}. This is because we used a monochromatic model for each energy bin in \cite{vazia2024material}, while in this work we used a realistic polychromatic model.

In addition, we showed that priors learned on material images are more important and impactful than priors learned on spectral images. Indeed, \ac{ProjTDPS} (resp. U-Net-I) delivers better results than \ac{ImTDPS} (resp. U-Net-P). As noted in \cite{abascal}, it is easier to learn a material image prior than a material projection prior (as U-Net-P does). Furthermore, in the case of two-step methods, a learned prior for \ac{MD} is better than a learned prior for image reconstruction, as the U-Net methods produce better results than \ac{ImTDPS}. 

Finally, we worked on \ac{2D} slices instead of full \ac{3D} volumes, as \ac{DM} are computationally expensive. Even on \ac{2D} slices, \ac{DPS} methods are time-consuming. Table~\ref{tab:TableLowCount} (full-view setting) suggests that \ac{ODPS} requires more computational time than other methods. However, this gap decreases in Table~\ref{tab:TableSparseView} (sparse-view setting). We believe that a smaller sinogram size decreases the computational time of the log-likelihood gradient. The \ac{ImTDPS} method is not impacted by this change, as it operates on the image domain, which is the same for both sparse and full-view settings. 

In addition, we performed all methods sequentially, one slice at a time. We believe that by leveraging parallelization, \ac{MD} on multiple slices could be performed simultaneously, significantly decreasing the computational time at the expense of higher memory cost. Methods such as wavelet \acp{DM} \cite{guth2022wavelet,friedrich2024wdm,phung2025joint2,depaepe2025adaptive}, cascaded \acp{DM} \cite{ho2022cascaded}, diffusion Schrödinger bridges \cite{de2021diffusion}, or \acp{DDIM} \cite{song2020denoising}, could also be considered to accelerate the methods or even work directly on \ac{3D} volumes. 

To our knowledge, the proposed Hybrid-\ac{ODPS} method is the first to combine a \ac{DM}-based learned prior with conventional regularization for out-of-database channels during inference. \citeauthor{barbano_steerable_2025}~\cite{barbano_steerable_2025} propose training a model on synthetic data and then adapting it during inference to solve inverse problems in cases where few high-quality data are available. As Hybrid-\ac{ODPS} is agnostic to the choice of material, we believe it can be applied to any scenario with materials (other than contrast agents) that are not present in the training data. 

For contrast-enhanced spectral \ac{CT}, we believe the large amount of data available for bones and soft tissues justifies the proposed Hybrid-\ac{ODPS}. The choices made to regularize the iodine image, however, need to be further evaluated. In particular, the Huber penalty used treats each material channel independently. Hybrid-\ac{ODPS} also allows for regularizations that take into account joint information (e.g., inner-product regularization \cite{innerproductreg}). Finding the optimal regularizers and weights for each specific task remains a challenge, and this particular problem requires further investigation.

%% file: content/conclusion.tex
\section{Conclusion}\label{sec:conclusion}

This study explored multiple \ac{DPS}-based approaches for \ac{MD} in \ac{PCCT}, evaluating their performance against traditional variational and \ac{DL}-based techniques. The results demonstrate that \ac{ODPS} provides the best overall performance, producing high-quality material maps with minimal noise and artifact contamination. Furthermore, the Hybrid-\ac{ODPS} method successfully extends the applicability of \ac{DPS} to materials absent from the training set, providing a practical solution for real-world clinical and research scenarios.

The findings highlight the potential of diffusion models as powerful regularizers for inverse problems in medical imaging. Given these promising results, future research should focus on improving computational efficiency and exploring the integration of \ac{DPS} with real-time imaging applications. Additionally, further validation on clinical datasets will be necessary to confirm the generalizability of these methods under diverse imaging conditions.